\documentclass[aps,graphicx,eqsecnum]{revtex4}
\usepackage{graphicx}
\usepackage{amsmath,amssymb}

\def\f{{\rm p}}
\def\vd{{5}}
\def \vu {{(5)}\!}
\def \D {\vec\nabla}
\def \curl {\mbox{curl}\,}
\def \ep {\varepsilon}
\def \cu{ \rho_{{\cal E}}}
\def \cq{ q^{{\cal E}}}
\def \cp{ \pi^{{\cal E}}}
\def \bcq { \bar{q}^{{\cal E}}}
\def \bcp { \bar{\pi}^{{\cal E}}}
\def\be{\begin{equation}}
\def\ee{\end{equation}}
\def\bea{\begin{eqnarray}}
\def\eea{\end{eqnarray}}

\begin{document}

\title{BRANE-WORLD GRAVITY}

\author{Roy Maartens}

\address{Institute of Cosmology \& Gravitation, Portsmouth University,
Portsmouth~PO1~2EG, Britain}


\begin{abstract}

The observable universe could be a 1+3-surface (the ``brane")
embedded in a 1+3+$d$-dimensional spacetime (the ``bulk"), with
Standard Model particles and fields trapped on the brane while
gravity is free to access the bulk. At least one of the $d$ extra
spatial dimensions could be very large relative to the Planck
scale, which lowers the fundamental gravity scale, possibly even
down to the electroweak ($\sim$TeV) level. This revolutionary
picture arises in the framework of recent developments in
M~theory. The 1+10-dimensional M~theory encompasses the known
1+9-dimensional superstring theories, and is widely considered to
be a promising potential route to quantum gravity. General
relativity cannot describe gravity at high enough energies and
must be replaced by a quantum gravity theory, picking up
significant corrections as the fundamental energy scale is
approached. At low energies, gravity is localized at the brane and
general relativity is recovered, but at high energies gravity
``leaks" into the bulk, behaving in a truly higher-dimensional
way. This introduces significant changes to gravitational dynamics
and perturbations, with interesting and potentially testable
implications for high-energy astrophysics, black holes and
cosmology. Brane-world models offer a phenomenological way to test
some of the novel predictions and corrections to general
relativity that are implied by M~theory. This review discusses the
geometry, dynamics and perturbations of simple brane-world models
for cosmology and astrophysics, mainly focusing on warped
5-dimensional brane-worlds based on the Randall-Sundrum models.

\end{abstract}
\maketitle

\section{Introduction}

At high enough energies, Einstein's theory of general relativity
breaks down, and will be superceded by a quantum gravity theory.
The classical singularities predicted by general relativity in
gravitational collapse and in the hot big bang will be removed by
quantum gravity. But even below the fundamental energy scale that
marks the transition to quantum gravity, significant corrections
to general relativity will arise. These corrections could have a
major impact on the behaviour of gravitational collapse, black
holes and the early universe, and they could leave a trace--a
``smoking gun"--in various observations and experiments. Thus it
is important to estimate these corrections and develop tests for
detecting them or ruling them out. In this way, quantum gravity
can begin to be subject to testing by astrophysical and
cosmological observations.

Developing a quantum theory of gravity and a unified theory of all
the forces and particles of nature are the two main goals of
current work in fundamental physics. There is as yet no generally
accepted (pre-)quantum gravity theory. Two of the main contenders
are M~theory~\cite{mtheory} and quantum geometry (loop quantum
gravity)~\cite{loop}. It is important to explore the astrophysical
and cosmological predictions of both these approaches. This review
considers only models that arise within the framework of M~theory,
and mainly the 5-dimensional warped brane-worlds.

\subsection{Heuristics of higher-dimensional gravity}

One of the fundamental aspects of string theory is the need for
extra spatial dimensions. This revives the original
higher-dimensional ideas of Kaluza and Klein in the 1920's, but in
a new context of quantum gravity. An important consequence of
extra dimensions is that the 4-dimensional Planck scale
$M_\f\equiv M_{4}$ is no longer the fundamental scale, which is
$M_{4+d}$, where $d$ is the number of extra dimensions. This can
be seen from the modification of the gravitational potential. For
an Einstein-Hilbert gravitational action we have,
 \bea
&& S_{\rm gravity}= {1\over 2\kappa_{4+d}^2}\int {\rm d}^4x\,{\rm
d}^dy\,\sqrt{-^{(4+d)\!}g}\left[\,{}^{(4+d)\!}R- 2\Lambda_{4+d}
\right]\,, \\ && ^{(4+d)\!}G_{AB}\equiv
^{(4+d)\!}R_{AB}-{1\over2}\,^{(4+d)\!}R\,\, ^{(4+d)\!}g_{AB}=
-\Lambda_{4+d}{}^{(4+d)\!}g_{AB}+
\kappa_{4+d}^2\,\,^{(4+d)\!}T_{AB}\,,\label{defe}
 \eea
where $X^A=(x^\mu,y^1, \cdots, y^d)$ and $\kappa_{4+d}^2$ is the
gravitational coupling constant,
 \be
\kappa_{4+d}^2=8\pi G_{4+d}={8\pi\over M_{4+d}^{2+d}}\,.
 \ee
The static weak field limit of the field equations leads to the
$4+d$-dimensional Poisson equation, whose solution is the
gravitational potential,
 \be\label{v}
V(r) \propto {\kappa_{4+d}^2\over r^{1+d}}\,.
 \ee
If the length scale of the extra dimensions is $L$, then on scales
$r\lesssim L$, the potential is $4+d$-dimensional, $V\sim
r^{-(1+d)}$. By contrast, on scales large relative to $L$, where
the extra dimensions do not contribute to variations in the
potential, $V$ behaves like a 4-dimensional potential, i.e.,
$r\sim L$ in the $d$ extra dimensions, and $V \sim L^{-d}r^{-1}$.
This means that the usual Planck scale becomes an effective
coupling constant, describing gravity on scales much larger than
the extra dimensions, and related to the fundamental scale via the
volume of the extra dimensions:
 \be
M_\f^2 \sim M_{4+d}^{2+d}\,L^d\,.
 \ee
If the extra-dimensional volume is Planck scale, i.e. $L\sim
M_\f^{-1}$, then $M_{4+d}\sim M_\f$. But if the extra-dimensional
volume is significantly above Planck-scale, then the true
fundamental scale $M_{4+d}$ can be much less than the effective
scale $M_\f \sim 10^{19}~{\rm GeV}$. In this case, we understand
the weakness of gravity as due to the fact that it ``spreads" into
extra dimensions and only a part of it is felt in 4 dimensions.

A lower limit on $M_{4+d}$ is given by null results in table-top
experiments to test for deviations from Newton's law in 4
dimensions, $V\propto r^{-1}$. These experiments
currently~\cite{exp} probe sub-millimetre scales, so that
 \be \label{tt}
L \lesssim 10^{-1}~{\rm mm} \sim  (10^{-15}~{\rm TeV})^{-1}
~\Rightarrow~ M_{4+d}\gtrsim 10^{(32-15d)/(d+2)}~{\rm TeV}\,.
 \ee
Stronger bounds for brane-worlds with compact flat extra
dimensions can be derived from null results in particle
accelerators and in high-energy
astrophysics~\cite{cav,cheung,hanraf}.

\subsection{Brane-worlds and M~theory}

String theory thus incorporates the possibility that the
fundamental scale is much less than the Planck scale felt in 4
dimensions. There are five distinct 1+9-dimensional superstring
theories, all giving quantum theories of gravity. Discoveries in
the mid-90's of duality transformations that relate these
superstring theories and the 1+10-dimensional supergravity theory,
led to the conjecture that all of these theories arise as
different limits of a single theory, which has come to be known as
M~theory. The 11th dimension in M~theory is related to the string
coupling strength; the size of this dimension grows as the
coupling becomes strong. At low energies, M~theory can be
approximated by 1+10-dimensional supergravity.

It was also discovered that p-branes, which are extended objects
of higher dimension than strings (1-branes), play a fundamental
role in the theory. In the weak coupling limit, p-branes ($p>1$)
become infinitely heavy, so that they do not appear in the
perturbative theory. Of particular importance among p-branes are
the D-branes, on which open strings can end. Roughly speaking,
open strings, which describe the non-gravitational sector, are
attached at their endpoints to branes, while the closed strings of
the gravitational sector can move freely in the bulk. Classically,
this is realised via the localization of matter and radiation
fields on the brane, with gravity propagating in the bulk (see
Fig.~1).

\begin{figure}[!bth]\label{brane}
\begin{center}
\includegraphics[height=3in,width=4in]{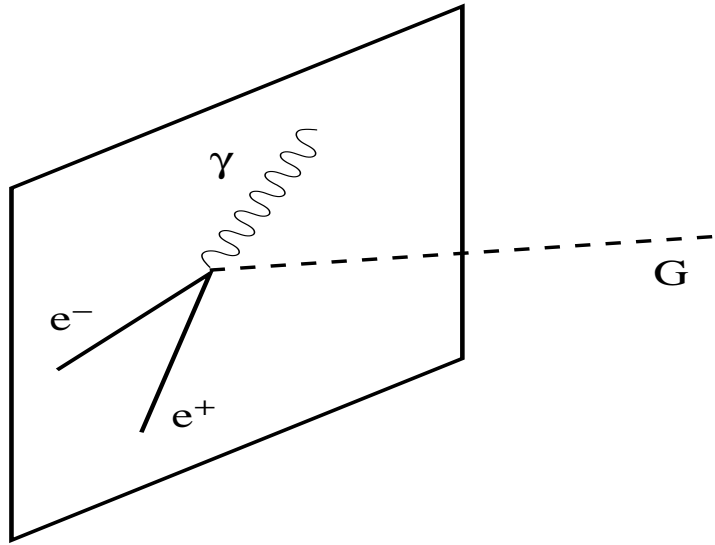}
\caption{Schematic of confinement of matter to the brane, while
gravity propagates in the bulk (from~\cite{cav}).}
\end{center}
\end{figure}

In the Horava-Witten solution~\cite{hv}, gauge fields of the
standard model are confined on two 1+9-branes located at the end
points of an $S^1/Z_2$ orbifold, i.e., a circle folded on itself
across a diameter. The 6 extra dimensions on the branes are
compactified on a very small scale, close to the fundamental
scale, and their effect on the dynamics is felt through ``moduli"
fields, i.e. 5D scalar fields. A 5D realization of the
Horava-Witten theory and the corresponding brane-world cosmology
is given in~\cite{low}.

These solutions can be thought of as effectively 5-dimensional,
with an extra dimension that can be large relative to the
fundamental scale. They provide the basis for the Randall-Sundrum
2-brane models of 5-dimensional gravity~\cite{rs1} (see Fig.~2).
The single-brane Randall-Sundrum models~\cite{rs2} with infinite
extra dimension arise when the orbifold radius tends to infinity.
The RS models are not the only phenomenological realizations of
M~theory ideas. They were preceded by the
Arkani-Hamed-Dimopoulos-Dvali (ADD)~\cite{add} brane-world models,
which put forward the idea that a large volume for the compact
extra dimensions would lower the fundamental Planck scale,
 \be\label{scales}
M_{\rm ew}\sim 1~{\rm TeV}\,\lesssim M_{4+d} \leq M_\f \sim
10^{16}~{\rm TeV}\,,
 \ee
where $M_{\rm ew}$ is the electroweak scale. If $M_{4+d}$ is close
to the lower limit in Eq.~(\ref{scales}), then this would address
the long-standing ``hierarchy" problem, i.e. why there is such a
large gap between $M_{\rm ew}$ and $M_\f$.

In the ADD models, more than one extra dimension is required for
agreement with experiments, and there is ``democracy" amongst the
equivalent extra dimensions, which are typically flat. By
contrast, the RS models have a ``preferred" extra dimension, with
other extra dimensions treated as ignorable (i.e., stabilized
except at energies near the fundamental scale). Furthermore, this
extra dimension is curved or ``warped" rather than flat: the bulk
is a portion of anti de Sitter (AdS$_5$) spacetime. As in the
Horava-Witten solutions, the RS branes are $Z_2$-symmetric (mirror
symmetry), and have a tension, which serves to counter the
influence of the negative bulk cosmological constant on the brane.
This also means that the self-gravity of the branes is
incorporated in  the RS models. The novel feature of the RS models
compared to previous higher-dimensional models is that the
observable 3 dimensions are protected from the large extra
dimension (at low energies) by curvature rather than
straightforward compactification.

The RS brane-worlds and their generalizations (to include matter
on the brane, scalar fields in the bulk, etc.) provide
phenomenological models that reflect at least some of the features
of M~theory, and that bring exciting new geometric and particle
physics ideas into play. The RS2 models also provide a framework
for exploring holographic ideas that have emerged in M~theory.
Roughly speaking, holography suggests that higher-dimensional
gravitational dynamics may be determined from knowledge of the
fields on a lower-dimensional boundary. The AdS/CFT correspondence
is an example, in which the classical dynamics of the
higher-dimensional gravitational field are equivalent to the
quantum dynamics of a conformal field theory (CFT) on the
boundary. The RS2 model with its AdS$_5$ metric satisfies this
correspondence to lowest perturbative order~\cite{acft} (see
also~\cite{acftcosmo} for the AdS/CFT correspondence in a
cosmological context).

\begin{figure}[!bth]\label{rs2}
\begin{center}
\includegraphics[height=3in,width=4in]{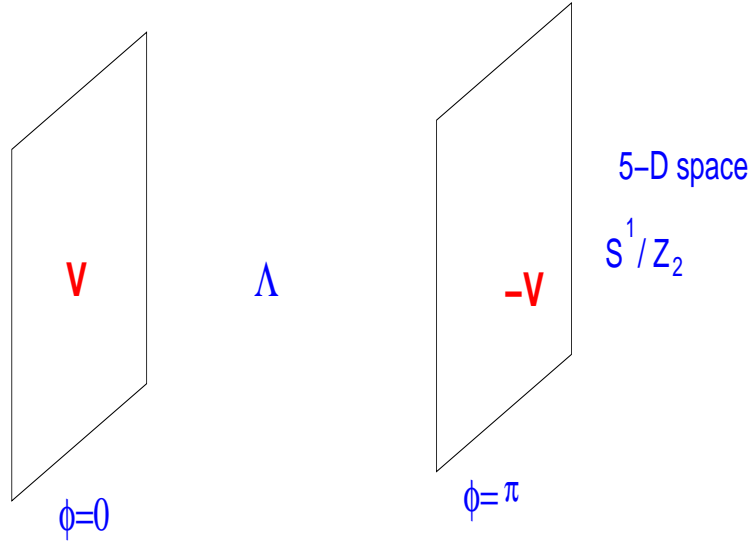}
\caption{The RS 2-brane model
 (from~\cite{cheung}).}
\end{center}
\end{figure}

In this review, I focus on RS brane-worlds (mainly RS 1-brane) and
their generalizations, with the emphasis on geometry and
gravitational dynamics (see~\cite{m2,rev,lan} for previous reviews
with a broadly similar approach). Other recent reviews focus on
string-theory aspects, e.g.~\cite{que}, or on particle physics
aspects, e.g.~\cite{r,cav}. Before turning to a more detailed
analysis of RS brane-worlds, I discuss the notion of Kaluza-Klein
(KK) modes of the graviton.

\subsection{Heuristics of KK modes}

The dilution of gravity via extra dimensions not only weakens
gravity on the brane, it also extends the range of graviton modes
felt on the brane beyond the massless mode of 4-dimensional
gravity. For simplicity, consider a flat brane with one flat extra
dimension, compactified through the identification
$y\leftrightarrow y+2\pi n L$, where $n=0,1,2,\cdots$. The
perturbative 5D graviton amplitude can be Fourier expanded as
 \be
f(x^a,y)=\sum_n e^{iny/L}\,f_n(x^a)\,,
 \ee
where $f_n$ are the amplitudes of the KK modes, i.e. the effective
4D modes of the the 5D graviton. To see that these KK modes are
massive from the brane viewpoint, we start from the 5D wave
equation that the massless 5D field $f$ satisfies (in a suitable
gauge):
 \be
^\vu\Box f=0 ~\Rightarrow~ \Box f+\partial_y^2 f=0\,.
 \ee
It follows that the KK modes satisfy a 4D Klein-Gordon equation
with an effective 4D mass, $m_n$,
 \be
\Box f_n=m_n^2\,f_n\,,~~m_n={n\over L}\,.
 \ee
The massless mode, $f_0$, is the usual 4D graviton mode. But there
is a tower of massive modes, $L^{-1},2L^{-1},\cdots$, which
imprint the effect of the 5D gravitational field on the 4D brane.
Compactness of the extra dimension leads to discreteness of the
spectrum. For an infinite extra dimension, $L\to\infty$, the
separation between the modes disappears and the tower forms a
continuous spectrum. In this case, the coupling of the KK modes to
matter must be very weak in order to avoid exciting the lightest
massive modes with $m\gtrsim 0$.

From a geometric viewpoint, the KK modes can also be understood
via the fact that the projection of the null graviton 5-momentum
$^\vu p_A$ onto the brane is timelike. If the unit normal to the
brane is $n_A$, then the induced metric on the brane is
 \be
g_{AB}= \, ^\vu g_{AB}-n_An_B\,,~  ^\vu g_{AB}n^An^B=1\,,~
g_{AB}n^B=0 \,,
 \ee
and the 5-momentum may be decomposed as
 \be
^\vu p_A=mn_A+ p_A\,,~ p_An^A=0\,,~m=\, ^\vu p_A\, n^A \,,
 \ee
where $p_A=g_{AB}\,^\vu p^B$ is the projection along the brane,
depending on the orientation of the 5-momentum relative to the
brane. The effective 4-momentum of the 5D graviton is thus $p_A$.
Expanding $^\vu g_{AB}{}{}\,^\vu p^A\,^\vu p^B=0$, we find that
 \be
g_{AB}p^Ap^B=-m^2\,.
 \ee
It follows that the 5D graviton has an effective mass $m$ on the
brane. The usual 4D graviton corresponds to the zero mode, $m=0$,
when $^\vu p_A$ is tangent to the brane.

The extra dimensions lead to new scalar and vector degrees of
freedom on the brane. In 5D, the spin-2 graviton is represented by
a metric perturbation $^\vu h_{AB}$ that is transverse traceless:
 \be
^\vu h^A{}_{A}=0=\partial_B\,^\vu h_{A}{}^{B}\,.
 \ee
In a suitable gauge, $^\vu h_{AB}$ contains a 3D transverse
traceless perturbation $h_{ij}$, a 3D transverse vector
perturbation $\Sigma_i$ and a scalar perturbation $\beta$, which
each satisfy the 5D wave equation~\cite{durkoc}:
 \bea \label{5dg}
&& ^\vu h_{AB}~\longrightarrow~ h_{ij}\,,~\Sigma_i\,,~\beta\,,\\
&& h^i{}_i=0=\partial_j h^{ij}\,,~ \partial_i \Sigma^i=0\,,\\
 &&(\Box+\partial_y^2)\left( \begin{array}{c} \beta\\ \Sigma_i \\
h_{ij} \end{array} \right)=0\,.
 \eea
The other components of $^\vu h_{AB}$ are determined via
constraints once these wave equations are solved. The 5 degrees of
freedom (polarizations) in the 5D graviton are thus split into 2
($h_{ij}$) + 2 ($\Sigma_i$) +1 ($\beta$) degrees of freedom in 4D.
On the brane, the 5D graviton field is felt as
\begin{itemize}
\item a 4D spin-2 graviton $h_{ij}$ (2 polarizations) \item a 4D
spin-1 gravi-vector (gravi-photon) $\Sigma_i$ (2 polarizations)
\item a 4D spin-0 gravi-scalar $\beta$.
\end{itemize}
The massive modes of the 5D graviton are represented via massive
modes in all 3 of these fields on the brane. The standard 4D
graviton corresponds to the massless zero-mode of $h_{ij}$.

In the general case of $d$ extra dimensions, the number of degrees
of freedom in the graviton follows from the irreducible tensor
representations of the isometry group as ${1\over2}(d+1)(d+4)$.

\section{Randall-Sundrum brane-worlds}

RS brane-worlds do not rely on compactification to localize
gravity at the brane, but on the curvature of the bulk (sometimes
called ``warped compactification"). What prevents gravity from
`leaking' into the extra dimension at low energies is a negative
bulk cosmological constant,
 \be
 \Lambda_\vd=-{6\over \ell^2}=-6\mu^2,
 \ee
where $\ell$ is the curvature radius of AdS$_5$ and $\mu$ is the
corresponding energy scale. The curvature radius determines the
magnitude of the Riemann tensor:
 \be
^\vu R_{ABCD}=-{1\over \ell^2}\left[\,^\vu g_{AC} \,^\vu g_{BD} -
\,^\vu g_{AD} \,^\vu g_{BC} \right].
 \ee
The bulk cosmological constant acts to ``squeeze" the
gravitational field closer to the brane. We can see this clearly
in Gaussian normal coordinates $X^A=(x^\mu,y)$ based on the brane
at $y=0$, for which the AdS$_5$ metric takes the form
 \be
^\vu ds^2=e^{-2|y|/\ell} \eta_{\mu\nu}dx^\mu dx^\nu + dy^2\,,
 \ee
with $\eta_{\mu\nu}$ the Minkowski metric. The exponential warp
factor reflects the confining role of the bulk cosmological
constant. The $Z_2$-symmetry about the brane at $y=0$ is
incorporated via the $|y|$ term. In the bulk, this metric is a
solution of the 5D Einstein equations,
 \be\label{rsefe}
^\vu G_{AB}=- \Lambda_\vd\,\, ^\vu g_{AB}\,,
 \ee
i.e., $^\vu T_{AB}=0$ in Eq.~(\ref{defe}). The brane is a flat
Minkowski spacetime,
$g_{AB}(x^\alpha,0)=\eta_{\mu\nu}\delta^\mu{}_A \delta^\nu{}_B$,
with self-gravity in the form of brane tension. One can also use
Poincare coordinates, which bring the metric into manifestly
conformally flat form:
 \be\label{poinc}
^\vu ds^2={\ell^2\over z^2}\left[ \eta_{\mu\nu}dx^\mu dx^\nu +
dz^2\right]\,,
 \ee
where $z=\ell e^{y/\ell}$.

The two RS models are distinguished as follows:

\begin{itemize}

\item[]{\bf RS 2-BRANE:}

There are two branes in this model~\cite{rs1}, at $y=0$ and $y=L$,
with $Z_2$-symmetry identifications
 \be
y \leftrightarrow -y\,,~~ y+L \leftrightarrow L-y\,.
 \ee
The branes have equal and opposite tensions, $\pm\lambda$, where
 \be\label{rst}
\lambda={3M_\f^2 \over 4\pi \ell^2}\,.
 \ee
The positive-tension brane has fundamental scale $M_\vd$ and is
``hidden". Standard model fields are confined on the negative
tension (or ``visible") brane. Because of the exponential warping
factor, the effective scale on the visible brane at $y=L$ is
$M_\f$, where
 \be
M_\f^2=M_\vd^3\, \ell\left[1-e^{-2L/\ell}\right]\,.
 \ee
So the RS 2-brane model gives a new approach to the hierarchy
problem. Because of the finite separation between the branes, the
KK spectrum is discrete. Furthermore, at low energies gravity on
the branes becomes Brans-Dicke-like, with the sign of the
Brans-Dicke parameter equal to the sign of the brane
tension~\cite{gt}. In order to recover 4D general relativity at
low energies, a mechanism is required to stabilize the inter-brane
distance, which corresponds to a scalar field degree of freedom
known as the radion~\cite{goldwise,2b}.

\item[]{\bf RS 1-BRANE:}

In this model~\cite{rs2}, there is only one, positive tension,
brane. It may be thought of as arising from sending the negative
tension brane off to infinity, $L\to\infty$. Then the energy
scales are related via
 \be\label{tt2}
M_\vd^3={M_\f^2 \over \ell}\,.
 \ee
The infinite extra dimension makes a finite contribution to the 5D
volume because of the warp factor:
 \be
\int\,d^5X\sqrt{-\,^\vu g}=2\int\,d^4x\,\int_0^\infty\,dy
e^{-4y/\ell}=  {\ell \over 2}\,\int\,d^4x\,.
 \ee
Thus the effective size of the extra dimension probed by the 5D
graviton is $\ell$.

\end{itemize}~\\

I will concentrate mainly on RS 1-brane from now on, referring to
RS 2-brane occasionally. The RS 1-brane models are in some sense
the most simple and geometrically appealing form of brane-world
model, while at the same time providing a framework for the
AdS/CFT correspondence~\cite{acft,acftcosmo}. The RS 2-brane
models introduce the added complication of radion stabilization,
as well as possible complications arising from negative tension.
However, they remain important and will occasionally be discussed.

In RS 1-brane, the negative $\Lambda_\vd$ is offset by the
positive brane tension $\lambda$. The fine-tuning in
Eq.~(\ref{rst}) ensures that there is zero effective cosmological
constant on the brane, so that the brane has the induced geometry
of Minkowski spacetime. To see how gravity is localized at low
energies, we consider the 5D graviton perturbations of the
metric~\cite{rs2,gt,morepert}:
 \be
^\vu g_{AB} \to \, ^\vu g_{AB} +e^{-2|y|/\ell}\,^\vu h_{AB}\,,~
\,^\vu h_{Ay}=0= \,^\vu h^\mu{}_\mu= \,^\vu
h^{\mu\nu}{}{}_{,\nu}\,.
 \ee
(See Fig.~3.) This is the RS gauge, which is different from the
gauge used in Eq.~(\ref{5dg}), but which also has no remaining
gauge freedom. The 5 polarizations of the 5D graviton are
contained in the 5 independent components of $^\vu h_{\mu\nu}$ in
the RS gauge.

\begin{figure}[!bth]\label{rs}
\begin{center}
\includegraphics[height=3.5in,width=4.5in]{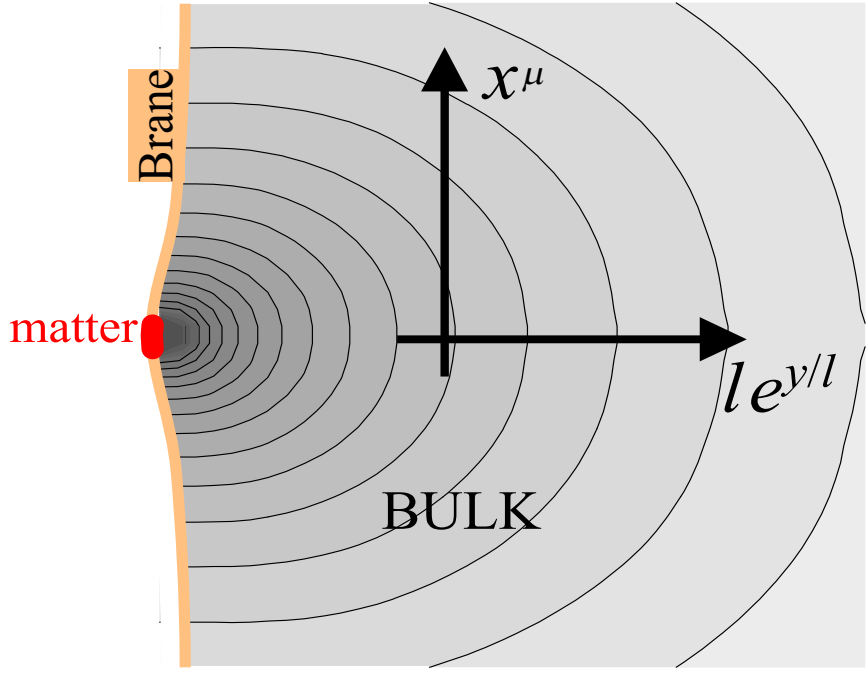}
\caption{Gravitational field of small point particle on the brane
in RS gauge (from~\cite{gt}).}
\end{center}
\end{figure}

We split the amplitude $f$ of $^\vu h_{AB}$ into 3D Fourier modes,
and the linearized 5D Einstein equations lead to the wave equation
($y>0$)
 \be
\ddot f+k^2f=e^{-2y/\ell}\left[f''- {4\over \ell}\, f'\right]\,.
 \ee
Separability means we can write
 \be
f(t,y)=\sum_m \varphi_m(t)\,f_m(y)\,,
 \ee
and the wave equation reduces to
 \bea
\ddot{\varphi}_m+(m^2+k^2)\varphi_m &=& 0 \,,\\ f_m''- {4\over
\ell}\, f_m'+e^{2y/\ell}m^2f_m&=& 0\,.
 \eea
The zero mode solution is
 \bea
\varphi_0(t)&=& A_{0+}e^{+ikt}+ A_{0-}e^{-ikt}\,, \\ f_0(y)&=&
B_0+C_0e^{4 y/\ell}\,,
 \eea
and the $m>0$ solutions are
 \bea
\varphi_m(t)&=& A_{m+}\exp\left(+i\sqrt{m^2+k^2}\,t\right)+
A_{m-}\exp \left(-i\sqrt{m^2+k^2}\,t \right)\,,\\ f_m(y) &=&
e^{2y/\ell} \left[ B_mJ_2\left(m\ell e^{y/\ell}\right) +C_m
Y_2\left(m\ell e^{y/\ell} \right)\right]\,.
 \eea

The boundary condition for the perturbations arises from the
junction conditions, Eq.~(\ref{ext}), discussed below, and leads
to $f'(t,0)=0$, since the transverse traceless part of the
perturbed energy-momentum tensor on the brane vanishes. This
implies
 \be \label{rsbc}
C_0=0\,,~~C_m=-{J_1(m\ell) \over Y_1(m\ell)}\,B_m\,.
 \ee
The zero mode is normalizable, since
 \be
\left| \int_0^\infty B_0 e^{-2y/\ell} dy \right|<\infty\,.
 \ee
Its contribution to the gravitational potential $V={1\over2}\,^\vu
h_{00}$ gives the 4D result, $V \propto r^{-1}$. The contribution
of the massive KK modes sums to a correction of the 4D potential.
For $r\ll \ell$, one obtains
 \be\label{newt2}
V(r)\approx {GM\ell \over r^2}\,,
 \ee
which simply reflects the fact that the potential becomes truly 5D
on small scales. For $r\gg\ell$,
 \be\label{newt}
V(r)\approx  {GM \over r}\left(1+{2\ell^2 \over 3r^2}\right)\,,
 \ee
which gives the small correction to 4D gravity at low energies
from extra-dimensional effects. These effects serve to slightly
strengthen the gravitational field, as expected.

Table-top tests of Newton's laws currently find no deviations down
to $O(10^{-1}~{\rm mm})$, so that $\ell\lesssim 0.1~$mm in
Eq.~(\ref{newt}). Then by Eqs.~(\ref{rst}) and (\ref{tt2}), this
leads to lower limits on the brane tension and the fundamental
scale of the  RS 1-brane model:
 \be\label{rslimit}
\lambda > (1~{\rm TeV})^4\,,~ M_\vd > 10^5~{\rm TeV}\,.
 \ee
These limits do not apply to the 2-brane case.

For the 1-brane model, the boundary condition, Eq.~(\ref{rsbc}),
admits a continuous spectrum $m>0$ of KK modes. In the 2-brane
model, $f'(t,L)=0$ must hold in addition to Eq.~(\ref{rsbc}). This
leads to conditions on $m$, so that the KK spectrum is discrete:
 \be
m_n= {x_n \over \ell}\,e^{-L/\ell} \mbox{ where } Y_1(m\ell)
J_1(x_n)= J_1(m\ell)Y_1(x_n) \,.
 \ee
The limit Eq.~(\ref{rslimit}) indicates that there are no
observable collider, i.e. $O({\rm TeV})$, signatures for the RS
1-brane model. The 2-brane model by contrast, for suitable choice
of $L$ and $\ell$ so that $m_1=O({\rm TeV})$, does predict
collider signatures that are distinct from those of the ADD
models~\cite{hanraf}.

\section{Covariant approach to brane-world geometry and dynamics}

The RS models and the subsequent generalization from a Minkowski
brane to a Friedmann-Robertson-Walker (FRW)
brane~\cite{bdel,morers2,gs}, were derived as solutions in
particular coordinates of the 5D Einstein equations, together with
the junction conditions at the $Z_2$-symmetric brane. A broader
perspective, with useful insights into the inter-play between 4D
and 5D effects, can be obtained via the covariant
Shiromizu-Maeda-Sasaki approach~\cite{sms}, in which the brane and
bulk metrics remain general. The basic idea is to use the
Gauss-Codazzi equations to project the 5D curvature along the
brane. (The general formalism for relating the geometries of a
spacetime and of hypersurfaces within that spacetime is given
in~\cite{wald}.)

The 5D field equations determine the 5D curvature tensor; in the
bulk, they are
 \be \label{5efe}
^{(5)}\!G_{AB}=-\Lambda_\vd\, ^{(5)}\!g_{AB}+ \kappa_\vd^2\, ^\vu
T_{AB}\,,
 \ee
where $^\vu T_{AB}$ represents any 5D energy-momentum of the
gravitational sector (e.g., dilaton and moduli scalar fields, form
fields).

Let $y$ be a Gaussian normal coordinate orthogonal to the brane
(which is at $y=0$ without loss of generality), so that
$n_AdX^A=dy$, with $n^A$ the unit normal. The 5D metric in terms
of the induced metric on $\{y=\mbox{ const}\}$ surfaces is locally
given by
 \be\label{gn}
{}^{(5)}\!g_{AB}=g_{AB}+ n_An_B\,,~{}^\vu
ds^2=g_{\mu\nu}(x^\alpha,y)dx^\mu dx^\nu +dy^2 \,.
 \ee
The extrinsic curvature of $\{y=\mbox{ const}\}$ surfaces
describes the embedding of these surfaces. It can be defined via
the Lie derivative or via the covariant derivative:
 \be\label{exc}
K_{AB}={1\over2}\mbox {\bf \pounds}_{\bf n}\,g_{AB}=g_A{}^C \,\,
^\vu{\nabla}_Cn_B\,,
 \ee
so that
 \be
K_{[AB]}=0=K_{AB}n^B\,,
 \ee
where square brackets denote anti-symmetrization. The Gauss
equation gives the 4D curvature tensor in terms of the projection
of the 5D curvature, with extrinsic curvature corrections:
 \be\label{gauss}
R_{ABCD}= {}^\vu {R}_{EFGH}\,g_A{}^Eg_B{}^Fg_C{}^Gg_D{}^H
+2K_{A[C}K_{D]B}\,,
 \ee
and the Codazzi equation determines the change of $K_{AB}$ along
$\{y=\mbox{ const}\}$ via
 \be\label{cod}
\nabla_BK^B{}_A-\nabla_AK= {}^\vu {R}_{BC}\,g_A{}^Bn^C\,,
 \ee
where $K=K^A{}_A$.

Some other useful projections of the 5D curvature are:
 \bea
&& {}^\vu {R}_{EFGH}\,g_A{}^Eg_B{}^Fg_C{}^Gn^H =
2\nabla_{[A}K_{B]C} \,,\\ && {}^\vu {R}_{EFGH}\,g_A{}^E n^Fg_B{}^G
n^H =-\mbox {\bf \pounds}_{\bf n} K_{AB}+K_{AC}K^C{}_B\,, \\ &&
{}^\vu {R}_{CD} \,g_A{}^Cg_B{}^D = R_{AB} -\mbox {\bf
\pounds}_{\bf n} K_{AB}-KK_{AB}+2K_{AC}K^C{}_B \,.
 \eea
The 5D curvature tensor has Weyl (trace-free) and Ricci parts:
 \be
{}^\vu {R}_{ABCD}={}^\vu{C}_{ACBD}+{2\over3}\left\{ {}^\vu g_{A[C}
{}^\vu R_{D]B}- {}^\vu g_{B[C} \,{}^\vu R_{D]A} \right\} -{1\over
6}{}^\vu g_{A[C}{}^\vu g_{D]B}{}^\vu R\,.
 \ee

\subsection{Field equations on the brane}

Using Eqs.~(\ref{5efe}) and (\ref{gauss}), it follows that
 \bea
{G}_{\mu\nu}&=&-{1\over2}{\Lambda}_\vd g_{\mu\nu}+{2\over3}
\kappa_\vd^2 \left[{}^\vu T_{AB}g_\mu {}^A g_\nu {}^B+
\left\{{}^\vu T_{AB}n^An^B-{1\over 4}\,{}^\vu T   \right\}
g_{\mu\nu} \right] \nonumber\\&&~{}+ K K_{\mu\nu}-K_\mu {}^\alpha
K_{\alpha\nu} +{1\over2}\left[K^{\alpha\beta}K_{\alpha\beta}-K^2
\right]g_{\mu\nu} - {\cal E}_{\mu\nu}\,,\label{ein}
 \eea
where ${}^\vu T={}^\vu T^A{}_{A}$, and where
 \be
{\cal E}_{\mu\nu} = {}^\vu{C}_{ACBD}\,n^Cn^Dg_\mu {}^A g_\nu
{}^B\,,
 \ee
is the projection of the bulk Weyl tensor orthogonal to $n^A$.
This tensor satisfies
 \be
{\cal E}_{AB}n^B= 0 ={\cal E}_{[AB]}={\cal E}_{A}{}^A\,,
 \ee
by virtue of the Weyl tensor symmetries. Evaluating
Eq.~(\ref{ein}) on the brane (strictly, as $y\to\pm 0$, since
${\cal E}_{AB}$ is not defined on the brane~\cite{sms}) will give
the field equations on the brane.

First, we need to determine $K_{\mu\nu}$ at the brane from the
junction conditions. The total energy-momentum tensor on the brane
is
 \be
T_{\mu\nu}^{\rm brane} =T_{\mu\nu}-\lambda g_{\mu\nu}\,,
 \ee
where $T_{\mu\nu}$ is the energy-momentum tensor of particles and
fields confined to the brane (so that $T_{AB}n^B=0$). The 5D field
equations, including explicitly the contribution of the brane, are
then
 \be \label{feb}
^{(5)}\!G_{AB}=-\Lambda_\vd\, ^{(5)}\!g_{AB}+ \kappa_\vd^2\left[
{}^\vu T_{AB}+ T_{AB}^{\rm brane}\delta(y)\right] \,.
 \ee
Here the delta function enforces in the classical theory the
string theory idea that Standard Model fields are confined to the
brane. This is not a gravitational confinement, since there is in
general a nonzero acceleration of particles normal to the
brane~\cite{m1}.

Integrating Eq.~(\ref{feb}) along the extra dimension from
$y=-\epsilon$ to $y=+\epsilon$, and taking the limit $\epsilon\to
0$, leads to the Israel-Darmois junction conditions at the brane,
 \bea
g^+_{\mu\nu}-g^-_{\mu\nu}&=&0\,, \\
K_{\mu\nu}^{+}-K_{\mu\nu}^{-}&=& -\kappa_\vd^2
\left[T_{\mu\nu}^{\rm brane}- {1\over3} T^{\rm brane}g_{\mu\nu}
\right]\,, \label{jun}
 \eea
where $T^{\rm brane}=g^{\mu\nu}T_{\mu\nu}^{\rm brane}$. The $Z_2$
symmetry means that when you approach the brane from one side and
go through it, you emerge into a bulk that looks the same, but
with the normal reversed, $n^A \to -n^A$. Then Eq.~(\ref{exc})
implies that
 \be\label{z2}
K_{\mu\nu}^{-}=-K_{\mu\nu}^{+}\,,
 \ee
so that we can use the junction condition Eq.~(\ref{jun}) to
determine the extrinsic curvature on the brane:
 \be\label{ext}
K_{\mu\nu}=-{1\over2}\kappa_\vd^2 \left[T_{\mu\nu}+ {1\over3}
\left(\lambda-T\right)g_{\mu\nu} \right] \,,
 \ee
where $T=T^\mu{}_\mu $, we have dropped the $(+)$ and we evaluate
quantities on the brane by taking the limit $y\to+0$.

Finally we arrive at the induced field equations on the brane, by
substituting Eq.~(\ref{ext}) into Eq.~(\ref{ein}):
\begin{equation} \label{e:einstein1}
G_{\mu\nu} = - \Lambda g_{\mu\nu} + \kappa^2 T_{\mu\nu} +
6\frac{\kappa^2}{\lambda} {\cal S}_{\mu\nu} - {\cal E}_{\mu\nu}+
4\frac{\kappa^2}{\lambda}{\cal F}_{\mu\nu}\,.
\end{equation}
The 4D gravitational constant is an effective coupling constant
inherited from the fundamental coupling constant, and the 4D
cosmological constant is nonzero when the RS balance between the
bulk cosmological constant and the brane tension is broken:
 \bea
\kappa^2 &\equiv & \kappa^2_{4}={1\over6}\lambda\kappa^4_\vd \,,\\
\Lambda &=& {1\over 2}\left[ \Lambda_\vd+\kappa^2\lambda
\right]\,,\,.
 \eea

The first correction term relative to Einstein's theory is
quadratic in the energy-momentum tensor, arising from the
extrinsic curvature terms in the projected Einstein tensor:
 \be
{\cal S}_{\mu\nu}= {{1\over12}}T T_{\mu\nu}
-{{1\over4}}T_{\mu\alpha}T^\alpha{}_\nu + {{1\over24}}g_{\mu\nu}
\left[3 T_{\alpha\beta} T^{\alpha\beta}-T^2 \right]\,.
 \ee
The second correction term is the projected Weyl term. The last
correction term on the right of Eq.~(\ref{e:einstein1}), which
generalizes the field equations in~\cite{sms}, is
 \be
{\cal F}_{\mu\nu}= {}^\vu T_{AB}g_\mu {}^A g_\nu {}^B+
\left[{}^\vu T_{AB}n^An^B-{1\over 4}\,{}^\vu T \right]
g_{\mu\nu}\,,
 \ee
where ${}^\vu T_{AB}$ describes any stresses in the bulk apart
from the cosmological constant (see~\cite{mw} for the case of a
scalar field).

What about the conservation equations? Using Eqs.~(\ref{5efe}),
(\ref{cod}) and (\ref{ext}), one obtains
 \be \label{cong}
\nabla^\nu T_{\mu\nu}=-2\,{}^\vu T_{AB}n^Ag^B{}_\mu \,.
 \ee
Thus in general there is exchange of energy-momentum between the
bulk and the brane. From now on, we will assume that
 \be
{}^\vu T_{AB}=0~\Rightarrow~{\cal F}_{\mu\nu}=0\,,
 \ee
so that
 \bea
^{(5)}\!G_{AB} &=& -\Lambda_\vd\, ^{(5)}\!g_{AB}\,,~(\mbox{in the bulk})\\
G_{\mu\nu} &=& - \Lambda g_{\mu\nu} + \kappa^2 T_{\mu\nu} +
6\frac{\kappa^2}{\lambda} {\cal S}_{\mu\nu} - {\cal E}_{\mu\nu}\,,
~\mbox{(on the brane)}\label{bife}
 \eea
and one then recovers from Eq.~(\ref{cong}) the standard 4D
conservation equations,
 \be\label{lc}
\nabla^\nu T_{\mu\nu}=0\,.
 \ee
This means that there is no exchange of energy-momentum between
the bulk and the brane; their interaction is purely gravitational.
Then the 4D contracted Bianchi identities ($\nabla^\nu
G_{\mu\nu}=0$), applied to Eq.~(\ref{e:einstein1}), lead to
 \be \label{nlc}
\nabla^\mu{\cal
E}_{\mu\nu}={6\kappa^2\over\lambda}\,\nabla^\mu{\cal
S}_{\mu\nu}\,,
 \ee
which shows qualitatively how 1+3 spacetime variations in the
matter-radiation on the brane can source KK modes.

The induced field equations~(\ref{bife}) show two key
modifications to the standard 4D Einstein field equations arising
from extra-dimensional effects:
\begin{itemize}
\item ${\cal S}_{\mu\nu}\sim (T_{\mu\nu})^2$ is the high-energy
correction term, which is negligible for $\rho\ll\lambda$, but
dominant for $\rho\gg\lambda$:
 \be
{|\kappa^2{\cal S}_{\mu\nu}/\lambda |\over |\kappa^2
T_{\mu\nu}|}\sim {|T_{\mu\nu}|\over\lambda} \sim
{\rho\over\lambda}\,.
 \ee

\item ${\cal E}_{\mu\nu}$, the projection of the bulk Weyl tensor
on the brane, encodes corrections from 5D graviton effects (the KK
modes in the linearized case). From the brane-observer viewpoint,
the energy-momentum corrections in ${\cal S}_{\mu\nu}$ are local,
whereas the KK corrections in ${\cal E}_{\mu\nu}$ are nonlocal,
since they incorporate 5D gravity wave modes. These nonlocal
corrections cannot be determined purely from data on the brane. In
the perturbative analysis of RS 1-brane which leads to the
corrections in the gravitational potential, Eq.~(\ref{newt}), the
KK modes that generate this correction are responsible for a
nonzero ${\cal E}_{\mu\nu}$; this term is what carries the
modification to the weak-field field equations. The 9 independent
components in the tracefree ${\cal E}_{\mu\nu}$ are reduced to 5
degrees of freedom by Eq.~(\ref{nlc}); these arise from the 5
polarizations of the 5D graviton.

Note that the covariant formalism applies also to the two-brane
case. In that case, the gravitational influence of the second
brane is felt via its contribution to ${\cal E}_{\mu\nu}$.

\end{itemize}

\subsection{5-dimensional equations and the initial-value problem}

The effective field equations are not a closed system. One needs
to supplement them by 5D equations governing ${\cal E}_{\mu\nu}$,
which are obtained from the 5D Einstein and Bianchi equations.
This leads to the coupled system~\cite{ssm}:
\begin{eqnarray}
\mbox {\bf \pounds}_{\bf n} K_{\mu\nu}&=& K_{\mu\alpha}K^\alpha
{}_\nu - {\cal E}_{\mu\nu}-{1\over6}\Lambda_\vd g_{\mu\nu}\, \\
\mbox {\bf \pounds}_{\bf n} {\cal E}_{\mu\nu}  &=& \nabla^\alpha
{\cal B}_{\alpha(\mu\nu)} + \frac{1}{6}
\Lambda_\vd\left(K_{\mu\nu}-g_{\mu\nu}K\right)
+K^{\alpha\beta}R_{\mu\alpha\nu\beta} +3K^\alpha{}_{(\mu}{\cal
E}_{\nu)\alpha}-K{\cal E}_{\mu\nu} \nonumber \\ & & {}
+\left(K_{\mu\alpha}K_{\nu\beta}
-K_{\alpha\beta}K_{\mu\nu}\right)K^{\alpha\beta}\,, \label{eq:EEE}
\\ \mbox {\bf \pounds}_{\bf n} {\cal B}_{\mu\nu\alpha}&=&-
2\nabla_{[\mu}{\cal E}_{\nu]\alpha}+K_\alpha{}^\beta {\cal
B}_{\mu\nu\beta} -2{\cal B}_{\alpha\beta [\mu }K_{\nu]}{}^\beta\,,
\label{eq:BBB}
\\\mbox {\bf \pounds}_{\bf n} R_{\mu\nu\alpha\beta}
&=&-2R_{\mu\nu\gamma [\alpha}K_{\beta]}{}^\gamma
-\nabla_{\mu}{\cal B}_{\alpha\beta\nu} + \nabla_{\mu}{\cal
B}_{\beta\alpha\nu} \,, \label{eq:bianchi3}
\end{eqnarray}
where the ``magnetic" part of the bulk Weyl tensor, counterpart to
the ``electric" part ${\cal E}_{\mu\nu}$, is
\begin{equation}
{\cal B}_{\mu\nu\alpha}= g_\mu {}^A g_\nu {}^B
g_\alpha{}^C\,\,{}^\vu C_{ABCD}n^D\,.
\end{equation}
These equations are to be solved subject to the boundary
conditions at the brane,
\begin{eqnarray}
\nabla^\mu {\cal E}_{\mu\nu} &\doteq& \kappa_\vd^4 \nabla^\mu
{\cal S}_{\mu\nu} \,,
\\ {\cal B}_{\mu\nu\alpha} &\doteq & 2\nabla_{[\mu}K_{\nu]\alpha}
\doteq-\kappa_\vd^2\nabla_{[\mu}\Bigl( T_{\nu] \alpha}
-\frac{1}{3}g_{\nu] \alpha}T \Bigr), \label{bcwall}
\end{eqnarray}
where $A\doteq B$ denotes $A|_{\rm brane}=B|_{\rm brane}$.

The above equations have been used to develop a covariant analysis
of the weak field~\cite{ssm}. They can also be used to develop a
Taylor expansion of the metric about the brane. In Gaussian normal
coordinates, Eq.~(\ref{gn}), we have $\mbox {\bf \pounds}_{\bf
n}=\partial/\partial y$. Then we find
 \bea
&&g_{\mu\nu}(x,y)= g_{\mu\nu}(x,0)-\kappa_\vd^2\left[
T_{\mu\nu}+{1\over
3}(\lambda-T)g_{\mu\nu}\right]_{y=0+}\,|y| \nonumber\\
&&~~{}+\left[-{\cal E}_{\mu\nu} +{1\over4}\kappa_\vd^4\left\{
T_{\mu\alpha}T^\alpha{}_\nu +{2\over3} (\lambda-T)T_{\mu\nu}
\right\} +{1\over6}\left\{ {1\over6}
\kappa_\vd^4(\lambda-T)^2-\Lambda_\vd
\right\}g_{\mu\nu}\right]_{y=0+}\, y^2+ \cdots \label{tay}
 \eea

In a non-covariant approach based on a specific form of the bulk
metric in particular coordinates, the 5D Bianchi identities would
be avoided and the equivalent problem would be one of solving the
5D field equations, subject to suitable 5D initial conditions and
to the boundary conditions Eq.~(\ref{ext}) on the metric. The
advantage of the covariant splitting of the field equations and
Bianchi identities along and normal to the brane is the clear
insight that it gives into the interplay between the 4D and 5D
gravitational fields. The disadvantage is that the splitting is
not well suited to dynamical evolution of the equations. Evolution
off the timelike brane in the spacelike normal direction does not
in general constitute a well-defined initial value
problem~\cite{antav}. One needs to specify initial data on a 4D
spacelike (or null) surface, with boundary conditions at the
brane(s) ensuring a consistent evolution~\cite{ichnak}. Clearly
the evolution of the observed universe is dependent upon initial
conditions which are inaccessible to brane-bound observers; this
is simply another aspect of the fact that the brane dynamics is
not determined by 4D but by 5D equations. The initial conditions
on a 4D surface could arise from models for creation of the 5D
universe~\cite{gs,ksb}, from dynamical attractor
behaviour~\cite{mukcol} or from suitable conditions (such as no
incoming gravitational radiation) at the past Cauchy horizon if
the bulk is asymptotically AdS.

\subsection{The brane viewpoint: a 1+3-covariant analysis}

A systematic analysis can be developed from the viewpoint of a
brane-bound observer, following~\cite{m1}. The effects of bulk
gravity are conveyed, from a brane observer viewpoint, via the
local (${\cal S}_{\mu\nu}$) and nonlocal (${\cal E}_{\mu\nu}$)
corrections to Einstein's equations. (In the more general case,
bulk effects on the brane are also carried by ${\cal F}_{\mu\nu}$,
which describes any 5D fields.) The ${\cal E}_{\mu\nu}$ term
cannot in general be determined from data on the brane, and the 5D
equations above (or their equivalent) need to be solved in order
to find ${\cal E}_{\mu\nu}$.

The general form of the brane energy-momentum tensor for any
matter fields (scalar fields, perfect fluids, kinetic gases,
dissipative fluids, etc.), including a combination of different
fields, can be covariantly given in terms of a chosen 4-velocity
$u^\mu$ as
\begin{equation}
T_{\mu\nu}=\rho u_\mu  u_\nu
+ph_{\mu\nu}+\pi_{\mu\nu}+q_{\mu}u_{\nu}+q_\nu  u_\mu \,.
 \label{3''}
\end{equation}
Here $\rho$ and $p$ are the energy density and isotropic pressure,
and
 \be
h_{\mu\nu}=g_{\mu\nu}+u_\mu  u_\nu = \, ^\vu g_{\mu\nu}-n_\mu
n_\nu +u_\mu  u_\nu \,,
 \ee
projects into the comoving rest space orthogonal to $u^\mu$ on the
brane. The momentum density and anisotropic stress obey
 \be
q_{\mu}=q_{\langle \mu \rangle}\,,~~ \pi_{\mu\nu}=\pi_{\langle
\mu\nu \rangle}\,,
 \ee
where angled brackets denote the spatially projected, symmetric
and tracefree part:
 \be
V_{\langle \mu \rangle}=h_\mu {}^\nu V_\nu \,,~~ W_{\langle \mu\nu
\rangle}=\left[h_{(\mu}{}^\alpha h_{\nu)}{}^\beta- {{1\over3}}h^{
\alpha\beta}h_{\mu\nu}\right]W_{\alpha\beta}\,.
 \ee
In an inertial frame at any point on the brane, we have
 \be
u^\mu=(1,\vec 0)\,,~h_{\mu\nu}={\rm diag}(0,1,1,1)\,,~V_\mu
=(0,V_i)\,,~W_{\mu 0} =0= \sum W_{ii}=W_{ij}- W_{ji} \,,
 \ee
where $i,j = 1,2,3$.

The tensor ${\cal S}_{\mu\nu}$, which carries local bulk effects
onto the brane, may then be irreducibly decomposed as
\begin{eqnarray}
{\cal S}_{\mu\nu}&=&{{1\over24}}\left[2\rho^2-3\pi_{\alpha\beta}
\pi^{\alpha\beta}\right]u_\mu  u_\nu
+{{1\over24}}\left[2\rho^2+4\rho p+\pi_{\alpha\beta}
\pi^{\alpha\beta}-4q_\alpha q^\alpha\right]h_{\mu\nu} \nonumber\\
&&{}~~- {{1\over12}}(\rho+2p)\pi_{\mu\nu}+\pi_{ \alpha \langle
\mu} \pi_{\nu \rangle}{}^\alpha +q_{\langle \mu}q_{\nu \rangle}+
{{1\over3}}\rho q_{(\mu}u_{\nu)}- {{1\over12}} q^\alpha \pi_{
\alpha(\mu}u_{\nu)} \,. \label{3'''}
\end{eqnarray}
This simplifies for a perfect fluid or minimally-coupled scalar
field:
 \be
{\cal S}_{\mu\nu}={{1\over12}}\rho\left[\rho u_\mu  u_\nu
+\left(\rho+2 p\right)h_{\mu\nu}\right]\,.
 \ee

The trace free ${\cal E}_{\mu\nu}$ carries nonlocal bulk effects
onto the brane, and contributes an effective ``dark" radiative
energy-momentum on the brane, with energy density $\rho_{\cal E}$,
pressure $\rho_{\cal E}/3$, momentum density $q^{\cal E}_\mu $ and
anisotropic stress $\pi^{\cal E}_{\mu\nu}$:
 \be
-{1\over\kappa^2} {\cal E}_{\mu\nu} = \cu\left(u_\mu  u_\nu +{
{1\over3}} h_{\mu\nu}\right)+ {\cq_\mu } u_{\nu} + {\cq_\nu }
u_{\mu}+\cp_{\mu\nu}\,.
 \ee
We can think of this as a KK or Weyl ``fluid". The brane ``feels"
the bulk gravitational field through this effective fluid. More
specifically:
\begin{itemize}
\item The KK (or Weyl) anisotropic stress $\cp_{\mu\nu}$
incorporates the scalar or spin-0 (``Coulomb"), the vector
(transverse) or spin-1 (gravimagnetic) and the tensor (transverse
traceless) or spin-2 (gravitational wave) 4D modes of the spin-2
5D graviton. \item The KK momentum density $\cq_\mu  $
incorporates spin-0 and spin-1 modes, and defines a velocity
$v^{\cal E}_\mu $ of the Weyl fluid relative to $u^\mu$ via
$\cq_\mu =\cu v^{\cal E}_\mu $. \item The KK energy density $\cu
$, often called the ``dark radiation", incorporates the spin-0
mode.
\end{itemize}

In special cases, symmetry will impose simplifications on this
tensor. For example, it must vanish for a conformally flat bulk,
including AdS$_5$,
 \be
{}^\vu g_{AB}~\mbox{conformally flat}~\Rightarrow~ {\cal
E}_{\mu\nu}=0\,.
 \ee
The RS models have a Minkowski brane in an AdS$_5$ bulk. This bulk
is also compatible with an FRW brane. However, the most general
vacuum bulk with a Friedmann brane is Schwarzschild-anti de Sitter
spacetime~\cite{birk}. Then it follows from the FRW symmetries
that
 \be
\mbox{Schwarzschild AdS$_5$ bulk, FRW brane:}~~\cq_\mu
=0=\cp_{\mu\nu}\,,
 \ee
where $\cu=0$ only if the mass of the black hole in the bulk is
zero. The presence of the bulk black hole generates via Coulomb
effects the dark radiation on the brane.

For a static spherically symmetric brane (e.g. the exterior of a
static star or black hole)~\cite{dmpr},
 \be
\mbox{static spherical brane:}~~\cq_\mu =0\,.
 \ee
This condition also holds for a Bianchi~I brane~\cite{mss}. In
these cases, $\cp_{\mu\nu}$ is not determined by the symmetries,
but by the 5D field equations. By contrast, the symmetries of a
G\"odel brane fix $\cp_{\mu\nu}$~\cite{tsab}.

The brane-world corrections can conveniently be consolidated into
an effective total energy density, pressure, momentum density and
anisotropic stress.
\begin{eqnarray}
\rho_{\rm tot} &=& \rho+{1\over 4\lambda}\left(2\rho^2 -3
\pi_{\mu\nu} \pi^{\mu\nu}\right) +\cu\,, \label{a}\\ p_{\rm tot}
&=& p+ {1\over 4\lambda}\left(2\rho^2+4\rho p+
\pi_{\mu\nu}\pi^{\mu\nu}-4q_\mu q^\mu\right) +{\cu\over 3}\,,
\label{b}\\ q^{\rm tot}_\mu &=&q_\mu +{1\over 2\lambda}
\left(2\rho q_\mu-3\pi_{\mu\nu}q^\nu\right)+ \cq_\mu \,,\label{d}\\
\pi^{\rm tot}_{\mu\nu} &=& \pi_{\mu\nu}+{1\over 2\lambda}
\left[-(\rho+3p)\pi_{\mu\nu}+ 3\pi_{\alpha\langle \mu}\pi_{\nu
\rangle}{}^\alpha+ 3q_{\langle \mu}q_ { \nu \rangle}\right] +
\cp_{\mu\nu}\,.\label{c}
\end{eqnarray}
These general expressions simplify in the case of a perfect fluid
(or minimally coupled scalar field, or isotropic one-particle
distribution function), i.e., for $q_\mu=0=\pi_{\mu\nu}$:
\begin{eqnarray}
\rho_{\text{tot}} &=& \rho\left(1 +\frac{\rho}{2\lambda} +
\frac{\rho_{\cal E}}{\rho} \right)\,,\label{rtot} \\ \label{ptot}
p_{\text{tot }} &=& p  + \frac{\rho}{2\lambda}
(2p+\rho)+\frac{\rho_{\cal E}}{3}\;,
\\ q^{\text{tot }}_\mu  &=& q^{\cal E}_\mu \;, \\
\label{e:pressure2} \pi^{\text{tot }}_{\mu\nu} &=& \pi^{\cal
E}_{\mu\nu}\;.
\end{eqnarray}
Note that nonlocal bulk effects can contribute to effective
imperfect fluid terms even when the matter on the brane has
perfect fluid form: there is in general an effective momentum
density and anisotropic stress induced on the brane by massive KK
modes of the 5D graviton.

The effective total equation of state and sound speed follow from
Eqs.~(\ref{rtot}) and (\ref{ptot}) as
 \bea
w_{\rm tot} &\equiv & {p_{\rm tot}\over\rho_{\rm tot}} =
{w+(1+2w)\rho/2\lambda+ \cu/3\rho \over 1+\rho/2\lambda +\cu/\rho}
\,,\label{vh1}\\ c_{\rm tot}^2 &\equiv & {\dot{p}_{\rm
tot}\over\dot{\rho}_{\rm tot}}= \left[c_{\rm s}^2+{\rho+p \over
\rho+\lambda} +{4\cu\over 9(\rho+p)(1+\rho/\lambda)}\right]
\left[1+ {4\cu\over 3(\rho+p)(1+\rho/\lambda)}\right]^{-1}
\,,\label{vh2}
 \eea
where $w=p/\rho$ and $c_{\rm s}^2=\dot p/\dot\rho$. At very high
energies, i.e., $\rho\gg\lambda$, we can generally neglect $\cu$
(e.g., in an inflating cosmology), and the effective equation of
state and sound speed are stiffened:
 \be
w_{\rm tot}\approx 2w+1\,,~~ c_{\rm tot}^2 \approx c_{\rm s}^2
+w+1\,.
 \ee
This can have important consequences in the early universe and
during gravitational collapse. For example, in a very high-energy
radiation era, $w={1\over3}$, the effective cosmological equation
of state is ultra-stiff: $w_{\rm tot}\approx {5\over3}$. In
late-stage gravitational collapse of pressureless matter, $w=0$,
the effective equation of state is stiff, $w_{\rm tot}\approx 1$,
and the effective pressure is nonzero and dynamically important.

\subsection{Conservation equations}

Conservation of $T_{\mu\nu}$ gives the standard general relativity
energy and momentum conservation equations, in the general,
nonlinear case:
\begin{eqnarray}
&&\dot{\rho}+\Theta(\rho+p)+\D^\mu q_\mu+2A^\mu q_\mu
+\sigma^{\mu\nu }\pi_{\mu\nu}=0\,,\label{c1}\\ && \dot{q}_{\langle
\mu\rangle}+{{4\over3}}\Theta q_\mu+\D_\mu p+(\rho+p)A_\mu +
\D^\nu \pi_{\mu\nu}+A^\nu\pi_{\mu\nu} +\sigma_{\mu\nu}q^\nu-
\ep_{\mu\nu\alpha}\omega^\nu q^\alpha =0\,.\label{c2}
\end{eqnarray}
In these equations, an overdot denotes $u^\nu\nabla_\nu$, and

$\Theta=\nabla^\mu u_\mu$ is the volume expansion rate of the
$u^\mu$ worldlines,

$A_\mu=\dot{u}_\mu  =A_{\langle \mu\rangle}$ is their
4-acceleration,

$\sigma_{ \mu\nu}=\D_{\langle \mu}u_{ \nu\rangle}$ is their shear
rate,

$\omega_\mu =-{1\over2}\curl u_\mu =\omega_{\langle \mu\rangle}$
is their vorticity rate. \\ On a Friedmann brane,
 \be
A_\mu =\omega_\mu =\sigma_{\mu\nu}=0\,,~~\Theta=3H\,,
 \ee
where $H=\dot a/a$ is the Hubble rate. The covariant spatial curl
is given by
 \be
\curl V_\mu =\ep_{\mu\alpha\beta}\D^\alpha V^\beta\,,~~ \curl W
_{\mu\nu}=\ep_{ \alpha\beta(\mu}\D^\alpha W^\beta{}_{\nu)}\,,
 \ee
where $\ep_{\mu\alpha\beta}$ is the projection orthogonal to $u^
\mu$ of the 4D brane alternating tensor, and $\D_\mu $ is the
projected part of the brane covariant derivative, defined by
 \be
\D_\mu  F^{\alpha\cdots}{}{}_{\cdots \beta}=\left(\nabla_\mu  F^{
\alpha\cdots}{}{}_{\cdots \beta}\right)_{\perp u}= h_\mu {}^ \nu
h^ \alpha{}_\gamma \cdots h_ \beta{}^\delta \nabla_\nu
F^{\gamma\cdots}{}{}_{\cdots \delta}\,.
 \ee
In a local inertial frame at a point on the brane, with $u^
\mu=\delta^ \mu{}_0$, we have: $0=A_0=\omega_0=\sigma_{0\mu}=
\ep_{0\alpha\beta}= \curl V_0 =\curl W_{0\mu}$ and
 \be
\D_\mu  F^{ \alpha\cdots}{}{}_{\cdots \beta}=\delta_\mu {}^i
\delta^ \alpha{}_j\cdots\delta_ \beta{}^k\nabla_i
F^{j\cdots}{}{}_{\cdots k}~\mbox{(local inertial frame)}\,,
 \ee
where $i,j,k=1,2,3$.

The absence of bulk source terms in the conservation equations is
a consequence of having $\Lambda_\vd$ as the only 5D source in the
bulk. For example, if there is a bulk scalar field, then there is
energy-momentum exchange between the brane and bulk (in addition
to the gravitational interaction)~\cite{mw,sca}.

Equation~(\ref{nlc}) may be called the ``nonlocal conservation
equation". Projecting along $u^\mu$ gives the nonlocal energy
conservation equation, which is a propagation equation for $\cu$.
In the general, nonlinear case, this gives
\begin{eqnarray}
&& \dot{\rho}_{\cal E}+{{4\over3}}\Theta{\cu}+\D^\mu \cq_\mu
+2A^\mu \cq_\mu +\sigma^{\mu\nu}\cp_{\mu\nu}\nonumber\\
&&~{}={{1\over 4\lambda}}\left[
6\pi^{\mu\nu}\dot{\pi}_{\mu\nu}+6(\rho+p)\sigma^{\mu\nu}
\pi_{\mu\nu}+2\Theta \left(2q^\mu  q_\mu +
\pi^{\mu\nu}\pi_{\mu\nu}
\right) +2A^\mu  q^\nu \pi_{\mu\nu}\right.\nonumber\\
&&~\left.{} -4q^\mu \D_\mu \rho+q^\mu \D^\nu \pi_{\mu\nu}
+\pi^{\mu\nu}\D_\mu  q_\nu -2\sigma^{\mu\nu}\pi_{
\alpha\mu}\pi_\nu {}^\alpha- 2\sigma^{\mu\nu}q_\mu  q_\nu
\right]\,. \label{c1'}
\end{eqnarray}
Projecting into the comoving rest space gives the nonlocal
momentum conservation equation, which is a propagation equation
for $\cq_\mu $:
\begin{eqnarray}
&& \dot{q}^{\cal E}_{\langle \mu\rangle}+{{4\over3}}\Theta\cq_\mu
+{{1\over3}}\D_\mu {\cu}+{{4\over3}}{\cu}A_\mu  +\D^\nu
\cp_{\mu\nu}+A^\nu \cp_{\mu\nu}+\sigma_{\mu}{}^\nu\cq_\nu
-\ep_{\mu}{}^{\nu\alpha}\omega_\nu \cq_\alpha \nonumber\\
&&~~{}={{1\over 4 \lambda}}\left[ -4(\rho+p)\D_\mu  \rho
+6(\rho+p)\D^\nu \pi_{\mu\nu} +q^\nu\dot{\pi}_{\langle \mu\nu
\rangle}+\pi_\mu {}^\nu \D_\nu (2\rho+5p)
\right.\nonumber\\
&&~~~\left.{}-{{2\over3}}\pi^{\alpha\beta} \left(\D_\mu
\pi_{\alpha\beta}+3\D_\alpha
\pi_{\beta\mu}\right)-3\pi_{\mu\alpha}\D_\beta
 \pi^{\alpha\beta}+{{28\over3}}q^\nu \D_\mu  q_\nu  \right.\nonumber\\
&&~~~\left.{}+4\rho A^\nu \pi_{\mu\nu}-3\pi_{\mu\alpha}A_\beta
\pi^{\alpha\beta} +{{8\over3}}A_\mu
\pi^{\alpha\beta}\pi_{\alpha\beta}
-\pi_{\mu\alpha}\sigma^{\alpha\beta}q_\beta \right.\nonumber\\
&&~~~\left.{}+\sigma_{\mu\alpha} \pi^{\alpha\beta}q_\beta+
\pi_{\mu\nu}\ep^{\nu\alpha\beta}\omega_\alpha q_\beta
-\ep_{\mu\alpha\beta}\omega^\alpha \pi^{\beta\nu}q_\nu
+4(\rho+p)\Theta q_\mu \right.\nonumber\\&&~~~\left.{}+ 6q_\mu
A^\nu  q_\nu +{{14\over3}}A_\mu  q^\nu  q_\nu +4q_\mu
\sigma^{\alpha\beta} \pi_{\alpha\beta}\right]\,.\label{c2'}
\end{eqnarray}
The 1+3-covariant decomposition shows two key features:

\begin{itemize}

\item inhomogeneous and anisotropic effects from the 4D
matter-radiation distribution on the brane are a source for the 5D
Weyl tensor, which nonlocally ``backreacts" on the brane via its
projection ${\cal E}_{\mu\nu}$;

\item there are evolution equations for the dark radiative
(nonlocal, Weyl) energy ($\cu$) and momentum ($\cq_\mu $)
densities (carrying scalar and vector modes from bulk gravitons),
but there is no evolution equation for the dark radiative
anisotropic stress ($\cp_{\mu\nu}$) (carrying tensor, as well as
scalar and vector, modes), which arises in both evolution
equations.

\end{itemize}

In particular cases, the Weyl anisotropic stress $\cp_{\mu\nu}$
may drop out of the nonlocal conservation equations, i.e., when we
can neglect $\sigma^{\mu\nu}\cp_{\mu\nu}$, $\D^\nu \cp_{\mu\nu}$
and $A^\nu \cp_{\mu\nu}$. This is the case when we consider
linearized perturbations about an FRW background (which remove the
first and last of these terms) and further when we can neglect
gradient terms on large scales (which removes the second term).
This case is discussed in Sec.~VI. But in general, and especially
in astrophysical contexts, the $\cp_{\mu\nu}$ terms cannot be
neglected. Even when we can neglect these terms, $\cp_{\mu\nu}$
arises in the field equations on the brane.

All of the matter source terms on the right of these two
equations, except for the first term on the right of
Eq.~(\ref{c2'}), are imperfect fluid terms, and most of these
terms are quadratic in the imperfect quantities $q_\mu $ and
$\pi_{\mu\nu}$. For a single perfect fluid or scalar field, only
the $\D_\mu \rho$ term on the right of Eq.~(\ref{c2'}) survives,
but in realistic cosmological and astrophysical models, further
terms will survive. For example, terms linear in $\pi_{\mu\nu}$
will carry the photon quadrupole in cosmology or the shear viscous
stress in stellar models. If there are two fluids (even if both
fluids are perfect), then there will be a relative velocity $v_\mu
$ generating a momentum density $q_\mu =\rho v_\mu $, which will
serve to source nonlocal effects.

In general, the 4 independent equations in Eqs.~(\ref{c1'}) and
(\ref{c2'}) constrain 4 of the 9 independent components of ${\cal
E}_{\mu\nu}$ on the brane. What is missing, is an evolution
equation for $\cp_{\mu\nu}$, which has up to 5 independent
components. These 5 degrees of freedom correspond to the 5
polarizations of the 5D graviton. Thus in general, the projection
of the 5-dimensional field equations onto the brane does not lead
to a closed system, as expected, since there are bulk degrees of
freedom whose impact on the brane cannot be predicted by brane
observers. The KK anisotropic stress $\cp_{\mu\nu}$ encodes the
nonlocality.

In special cases the missing equation does not matter. For
example, if $\cp_{\mu\nu}=0$ by symmetry, as in the case of an FRW
brane, then the evolution of ${\cal E}_{\mu\nu}$ is determined by
Eqs. (\ref{c1'}) and (\ref{c2'}). If the brane is stationary (with
Killing vector parallel to $u^\mu $), then evolution equations are
not needed for ${\cal E}_{\mu\nu}$, although in general
$\cp_{\mu\nu}$ will still be undetermined. However, small
perturbations of these special cases will immediately restore the
problem of missing information.

If the matter on the brane has a perfect-fluid or scalar-field
energy-momentum tensor, the local conservation
equations~(\ref{c1}) and (\ref{c2}) reduce to
\begin{eqnarray}
&&\dot{\rho}+\Theta(\rho+p)=0\,,\label{pc1}\\ && \D_\mu
p+(\rho+p)A_\mu  =0\,,\label{pc2}
\end{eqnarray}
while the nonlocal conservation equations~(\ref{c1'}) and
(\ref{c2'}) reduce to
\begin{eqnarray}
&& \dot{\cu}+{{4\over3}}\Theta{\cu}+\D^\mu \cq_\mu +2A^\mu \cq_\mu
+\sigma^{\mu\nu}\cp_{\mu\nu}=0\,, \label{pc1'}\\&& \dot{q}^{\cal
E}_{\langle \mu\rangle}+{{4\over3}}\Theta\cq_\mu
+{{1\over3}}\D_\mu {\cu}+{{4\over3}}{\cu}A_\mu  +\D^\nu
\cp_{\mu\nu}+A^\nu \cp_{\mu\nu}+\sigma_{\mu}{}^\nu\cq_\nu
-\ep_{\mu}{}^{\nu\alpha}\omega_\nu \cq_\alpha \nonumber\\&&~~{}
=-{(\rho+p)\over\lambda} \D_\mu \rho\,.\label{pc2'}
\end{eqnarray}

Equation (\ref{pc2'}) shows that~\cite{sms}

\begin{itemize}

\item if ${\cal E}_{\mu\nu}=0$ and the brane energy-momentum
tensor has perfect fluid form, then the density $\rho$ must be
homogeneous, $\D_\mu \rho=0$;

\item the converse does not hold, i.e., homogeneous density does
{\em not} in general imply vanishing ${\cal E}_{\mu\nu}$.

\end{itemize}

A simple example of the latter point is the FRW case:
Eq.~(\ref{pc2'}) is trivially satisfied, while Eq.~(\ref{pc1'})
becomes
 \be
\dot{\rho}_{\cal E}+4H{\cu}=0\,.
 \ee
This equation has the dark radiation solution
\begin{equation}\label{dr}
{\cu}=\rho_{{\cal E}\,0}\left({a_0\over a}\right)^4\,.
\end{equation}

If ${\cal E}_{\mu\nu}=0$, then the field equations on the brane
form a closed system. Thus for perfect fluid branes with
homogeneous density and ${\cal E}_{\mu\nu}=0$, the brane field
equations form a consistent closed system. However, this is
unstable to perturbations, and there is also no guarantee that the
resulting brane metric can be embedded in a regular bulk.

It also follows as a corollary that inhomogeneous density requires
nonzero ${\cal E}_{\mu\nu}$:
 \be
\D_\mu \rho\neq0~\Rightarrow~ {\cal E}_{\mu\nu}\neq0\,.
 \ee
For example, stellar solutions on the brane necessarily have
${\cal E}_{\mu\nu}\neq0$ in the stellar interior if it is
non-uniform. Perturbed FRW models on the brane also must have
${\cal E}_{\mu\nu}\neq0$. Thus a nonzero ${\cal E}_{\mu\nu}$, and
in particular a nonzero $\cp_{\mu\nu}$, is inevitable in realistic
astrophysical and cosmological models.

\subsection{Propagation and constraint equations on the brane}

The propagation equations for the local and nonlocal energy
density and momentum density are supplemented by further
1+3-covariant propagation and constraint equations for the
kinematic quantities $\Theta$, $A_\mu $, $\omega_\mu $,
$\sigma_{\mu\nu}$, and for the free gravitational field on the
brane. The kinematic quantities govern the relative motion of
neighbouring fundamental world-lines. The free gravitational field
on the brane is given by the brane Weyl tensor $C_{\mu\nu
\alpha\beta}$. This splits into the gravito-electric and
gravito-magnetic fields on the brane:
 \be
E_{ \mu\nu}=C_{\mu\alpha\nu\beta}u^\alpha  u^\beta =E_{\langle
\mu\nu\rangle }\,,~~ H_{\mu\nu}={{1\over2}}\ep_{\mu\alpha\beta}
C^{\alpha\beta}{}{}_{\nu\gamma}u^\gamma=H_{\langle \mu\nu\rangle}
\,,
 \ee
where $E_{ \mu\nu}$ is not to be confused with ${\cal
E}_{\mu\nu}$. The Ricci identity for $u^\mu$
 \be
\nabla_{[\mu}\nabla_{\nu]}u_\alpha={1\over2}
R_{\alpha\nu\mu\beta}u^\beta\,,
 \ee
and the Bianchi identities
 \be
\nabla^\beta C_{\mu\nu \alpha\beta} = \nabla_{[\mu}\left(-R_{\nu]
\alpha} + {1\over6}Rg_{ \nu] \alpha}\right)\,,
 \ee
produce the fundamental evolution and constraint equations
governing the above covariant quantities. The field equations are
incorporated via the algebraic replacement of the Ricci tensor
$R_{ \mu\nu}$ by the effective total energy-momentum tensor,
according to Eq.~(\ref{e:einstein1}). The brane equations are
derived directly from the standard general relativity versions by
simply replacing the energy-momentum tensor terms $\rho,\dots$ by
$\rho_{\rm tot},\dots$. For a general fluid source, the equations
are given in~\cite{m1}. In the case of a single perfect fluid or
minimally-coupled scalar field, the
equations reduce to the following nonlinear equations.\\

\noindent Generalized Raychaudhuri equation (expansion
propagation):
\begin{eqnarray}
&&\dot{\Theta}+{{1\over3}}\Theta^2+\sigma_{ \mu\nu} \sigma^{
\mu\nu} -2\omega_  \mu\omega^\mu  -\D^\mu  A_\mu +A_\mu A^\mu
+{{\kappa^2\over2}}(\rho + 3p) -\Lambda \nonumber\\&&~~{}=
-{{\kappa^2\over2}}(2\rho+3p){\rho\over\lambda}- \kappa^2\cu\,.
\label{pr}
\end{eqnarray}
Vorticity propagation:
\begin{equation}
\dot{\omega}_{\langle   \mu\rangle } +{{2\over3}}\Theta\omega_\mu
+{{1\over2}}\curl A_\mu  -\sigma_{ \mu\nu}\omega^\nu =0
\,.\label{pe4}
\end{equation}
Shear propagation:
\begin{equation}
\dot{\sigma}_{\langle   \mu\nu\rangle }
+{{2\over3}}\Theta\sigma_{\mu\nu} +E_{\mu\nu}-\D_{\langle\mu}A_{
\nu\rangle } +\sigma_{ \alpha\langle\mu}\sigma_{ \nu\rangle
}{}^\alpha + \omega_{\langle  \mu}\omega_{ \nu\rangle} -
A_{\langle
 \mu}A_{ \nu\rangle} ={\kappa^2\over 2}\cp_{ \mu\nu}\,.
\label{pe5}
\end{equation}
Gravito-electric propagation (Maxwell-Weyl E-dot equation):
\begin{eqnarray}
 && \dot{E}_{\langle   \mu\nu\rangle }
+\Theta E_{ \mu\nu} -\curl H_{ \mu\nu}
+{{\kappa^2\over2}}(\rho+p)\sigma_{ \mu\nu} \nonumber\\&&~{}
-2A^\alpha \ep_{\alpha\beta( \mu}H_{\nu)}{}^\beta  -3\sigma_{
\alpha\langle \mu}E_{ \nu\rangle }{}^\alpha  +\omega^\alpha
\ep_{\alpha\beta(\mu}E_{\nu)}{}^\beta  \nonumber\\&&~~{}=
-{{\kappa^2\over2}} (\rho+p){\rho\over\lambda}\sigma_{\mu\nu}
-{\kappa^2\over6}\left[4\cu \sigma_{\mu\nu}+3\dot{\pi}^{\cal
E}_{\langle \mu\nu\rangle} +\Theta \cp_{\mu\nu}
+3\D_{\langle\mu}\cq_{ \nu\rangle} \right.\nonumber\\&&~~~\left.{}
+6A_{\langle \mu}\cq_{ \nu\rangle}+ 3\sigma^\alpha {}_{\langle\mu}
\cp_{ \nu\rangle \alpha}+3 \omega^\alpha \ep_{\alpha\beta(\mu}
\cp_{\nu)}{}^\beta \right] \,. \label{pe6}
\end{eqnarray}
Gravito-magnetic propagation (Maxwell-Weyl H-dot equation):
\begin{eqnarray}
 &&\dot{H}_{\langle
  \mu\nu\rangle } +\Theta H_{ \mu\nu} +\curl E_{ \mu\nu}-
3\sigma_{\alpha\langle  \mu}H_{ \nu\rangle }{}^\alpha
+\omega^\alpha \ep_{\alpha\beta(\mu}H_{\nu)}{}^\beta  +2A^\alpha
\ep_{\alpha\beta(\mu}E_{\nu)}{}^\beta  \nonumber\\&&~~{}=
{\kappa^2\over2}\left[ \curl\cp_{\mu\nu}-3\omega_{\langle\mu}
\cq_{ \nu\rangle} +\sigma_{\alpha( \mu}\ep_{\nu)}{}^{\alpha\beta}
\cq_\beta\right] \,. \label{pe7}
\end{eqnarray}
Vorticity constraint:
\begin{equation}
\D^  \mu\omega_\mu  -A^\mu \omega_\mu  =0\,.\label{pcc1}
\end{equation}
Shear constraint:
\begin{equation}
\D^\nu \sigma_{ \mu\nu}-\curl\omega_\mu  -{{2\over3}}\D_\mu \Theta
+2\ep_{\mu\nu\alpha}\omega^\nu A^\alpha = -\kappa^2\cq_\mu
 \,.\label{pcc2}
\end{equation}
Gravito-magnetic constraint:
\begin{equation}
 \curl\sigma_{ \mu\nu}+\D_{\langle  \mu}\omega_{ \nu\rangle  }
 -H_{ \mu\nu}+2A_{\langle  \mu}
\omega_{ \nu\rangle  }=0 \,.\label{pcc3}
\end{equation}
Gravito-electric divergence (Maxwell-Weyl div-E equation) :
\begin{eqnarray}
 && \D^\nu  E_{ \mu\nu}
 -{{\kappa^2\over3}}\D_  \mu\rho
 -\ep_{\mu\nu\alpha}\sigma^\nu {}_\beta H^{\alpha\beta}
+3H_{ \mu\nu}\omega^\nu  \nonumber\\&&{}= {\kappa^2\over3}
{\rho\over \lambda} \D_  \mu\rho +{\kappa^2\over6}\left(2\D_\mu
\cu-2\Theta \cq_\mu -3\D^\nu \cp_{ \mu\nu} +3\sigma_{\mu}{}^\nu
\cq_\nu -9\ep_{\mu}{}^{\nu\alpha}\omega_\nu \cq_\alpha \right)\!.
\label{pcc4}
\end{eqnarray}
Gravito-magnetic divergence (Maxwell-Weyl div-H equation):
\begin{eqnarray}
 &&\D^\nu  H_{ \mu\nu}
-\kappa^2(\rho+p)\omega_\mu  +\ep_{\mu\nu\alpha}\sigma^\nu
{}_\beta E^{\alpha\beta}
 -3E_{ \mu\nu}\omega^\nu
\nonumber\\&&~~{}=\kappa^2(\rho+ p){\rho\over\lambda} \omega_\mu +
{\kappa^2\over6}\left(8 \cu \omega_\mu -3\curl\cq_\mu -3\ep_\mu
{}^{\nu\alpha}\sigma_\nu {}^\beta \cp_{\alpha\beta}-3\cp_{
\mu\nu}\omega^\nu \right) \,.\label{pcc5}
\end{eqnarray}
Gauss-Codazzi equations on the brane (with $\omega_\mu =0$):
 \bea
&&R^\perp_{\langle   \mu\nu\rangle}+\dot{\sigma}_{\langle
\mu\nu\rangle }+\Theta\sigma_{ \mu\nu} -\D_{\langle  \mu}A_{
\nu\rangle }
-A_{\langle  \mu}A_{ \nu\rangle}=\kappa^2\cp_{\mu\nu}\,, \label{gc1}\\
&&R^\perp+ {{2\over3}}\Theta^2-\sigma_{ \mu\nu} \sigma^{ \mu\nu}
-2\kappa^2\rho -2\Lambda = {\kappa^2}{\rho^2\over\lambda}+
2\kappa^2\cu\,, \label{gc2}
 \eea
where $R^\perp_{ \mu\nu}$ is the Ricci tensor for 3-surfaces
orthogonal to $u^\mu $ on the brane and $R^\perp=h^{
\mu\nu}R^\perp_{\mu\nu }$.\\

\begin{figure}[h]\label{bianchi}
\begin{center}
\includegraphics{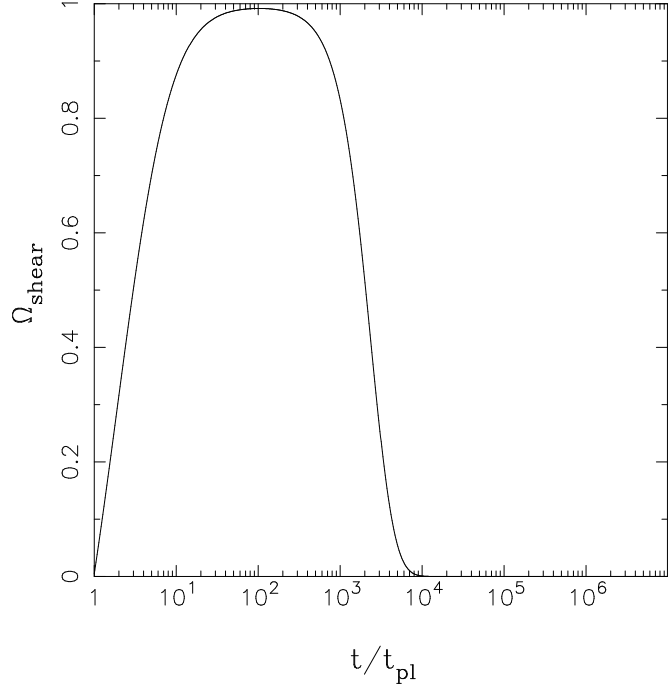}
\caption{The evolution of the dimensionless shear parameter
$\Omega_{\rm shear} = \sigma^2/6H^2$ on a Bianchi~I brane, for a
$V={1\over2}m^2\phi^2$ model. The early and late-time expansion of
the universe is isotropic, but the shear dominates during an
intermediate anisotropic stage (from~\cite{mss}).}
\end{center}
\end{figure}

The standard 4D general relativity results are regained when
$\lambda^{-1}\to0$ and ${\cal E}_{\mu\nu}=0$, which sets all right
hand sides to zero in Eqs.~(\ref{pr})--(\ref{gc2}). Together with
Eqs.~(\ref{pc1})--(\ref{pc2'}), these equations govern the
dynamics of the matter and gravitational fields on the brane,
incorporating both the local, high-energy (quadratic
energy-momentum) and nonlocal, KK (projected 5D Weyl) effects from
the bulk. High-energy terms are proportional to $\rho/\lambda$,
and are significant only when $\rho>\lambda$. The KK terms contain
$\cu$, $\cq_\mu $ and $\cp_{\mu\nu }$, with the latter two
quantities introducing imperfect fluid effects, even when the
matter has perfect fluid form.

Bulk effects give rise to important new driving and source terms
in the propagation and constraint equations. The vorticity
propagation and constraint, and the gravito-magnetic constraint
have no direct bulk effects, but all other equations do.
High-energy and KK energy density terms are driving terms in the
propagation of the expansion $\Theta$. The spatial gradients of
these terms provide sources for the gravito-electric field
$E_{\mu\nu}$. The KK anisotropic stress is a driving term in the
propagation of shear $\sigma_{\mu\nu}$ and the gravito-electric/
-magnetic fields, $E/H_{\mu\nu}$, and the KK momentum density is a
source for shear and the gravito-magnetic field. The 4D
Maxwell-Weyl equations show in detail the contribution to the 4D
gravito-electromagnetic field on the brane, i.e., $(E_{\mu\nu},H_{
\mu\nu})$, from the 5D Weyl field in the bulk.

An interesting example of how high-energy effects can modify
general relativistic dynamics arises in the analysis of
isotropization of Bianchi spacetimes. For a Binachi type~I brane,
Eq.~(\ref{gc2}) becomes~\cite{mss}
 \be
H^2={\kappa^2\over3}\rho\left(1+{\rho\over2\lambda}\right)+
{\Sigma^2\over a^6}\,,
 \ee
if we neglect the dark radiation, where $a$ and $H$ are the
average scale factor and expansion rate, and $\Sigma$ is the shear
constant. In general relativity, the shear term dominates as
$a\to0$, but in the brane-world, the high-energy $\rho^2$ term
will dominate if $w>0$, so that the matter-dominated early
universe is isotropic~\cite{mss,cs,b1}. This is illustrated in
Fig.~4.

Note that this conclusion is sensitive to the assumption that
$\cu\approx0$, which, by Eq.~(\ref{pc1'}) implies the restriction
 \be
\sigma^{\mu\nu}\cp_{\mu\nu} \approx 0\,.
 \ee
Relaxing this assumption can lead to non-isotropizing
solutions~\cite{ruth}.

The system of propagation and constraint equations, i.e.
Eqs.~(\ref{pc1})--(\ref{pc2'}) and (\ref{pr})--(\ref{gc2}), is
exact and nonlinear, applicable to both cosmological and
astrophysical modelling, including strong-gravity effects. In
general the system of equations is not closed: there is no
evolution equation for the KK anisotropic stress $\cp_{\mu\nu}$.

\section{Gravitational collapse and black holes on the brane}

The physics of brane-world compact objects and gravitational
collapse is complicated by a number of factors, especially the
confinement of matter to the brane, while the gravitational field
can access the extra dimension, and the nonlocal (from the brane
viewpoint) gravitational interaction between the brane and the
bulk. Extra-dimensional effects mean that the 4D matching
conditions on the brane, i.e., continuity of the induced metric
and extrinsic curvature across the 2-surface boundary, are much
more complicated to implement~\cite{germ,bgm}. High-energy
corrections increase the effective density and pressure of stellar
and collapsing matter. In particular this means that the effective
pressure does not in general vanish at the boundary 2-surface,
changing the nature of the 4D matching conditions on the brane.
The nonlocal KK effects further complicate the matching problem on
the brane, since they in general contribute to the effective
radial pressure at the boundary 2-surface. Gravitational collapse
inevitably produces energies high enough, i.e., $\rho\gg\lambda$,
to make these corrections significant.

We expect that extra-dimensional effects will be negligible
outside the high-energy, short-range regime. The corrections to
the weak-field potential, Eq.~(\ref{newt}), are at the second
post-Newtonian (2PN) level~\cite{gr1,inta}. However, modifications
to Hawking radiation may bring significant corrections even for
solar-sized black holes, as discussed below.

A vacuum on the brane, outside a star or black hole, satisfies the
brane field equations
 \be\label{vac}
R_{\mu\nu}=-{\cal E}_{\mu\nu}\,,~~ R^\mu {}_\mu =0={\cal E}^\mu
{}_\mu \,,~~ \nabla^\nu {\cal E}_{\mu\nu}=0\,.
 \ee
The Weyl term ${\cal E}_{\mu\nu}$ will carry an imprint of
high-energy effects that source KK modes (as discussed above).
This means that high-energy stars and the process of gravitational
collapse will in general lead to deviations from the 4D general
relativity problem. The weak-field limit for a static spherical
source, Eq.~(\ref{newt}), shows that ${\cal E}_{\mu\nu}$ must be
nonzero, since this is the term responsible for the corrections to
the Newtonian potential.

\subsection{The black string}

The projected Weyl term vanishes in the simplest candidate for a
black hole solution. This is obtained by assuming the exact
Schwarzschild form for the induced brane metric and ``stacking" it
into the extra dimension~\cite{chr}
 \bea
^\vu ds^2 &=& e^{-2|y|/\ell}\tilde{g}_{\mu\nu}dx^\mu dx^\nu +
dy^2\,, \label{bs1}
\\ \tilde{g}_{\mu\nu} &=&e^{2|y|/\ell}{g}_{\mu\nu}=
-(1-{2GM/ r})dt^2+ {dr^2\over 1-2GM/ r}+r^2d \Omega^2 \label{bs2}
\,.
 \eea
(Note that Eq.~(\ref{bs1}) is in fact a solution of the 5D field
equations~(\ref{rsefe}) if $\tilde{g}_{\mu\nu}$ is any 4D Einstein
vacuum solution, i.e., if $\tilde{R}_{\mu\nu}=0$, and this can be
generalized to the case $\tilde{R}_{\mu\nu}=-\tilde\Lambda
\tilde{g}_{\mu\nu}\,$~\cite{anlid,bmsv}.)

Each $\{y=\mbox{ const }\}$ surface is a 4D Schwarzschild
spacetime, and there is a line singularity along $r=0$ for all
$y$. This solution is known as the Schwarzschild black string,
which is clearly not localized on the brane $y=0$. Although
${}^\vu C_{ABCD}\neq0$, the projection of the bulk Weyl tensor
along the brane is zero, since there is no correction to the 4D
gravitational potential:
 \be
V(r)={GM \over r} ~\Rightarrow~ {\cal E}_{\mu\nu}=0\,.
 \ee
The violation of the perturbative corrections to the potential
signals some kind of non-AdS$_5$ pathology in the bulk. Indeed,
the 5D curvature is unbounded at the Cauchy horizon, as
$y\to\infty$~\cite{chr}:
 \be
{}^\vu R_{ABCD}\,{}^\vu R^{ABCD} = {40\over \ell^4}+{48G^2M^2
\over r^6}\, e^{4|y|/\ell}\,.
 \ee
Furthermore, the black string is unstable to large-scale
perturbations~\cite{g}.

Thus the ``obvious" approach to finding a brane black hole fails.
An alternative approach is to seek solutions of the brane field
equations with nonzero ${\cal E}_{\mu\nu}$~\cite{dmpr}. Brane
solutions of static black hole exteriors with 5D corrections to
the Schwarzschild metric have been found~\cite{dmpr,germ,bhsol},
but the bulk metric for these solutions has not been found.
Numerical integration into the bulk, starting from static black
hole solutions on the brane, is plagued with
difficulties~\cite{num}.

\subsection{Taylor expansion into the bulk}

One can use a Taylor expansion, as in Eq.~(\ref{tay}), in order to
probe properties of a static black hole on the brane~\cite{dms}.
(An alternative expansion scheme is discussed in~\cite{cas}.) For
a vacuum brane metric,
\begin{eqnarray}
\tilde{g}_{\mu\nu}(x,y) & = & \tilde {g}_{\mu\nu}(x,0)-{\cal
E}_{\mu\nu}(x,0+)y^2-\frac{2}{\ell} {\cal E}_{\mu\nu}(x,0+)|y|^3
\nonumber
\\ & & +  \frac{1}{12}\Bigl[
\Box {\cal E}_{\mu\nu}-\frac{32}{\ell^2} {\cal E}_{\mu\nu}
+2R_{\mu\alpha\nu\beta}{\cal E}^{\alpha\beta} +6{\cal E}_\mu
{}^\alpha {\cal E}_{\alpha\nu} \Bigr]_{y=0+}\,y^4+\cdots
\label{tayv}
\end{eqnarray}
This shows in particular that the propagating effect of 5D gravity
arises only at the fourth order of the expansion. For a static
spherical metric on the brane,
 \be\label{sss}
\tilde{g}_{\mu\nu}dx^\mu dx^\nu =-F(r)dt^2+{dr^2\over
H(r)}+r^2d\Omega^2\,,
 \ee
the projected Weyl term on the brane is given by
\begin{eqnarray}
{\cal E}_{00} & = & \frac{F}{r}\Bigl[H'-\frac{1-H}{r} \Bigr]\,,
\label{e00}\\ {\cal E}_{rr}
 & = & -\frac{1}{rH}\Bigl[{F'\over F}-\frac{1-H}{r} \Bigr]\,, \\
{\cal E}_{\theta\theta}&=&-1+H +\frac{r}{2}H\Bigl(\frac{F'}{F}
+\frac{H'}{H} \Bigr)\,.
\end{eqnarray}
These components allow one to evaluate the metric coefficients in
Eq.~(\ref{tayv}). For example, the area of the 5D horizon is
determined by $\tilde{g}_{\theta\theta}$; defining $\psi(r)$ as
the deviation from a Schwarzschild form for $H$, i.e.,
 \be
H(r)=1-{2m\over r} + \psi(r)\,,
 \ee
where $m$ is constant, we find
 \be
\tilde{g}_{\theta\theta}(r,y)=r^2-\psi'\left(1+{2\over\ell}|y|
\right)y^2 +{1\over 6r^2}\left[\psi'+{1\over2}
(1+\psi')(r\psi'-\psi)' \right]y^4+ \cdots
 \ee
This shows how $\psi$ and its $r$-derivatives determine the change
in area of the horizon along the extra dimension. For the black
string, $\psi=0$ and we have $\tilde{g}_{\theta\theta}(r,y)=r^2$.
For a large black hole, with horizon scale $\gg \ell$, we have
from Eq.~(\ref{newt}) that
 \be
\psi \approx -{4m\ell^2 \over 3r^3}\,.
 \ee
This implies that $\tilde{g}_{\theta\theta}$ is decreasing as we
move off the brane, consistent with a pancake-like shape of the
horizon. However, note that the horizon shape is tubular in
Gaussian normal coordinates~\cite{giaren}.

\subsection{The ``tidal charge" black hole}

The equations~(\ref{vac}) form a system of constraints on the
brane in the stationary case, including the static spherical case,
for which
 \be
\Theta=0=\omega_\mu =\sigma_{\mu\nu}\,,~~\dot{\rho}_{\cal E} = 0 =
\cq_\mu =\dot{\pi}^{\cal E}_{\mu\nu} \,.
 \ee
The nonlocal conservation equations $\nabla^\nu {\cal
E}_{\mu\nu}=0$ reduce to
 \be
{{1\over3}}{\D}_{\mu}\cu+{{4\over3}}\cu A_\mu  + \D^\nu
\cp_{\mu\nu}+A^\nu \cp_{\mu\nu}=0\,,
 \ee
where, by symmetry,
 \be\label{ass}
\cp_{\mu\nu}=\Pi_{\cal E}\left({1\over3}h_{\mu\nu}-r_\mu r_\nu
\right)\,,
 \ee
for some $\Pi_{\cal E}(r)$, with $r_\mu $ the unit radial vector.
The solution of the brane field equations requires the input of
${\cal E}_{\mu\nu}$ from the 5D solution. In the absence of a 5D
solution, one can make an assumption about ${\cal E}_{\mu\nu}$ or
$g_{\mu\nu}$ to close the 4D equations .

If we assume a metric on the brane of Schwarzschild-like form,
i.e., $H=F$ in Eq.~(\ref{sss}), then the general solution of the
brane field equations is~\cite{dmpr}
 \bea
F&=&1-{2GM\over  r} +{2G\ell Q\over r^2}\,,\label{bh}\\ {\cal
E}_{\mu\nu }&=&-{2G\ell Q\over r^4}\left[u_\mu  u_\nu -2r_\mu
r_\nu +h_{\mu\nu } \right]\,,
 \eea
where $Q$ is a constant. It follows that the KK energy density and
anisotropic stress scalar (defined via Eq.~(\ref{ass})) are given
by
 \be
\cu = {\ell Q \over 4\pi\, r^4} = {1\over 2} \Pi_{\cal E}\,.
 \ee

The solution Eq.~(\ref{bh}) has the form of the general relativity
Reissner-Nordstr\"om solution, but there is {\em no} electric
field on the brane. Instead, the nonlocal Coulomb effects
imprinted by the bulk Weyl tensor have induced a ``tidal" charge
parameter $Q$, where $Q=Q(M)$, since $M$ is the source of the bulk
Weyl field. We can think of the gravitational field of $M$ being
``reflected back" on the brane by the negative bulk cosmological
constant~\cite{dad}. If we impose the small-scale perturbative
limit ($r\ll\ell$) in Eq.~(\ref{newt2}), we find that
 \be
Q=-2M\,.
 \ee
Negative $Q$ is in accord with the intuitive idea that the tidal
charge strengthens the gravitational field, since it arises from
the source mass $M$ on the brane. By contrast, in the
Reissner-Nordstr\"om solution of general relativity, $Q\propto
+q^2$, where $q$ is the electric charge, and this weakens the
gravitational field. Negative tidal charge also preserves the
spacelike nature of the singularity, and it means that there is
only one horizon on the brane, outside the Schwarzschild horizon:
 \be
r_{\rm h}=GM\left[1+\sqrt {1 -{2\ell Q \over GM^2}}\,\right]=
GM\left[1+\sqrt {1 +{4\ell \over GM}}\,\right]\,.
 \ee

The tidal-charge black hole metric does not satisfy the far-field
$r^{-3}$ correction to the gravitational potential, as in
Eq.~(\ref{newt}), and therefore cannot describe the end-state of
collapse. However, Eq.~(\ref{bh}) shows the correct 5D behaviour
of the potential ($\propto r^{-2}$) at short distances, so that
the tidal-charge metric could be a good approximation in the
strong-field regime for small black holes.

\subsection{Realistic black holes}

Thus a simple brane-based approach, while giving useful insights,
does not lead to a realistic black hole solution. There is no
known solution representing a realistic black hole localized on
the brane, which is stable and without naked singularity. This
remains a key open question of nonlinear brane-world gravity.
(Note that an exact solution is known for a black hole on a
1+2-brane in a 4D bulk~\cite{ehm}, but this is a very special
case.) Given the nonlocal nature of ${\cal E}_{\mu\nu}$, it is
possible that the process of gravitational collapse itself leaves
a signature in the black hole end-state, in contrast with general
relativity and its no-hair theorems. There are contradictory
indications about the nature of the realistic black hole solution
on the brane:

\begin{itemize}

\item numerical simulations of highly relativistic static stars on
the brane~\cite{w} indicate that general relativity remains a good
approximation;

\item exact analysis of Oppenheimer-Snyder collapse on the brane
shows that the exterior is non-static~\cite{bgm}, and this is
extended to general collapse by arguments based on a generalized
AdS/CFT correspondence~\cite{t,efk}.

\end{itemize}

The first result suggests that static black holes could exist as
limits of increasingly compact static stars, but the second result
and conjecture suggest otherwise. This remains an open question.
More recent numerical evidence is also not conclusive, and it
introduces further possible subtleties to do with the size of the
black hole~\cite{ktn}.

On very small scales relative to the AdS$_5$ curvature scale,
$r\ll\ell$, the gravitational potential becomes 5D, as shown in
Eq.~(\ref{newt2}),
 \be
V(r) \approx {G\ell M \over r^2}={G_\vd M \over r^2}\,.
 \ee
In this regime, the black hole is so small that it does not ``see"
the brane, so that it is approximately a 5D Schwarzschild (static)
solution. However, this is always an approximation because of the
self-gravity of the brane (the situation is different in ADD-type
brane-worlds where there is no brane tension). As the black hole
size increases, the approximation breaks down. Nevertheless, one
might expect that static solutions exist on sufficiently small
scales. Numerical investigations appear to confirm
this~\cite{ktn}: static metrics satisfying the asymptotic AdS$_5$
boundary conditions are found if the horizon is small compared to
$\ell$, but no numerical convergence can be achieved close to
$\ell$. The numerical instability that sets in may mask the fact
that even the very small black holes are not strictly static. Or
it may be that there is a transition from static to non-static
behaviour. Or it may be that static black holes do exist on all
scales.

The 4D Schwarzschild metric cannot describe the final state of
collapse, since it cannot incorporate the 5D behaviour of the
gravitational potential in the strong-field regime (the metric is
incompatible with massive KK modes). A non-perturbative exterior
solution should have nonzero ${\cal E}_{\mu\nu}$ in order to be
compatible with massive KK modes in the strong-field regime. In
the end-state of collapse, we expect a ${\cal E}_{\mu\nu}$ which
goes to zero at large distances, recovering the Schwarzschild
weak-field limit, but which grows at short range. Furthermore,
${\cal E}_{\mu\nu}$ may carry a Weyl ``fossil record" of the
collapse process.

\subsection{Oppenheimer-Snyder collapse gives a non-static black
hole}

The simplest scenario in which to analyze gravitational collapse
is the Oppenheimer-Snyder model, i.e., collapsing homogeneous and
isotropic dust~\cite{bgm}. The collapse region on the brane has
FRW metric, while the exterior vacuum has an unknown metric. In 4D
general relativity, the exterior is a Schwarzschild spacetime; the
dynamics of collapse leaves {\em no} imprint on the exterior.

The collapse region has metric
\begin{equation}\label{os1}
ds^2=-d\tau^2+ {a(\tau)^2\left[dr^2+r^2d\Omega^2\right] \over (1+
 kr^2/4)^{2}}\,,
\end{equation}
where the scale factor satisfies the modified Friedmann equation
(see below),
\begin{eqnarray} \label{evol}
{\dot{a}^2 \over a^2}= {8\pi G\over 3}\rho\left(1 + {\rho \over
2\lambda}+ {\cu \over \rho} \right) \,.
\end{eqnarray}
The dust matter and the dark radiation evolve as
 \be
\rho=\rho_0\left({a_0\over\mu}\right)^3\,,~~ \cu=\rho_{{\cal
E}\,0}\left({a_0\over\mu}\right)^4\,,
 \ee
where $a_0$ is the epoch when the cloud started to collapse. The
proper radius from the centre of the cloud is $R(\tau)=r
a(\tau)/(1+ { {1\over4}} kr^2)$. The collapsing boundary surface
$\Sigma$ is given in the interior comoving coordinates as a
free-fall surface, i.e.\ $r=r_0=$~const, so that $R_\Sigma(\tau)=
r_0 a(\tau)/(1+ { {1\over4}} kr_0^2)$.

We can rewrite the modified Friedmann equation on the interior
side of $\Sigma$ as
\begin{equation}\label{geo1}
\dot{R}^2= {2GM\over R}+ {3GM^2\over 4\pi\lambda R^4}+ {Q \over
R^2}+ E\,,
\end{equation}
where the ``physical mass" $M$ (total energy per proper star
volume), the total ``tidal charge" $Q$ and the ``energy" per unit
mass are given by
 \bea
M &=&{4\pi a_0^3 r_0^3\rho_0 \over 3(1+ {{1\over4}}kr_0^2)^3}\,,\\
Q &=& {\rho_{{\cal E}\,0}a_0^4r_0^4 \over(1+ { {1\over4}}kr_0^2)^4}\,,\\
E&=&-{kr_0^2 \over(1+ { {1\over4}}kr_0^2)^2}>-1 \,.
 \eea

Now we {\em assume} that the exterior is static, and satisfies the
standard 4D junction conditions. Then we check whether this
exterior is physical by imposing the modified Einstein
equations~(\ref{vac}). We will find a contradiction.

The standard 4D Darmois-Israel matching conditions, which we
assume hold on the brane, require that the metric and the
extrinsic curvature of $\Sigma$ be continuous (there are no
intrinsic stresses on $\Sigma$). The extrinsic curvature is
continuous if the metric is continuous and if $\dot R$ is
continuous. We therefore need to match the metrics and $\dot R$
across $\Sigma$.

The most general static spherical metric that could match the
interior metric on $\Sigma$ is
\begin{eqnarray}
ds^2&=& -F(R)^2\left[1-{2Gm(R)\over R}\right]dt^2 +{dR^2 \over
1-{2Gm(R)/ R}} +R^2d\Omega^2 \,.\label{s1}
\end{eqnarray}
We need two conditions to determine the functions $F(R)$ and
$m(R)$. Now $\Sigma$ is a freely falling surface in both metrics,
and the radial geodesic equation for the exterior metric gives
$\dot{R}^2=-1+2Gm(R)/R+{\tilde{E}/ F(R)^2}\,,$ where $\tilde{E}$
is a constant and the dot denotes a proper time derivative, as
above. Comparing this with Eq.~(\ref{geo1}) gives one condition.
The second condition is easier to derive if we change to null
coordinates. The exterior static metric, with
$dv=dt+dR/[(1-2Gm/R)F]$, becomes
\begin{eqnarray}
ds^2&=& -F^2(1-2Gm/R)dv^2 +2FdvdR+R^2d\Omega^2\,.\label{s1'}
\end{eqnarray}
The interior Robertson-Walker metric takes the form~\cite{bgm}
\begin{eqnarray}
ds^2&=& -{\tau_{,v}^2\left[ 1- (k+\dot{a}^2)R^2/a^2 \right] dv^2
\over (1-kR^2/a^2)}+ {2 \tau_{,v} dvdR \over \sqrt{1-kR^2/a^2}}
+R^2d\Omega^2\,,\label{s1''}
\end{eqnarray}
where $d\tau=\tau_{,v}dv+(1+{1\over4}kr^2) dR/[r\dot
a-1+{1\over4}kr^2]$. Comparing Eqs.~(\ref{s1'}) and (\ref{s1''})
on $\Sigma$ gives the second condition. The two conditions
together imply that $F$ is a constant, which we can take as
$F(R)=1$ without loss of generality (choosing $\tilde E=E+1$), and
that
\begin{equation}\label{s3}
m(R)= M+{3M^2 \over 8\pi\lambda R^3}+ {Q\over 2G R}\,.
\end{equation}
In the limit $\lambda^{-1}=0=Q$, we recover the 4D Schwarzschild
solution. In the general brane-world case, Eqs.~(\ref{s1}) and
(\ref{s3}) imply that the brane Ricci scalar is
\begin{eqnarray}\label{m1}
R^\mu{}_\mu ={9GM^2 \over 2\pi\lambda R^6}\,.
\end{eqnarray}
However, this contradicts the field equations~(\ref{vac}), which
require
\begin{equation}\label{vac2}
R^\mu{}_\mu =0\,.
\end{equation}
It follows that a static exterior is only possible if
$M/\lambda=0\,,$ which is the 4D general relativity limit. In the
brane-world, collapsing homogeneous and isotropic dust leads to a
{\em non-static} exterior. Note that this no-go result does not
require any assumptions on the nature of the bulk spacetime, which
remains to be determined.

Although the exterior metric is not determined (see~\cite{govdad}
for a toy model), we know that its non-static nature arises from
\begin{itemize}
\item 5D bulk graviton stresses, which transmit effects nonlocally
from the interior to the exterior, and \item the non-vanishing of
the effective pressure at the boundary, which means that dynamical
information from the interior can be conveyed outside via the 4D
matching conditions.
\end{itemize}

The result suggests that gravitational collapse on the brane may
leave a signature in the exterior, dependent upon the dynamics of
collapse, so that astrophysical black holes on the brane may in
principle have KK ``hair". It is possible that the non-static
exterior will be transient, and will tend to a static geometry at
late times, close to Schwarzschild at large distances.

\subsection{AdS/CFT and black holes on 1-brane RS-type models}

Oppenheimer-Snyder collapse is very special; in particular, it is
homogeneous. One could argue that the non-static exterior arises
because of the special nature of this model. However, the
underlying reasons for non-static behaviour are not special to
this model; on the contrary, the role of high-energy corrections
and KK stresses will if anything be enhanced in a general,
inhomogeneous collapse. There is in fact independent heuristic
support for this possibility, arising from the AdS/CFT
correspondence.

The basic idea of the correspondence is that the classical
dynamics of the AdS$_5$ gravitational field correspond to the
quantum dynamics of a 4D conformal field theory on the brane. This
correspondence holds at linear perturbative order~\cite{acft}, so
that the RS 1-brane infinite AdS$_5$ brane-world (without matter
fields on the brane) is equivalently described by 4D general
relativity coupled to conformal fields,
 \be\label{cft}
G_{\mu\nu}=8\pi G T^{\rm (cft)}_{\mu\nu}\,.
 \ee
According to a conjecture~\cite{t}, the correspondence holds also
in the case where there is strong gravity on the brane, so that
the classical dynamics of the bulk gravitational field of the
brane black hole are equivalent to the dynamics of a
quantum-corrected 4D black hole (in the dual CFT-plus-gravity
description). In other words~\cite{t,efk},
\begin{itemize}
\item quantum backreaction due to Hawking radiation in the 4D
picture is described as classical dynamics in the 5D picture;

\item the black hole evaporates as a classical process in the 5D
picture, and there is thus no stationary black hole solution in RS
1-brane.

\end{itemize}

A further remarkable consequence of this conjecture is that
Hawking evaporation is dramatically enhanced, because of the very
large number, of order $(\ell/\ell_{\rm p})^2$, of CFT modes. The
energy loss rate due to evaporation is
 \be
{\dot M \over M} \sim N \left({1 \over G^2M^3}\right)\,,
 \ee
where $N$ is the number of light degrees of freedom. Using $N\sim
\ell^2/G$, this gives an evaporation timescale~\cite{t}
 \be
t_{\rm evap} \sim \left({M\over M_\odot}\right)^3 \left({1~{\rm
mm} \over \ell}\right)^2~{\rm yr}\,.
 \ee
A more detailed analysis~\cite{egk} shows that this expression
should be multiplied by a factor $\approx 100$. Then the existence
of stellar-mass black holes on long time scales places limits on
the AdS$_5$ curvature scale that are more stringent than the
table-top limit, Eq.~(\ref{tt}). The existence of black hole X-ray
binaries implies
 \be
\ell \lesssim  10^{-2}~{\rm mm}\,,
 \ee
already an order of magnitude improvement on the table-top limit.

One can also relate the Oppenheimer-Snyder result to these
considerations. In the AdS/CFT picture, the non-vanishing of the
Ricci scalar, Eq.~(\ref{m1}) arises from the trace of the Hawking
CFT energy-momentum tensor, as in Eq.~(\ref{cft}). If we evaluate
the Ricci scalar at the black hole horizon, $R\sim 2GM$, using
$\lambda=6M_\vd^6/M_{\rm p}^2$, we find
 \be
R^\mu{}_\mu  \sim {M_\vd ^{12}\,\ell^6 \over M^4}\,.
 \ee
The CFT trace on the other hand is given by $T^{\rm (cft)}\sim N
T_{\rm h}^4/M_{\rm p}^2$, so that
 \be
8\pi GT^{\rm (cft)}\sim {M_\vd ^{12}\,\ell^6 \over M^4}\,.
 \ee
Thus the Oppenheimer-Snyder result is qualitatively consistent
with the AdS/CFT picture.

Clearly the black hole solution, and the collapse process that
leads to it, have a far richer structure in the brane-world than
in general relativity, and deserve further attention. In
particular, two further topics are of interest:
\begin{itemize}
\item Primordial black holes in 1-brane RS-type cosmology have
been investigated in~\cite{inta,clan}. High-energy effects in the
early universe (see the next section) can significantly modify the
evaporation and accretion processes, leading to a prolonged
survival of these black holes. Such black holes evade the enhanced
Hawking evaporation described above when they are formed, because
they are much smaller than $\ell$. \item Black holes will also be
produced in particle collisions at energies $\gtrsim M_\vd$,
possibly well below the Planck scale. In ADD brane-worlds, where
$M_{4+d} =O({\rm TeV})$ is not ruled out by current observations
if $d>1$, this raises the exciting prospect of observing black
hole production signatures in the next-generation colliders and
cosmic ray detectors (see~\cite{cav,gid}).
\end{itemize}

\section{Brane-world cosmology: dynamics}

A 1+4-dimensional spacetime with spatial 4-isotropy (4D spherical/
plane/ hyperbolic symmetry) has a natural foliation into the
symmetry group orbits, which are 1+3-dimensional surfaces with
3-isotropy and 3-homogeneity, i.e. FRW surfaces. In particular,
the AdS$_5$ bulk of the RS brane-world, which admits a foliation
into Minkowski surfaces, also admits an FRW foliation since it is
4-isotropic. Indeed this feature of 1-brane RS-type cosmological
brane-worlds underlies the importance of the AdS/CFT
correspondence in brane-world cosmology~\cite{acftcosmo}.

The generalization of AdS$_5$ that preserves 4-isotropy and solves
the vacuum 5D Einstein equation~(\ref{rsefe}), is
Schwarzschild-AdS$_5$, and this bulk therefore admits an FRW
foliation. It follows that an FRW brane-world, the cosmological
generalization of the RS brane-world, is a part of
Schwarzschild-AdS$_5$, with the $Z_2$-symmetric FRW brane at the
boundary. (Note that FRW branes can also be embedded in non-vacuum
generalizations, e.g. in Reissner-Nordstr\"om-AdS$_5$ and
Vaidya-AdS$_5$.)

In natural static coordinates, the bulk metric is
 \bea
{}^\vu ds^2 &=& -F(R)dT^2+{dR^2 \over F(R)}+R^2\left({dr^2 \over
1- Kr^2}+r^2d\Omega^2\right)\,,\label{sads}\\ F(R) &=& K+
{R^2\over \ell^2} -{m \over R^2}\,,\label{sads2}
 \eea
where $K=0,\pm1$ is the FRW curvature index and $m$ is the mass
parameter of the black hole at $R=0$ (recall that the 5D
gravitational potential has $R^{-2}$ behaviour). The bulk black
hole gives rise to dark radiation on the brane via its Coulomb
effect. The FRW brane moves radially along the 5th dimension, with
$R=a(T)$, where $a$ is the FRW scale factor, and the junction
conditions determine the velocity via the Friedmann equation for
$a$~\cite{birk}. Thus one can interpret the expansion of the
universe as motion of the brane through the static bulk. In the
special case $m=0$ and $da/dT=0$, the brane is fixed and has
Minkowski geometry, i.e., the original RS 1-brane brane-world is
recovered, in different coordinates.

The velocity of the brane is coordinate-dependent, and can be set
to zero. We can use Gaussian normal coordinates, in which the
brane is fixed but the bulk metric is not manifestly
static~\cite{bdel}:
 \be\label{gnm}
{}^\vu ds^2 = -N^2(t,y)dt^2+A^2(t,y)\left[{dr^2 \over 1-
Kr^2}+r^2d\Omega^2\right]+dy^2\,.
 \ee
Here $a(t)=A(t,0)$ is the scale factor on the FRW brane at $y=0$,
and $t$ may be chosen as proper time on the brane, so that
$N(t,0)=1$. In the case where there is no bulk black hole ($m=0$),
the metric functions are
 \bea
N &=& {\dot {A}(t,y) \over \dot {a}(t)}\,,\label{gnm1} \\ A &=&
a(t)\left[ \cosh \left({y \over \ell}\right) -
\left\{1+{\rho(t)\over\lambda}\right\}\sinh \left({|y| \over
\ell}\right)\right]\,.\label{gnm2}
 \eea
Again, the junction conditions determine the Friedmann equation.
The extrinsic curvature at the brane is
 \be
K^\mu{}_\nu=\mbox{diag}\left({N'\over N}, {A'\over A}, {A'\over
A}, {A'\over A}\right)_{\rm brane}.
 \ee
Then, by Eq.~(\ref{ext}),
 \bea
{N'\over N}\Big|_{\rm brane}&=&{\kappa_\vd^2 \over
6}(2\rho+3p-\lambda)\,,\\ {A'\over A}\Big|_{\rm brane}
&=&-{\kappa_\vd^2 \over 6}(\rho+\lambda)\,.\label{biga}
 \eea
The field equations yield the first integral~\cite{bdel}
 \be
(AA')^2-{A^2\over N^2}\dot{A}^2+{\Lambda_\vd \over6}A^4+m=0\,,
 \ee
where $m$ is constant. Evaluating this at the brane, using
Eq.~(\ref{biga}), gives the modified Friedmann
equation~(\ref{mf}).

The dark radiation carries the imprint on the brane of the bulk
gravitational field. Thus we expect that ${\cal E}_{\mu\nu}$ for
the Friedmann brane contains bulk metric terms evaluated at the
brane. In Gaussian normal coordinates (using the field equations
to simplify),
 \be
{\cal E}^0{}_0=3{A''\over A}\Big|_{\rm brane}+{\Lambda_\vd \over
2}\,,~~ {\cal E}^i{}_j=-\left({1\over 3} {\cal E}^0{}_0\right)\,
\delta^i{}_j\,.
 \ee

Either form of the cosmological metric, Eq.~(\ref{sads}) or
(\ref{gnm}), may be used to show that 5D gravitational wave
signals can take ``short-cuts" through the bulk in travelling
between points A and B on the brane~\cite{speed}. The travel time
for such a graviton signal is less than the time taken for a
photon signal (which is stuck to the brane) from A to B.

Instead of using the junction conditions, we can use the covariant
3D Gauss-Codazzi equation~(\ref{gc2}) to find the modified
Friedmann equation:
 \be\label{mf}
H^2 = \frac{\kappa^2}{3} \rho\left(1+{\rho\over 2\lambda}\right)
+{m\over a^4}+ \frac{1}{3} \Lambda - \frac{K}{a^2} \,,
 \ee
on using Eq.~(\ref{dr}), where
 \be
m= \frac{\kappa^2}{3}\rho_{{\cal E}\,0}a_0^4\,.
 \ee
The covariant Raychauhuri equation~(\ref{pr}) yields
 \be
\dot H= - {\kappa^2\over 2}(\rho+p)\left(1+ {\rho\over
\lambda}\right)-2{m\over a^4}+{K\over a^2}\,,
 \ee
which also follows from differentiating Eq.~(\ref{mf}) and using
the energy conservation equation.

When the bulk black hole mass vanishes, the bulk geometry reduces
to AdS$_5$ and $\cu=0$. In order to avoid a naked singularity, we
assume that the black hole mass is non-negative, so that
$\rho_{{\cal E}\,0}\geq0$. [By Eq.~(\ref{sads2}), it is possible
to avoid a naked singularity with negative $m$ when $K=-1$,
provided $|m| \leq \ell^2/4$.] This additional effective
relativistic degree of freedom is constrained by nucleosynthesis
and CMB observations to be no more than $\sim$5\% of the radiation
energy density~\cite{lmsw,bm,dr}:
 \be
\left. {\cu \over \rho_{\rm rad}}\right|_{\rm nuc} \lesssim 0.05
 \ee
The other modification to the Hubble rate is via the high-energy
correction $\rho/\lambda$. In order to recover the observational
successes of general relativity, the high-energy regime where
significant deviations occur must take place before
nucleosynthesis, i.e., cosmological observations impose the lower
limit
 \be
\lambda > (1~{\rm MeV})^4~\Rightarrow~ M_\vd> 10^4~{\rm GeV}\,.
 \ee
This is much weaker than the limit imposed by table-top
experiments, Eq.~(\ref{rslimit}). Since $\rho^2/\lambda$ decays as
$a^{-8}$ during the radiation era, it will rapidly become
negligible after the end of the high-energy regime,
$\rho=\lambda$.

If $\cu=0$ and $K=0=\Lambda$, then the exact solution of the
Friedmann equations for $w=p/\rho=\mbox{ const}$ is~\cite{bdel}
 \be\label{ex1}
a=\,{\rm const}\,[t(t+t_{\lambda})]^{1/3(w+1)}\,, ~~
t_{\lambda}={M_{\rm p}\over\sqrt{3\pi\lambda}(1+w)} \lesssim
(1+w)^{-1}10^{-9}\, {\rm sec}\,,
 \ee
where $w> -1$. If $\cu\neq0$ (but $K=0=\Lambda$), then the
solution for the radiation era ($w={1\over3}$) is~\cite{bm}
 \be\label{ex2}
a=\,{\rm const}\,[t(t+t_{\lambda})]^{1/4}\,, ~~
t_{\lambda}={\sqrt{3}\,M_{\rm p}\over
4\sqrt{\pi\lambda}\,(1+\cu/\rho)}\,.
 \ee
For $t\gg t_\lambda$ we recover from Eqs.~(\ref{ex1}) and
(\ref{ex2}) the standard behaviour, $a\propto t^{2/3(w+1)}$,
whereas for $t\ll t_\lambda$, we have the very different behaviour
of the high-energy regime,
 \be
\rho\gg\lambda ~ \Rightarrow~ a \propto t^{1/3(w+1)}\,.
 \ee

When $w=-1$ we have $\rho=\rho_0$ from the conservation equation.
If $K=0=\Lambda$, we recover the de Sitter solution for $\cu=0$
and an asymptotically de Sitter solution for $\cu>0$:
 \bea
\cu=0:&&~~a=a_0\exp[H_0(t-t_0)]\,,~~H_0=\kappa\sqrt{{\rho_0\over
3} \left(1+ {\rho_0\over 2\lambda}\right)}\,,\\
\cu>0:&&~~a^2=\sqrt{m \over H_0}\, \sinh[2H_0(t-t_0)]\,.
 \eea

A qualitative analysis of the Friedmann equations is given
in~\cite{cs}.

\subsection{Brane-world inflation}

In 1-brane RS-type brane-worlds, where the bulk has only a vacuum
energy, inflation on the brane must be driven by a 4D scalar field
trapped on the brane. In more general brane-worlds, where the bulk
contains a 5D scalar field, it is possible that the 5D field
induces inflation on the brane via its effective
projection~\cite{hs}.

More exotic possibilities arise from the interaction between two
branes, including possible collision, which is mediated by a 5D
scalar field and which can induce either inflation~\cite{kss} or a
hot big-bang radiation era, as in the ``ekpyrotic" or cyclic
scenario~\cite{ek}, or in colliding bubble scenarios~\cite{bub}.
(See also~\cite{ek5} for colliding branes in an M~theory
approach.) Here we discuss the simplest case of a 4D scalar field
$\phi$ with potential $V(\phi)$ (see~\cite{lidrev} for a review).

High-energy brane-world modifications to the dynamics of inflation
on the brane have been investigated~\cite{mwbh,inf,mss}.
Essentially, the high-energy corrections provide increased Hubble
damping, since $\rho\gg\lambda$ implies $H$ is larger for a given
energy than in 4D general relativity. This makes slow-roll
inflation possible even for potentials that would be too steep in
standard cosmology~\cite{mwbh,steep,hulid1}.

The field satisfies the Klein-Gordon equation
\begin{equation}
\ddot{\phi}+3H\dot{\phi}+V'(\phi)=0\,. \label{5}
\end{equation}
In 4D general relativity, the condition for inflation,
$\ddot{a}>0$, is $\dot{\phi}^2<V(\phi)$, i.e., $p<-{1\over3}\rho$,
where $\rho={1\over2}\dot{\phi}^2+V$ and
$p={1\over2}\dot{\phi}^2-V$. The modified Friedmann equation leads
to a stronger condition for inflation: using Eq.~(\ref{mf}), with
$m=0=\Lambda=K$, and Eq.~(\ref{5}), we find that
\begin{equation}
\ddot{a}>0 ~\Rightarrow~ w<-{1\over3}\left[{1+2\rho/\lambda \over
1+ \rho/\lambda}\right]\,, \label{6}
\end{equation}
where the square brackets enclose the brane correction to the
general relativity result. As $\rho/\lambda/\rightarrow 0$, the 4D
result $w<-{1\over3}$ is recovered, but for $\rho>\lambda$, $w$
must be more negative for inflation. In the very high-energy limit
$\rho/\lambda\rightarrow\infty$, we have $w<-{2\over3}$. When the
only matter in the universe is a self-interacting scalar field,
the condition for inflation becomes
\begin{equation}
\label{endinf} \dot\phi^2 - V + \left[{{1\over2}\dot\phi^2 + V
\over \lambda}\left({5\over4}\dot\phi^2-{1\over2}V\right)\right] <
0 \,,
\end{equation}
which reduces to $\dot{\phi}^2<V$ when $\rho_\phi =
{1\over2}\dot\phi^2+V \ll\lambda$.

In the the slow-roll approximation,
\begin{eqnarray}
H^2 &\approx&  {\kappa^2\over3} V\left[ 1+{V\over2\lambda}
\right]\,,
 \label{7}\\
\dot\phi &\approx & -{V'\over 3H}\,. \label{8}
\end{eqnarray}
The brane-world correction term $V/\lambda$ in Eq.~(\ref{7})
serves to enhance the Hubble rate for a given potential energy,
relative to general relativity. Thus there is enhanced Hubble
`friction' in Eq.~(\ref{8}), and brane-world effects will
reinforce slow-roll at the same potential energy. We can see this
by defining slow-roll parameters that reduce to the standard
parameters in the low-energy limit:
\begin{eqnarray}
\label{epsilon} \epsilon &\equiv& -{\dot H \over H^2}={M_{\rm p}^2
\over 16\pi} \left( {V' \over V} \right)^2 \left[ {1+V/\lambda
\over(1+V/2\lambda)^2} \right] \,,\label{10}\\ \label{eta} \eta
&\equiv & -{\ddot\phi \over H \dot\phi}={M_{\rm p}^2 \over 8\pi}
\left( {V'' \over V} \right) \left[ {1 \over 1+V/2\lambda} \right]
\,.\label{11}
\end{eqnarray}
Self-consistency of the slow-roll approximation then requires
$\epsilon,|\eta|\ll 1$. At low energies, $V\ll\lambda$, the
slow-roll parameters reduce to the standard form. However at high
energies, $V\gg\lambda$, the extra contribution to the Hubble
expansion helps damp the rolling of the scalar field and the new
factors in square brackets become $\approx\lambda/V$:
 \bea
\epsilon\approx\epsilon_{\rm gr}\left[{ {4\lambda\over
V}}\right]\,,~ \eta\approx\eta_{\rm gr}\left[{ {2\lambda\over
V}}\right],
 \eea
where $\epsilon_{\rm gr},\eta_{\rm gr}$ are the standard general
relativity slow-roll parameters. In particular, this means that
steep potentials which do not give inflation in general
relativity, can inflate the brane-world at high energy and then
naturally stop inflating when $V$ drops below $\lambda$. These
models can be constrained because they typically end inflation in
a kinetic-dominated regime and thus generate a blue spectrum of
gravitational waves, which can disturb
nucleosynthesis~\cite{steep}. They also allow for the novel
possibility that the inflaton could act as dark matter or
quintessence at low energies~\cite{steep,dq}.

The number of e-folds during inflation, $N = \int Hdt$, is, in the
slow-roll approximation,
\begin{equation}
\label{efold} N \approx - {8\pi  \over M_{\rm p}^2}\int_{\phi_{\rm
i}}^{\phi_{\rm f}}{V\over V'} \left[ 1+{V \over 2\lambda}
 \right]  d\phi \,.
\end{equation}
Brane-world effects at high energies increase the Hubble rate by a
factor $V/2\lambda$, yielding more inflation between any two
values of $\phi$ for a given potential. Thus we can obtain a given
number of e-folds for a smaller initial inflaton value $\phi_{\rm
i}$. For $V\gg\lambda$, Eq.~(\ref{efold}) becomes
 \be
N \approx - {128\pi^3\over 3 M_{\vd}^6}\int_{\phi_{\rm
i}}^{\phi_{\rm f}}{V^2\over  V'}\,d\phi\,.
 \ee

The key test of any modified gravity theory during inflation, will
be the spectrum of perturbations produced due to quantum
fluctuations of the fields about their homogeneous background
values. We will discuss brane-world cosmological perturbations in
the next section. In general, perturbations on the brane are
coupled to bulk metric perturbations, and the problem is very
complicated. However on large scales on the brane, the density
perturbations decouple from the bulk metric
perturbations~\cite{m1,lmsw,gm}. For 1-brane RS-type models, there
is no scalar zero-mode of the bulk graviton, and in the extreme
slow-roll (de Sitter) limit, the massive scalar modes are heavy
and stay in their vacuum state during inflation~\cite{fk}. Thus it
seems a reasonable approximation in slow-roll to neglect the KK
effects carried by ${\cal E}_{\mu\nu}$ when computing the density
perturbations.

\begin{figure}[!bth]\label{inf}
\begin{center}
\includegraphics[height=3.5in,width=4.5in]{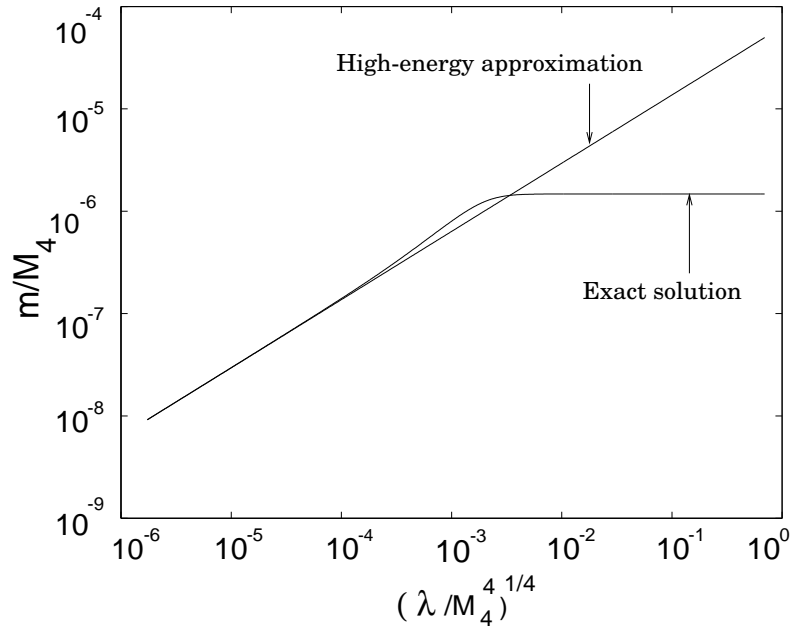}
\caption{The relation between the inflaton mass $m/M_4$
($M_4\equiv M_{\rm p}$) and the brane tension
$(\lambda/M_4^4)^{1/4}$ necessary to satisfy the COBE constraints.
The straight line is the approximation used in Eq.~(\ref{phi55}),
which at high energies is in excellent agreement with the exact
solution, evaluated numerically in slow-roll. (From~\cite{mwbh}.)}
\end{center}
\end{figure}

To quantify the amplitude of scalar (density) perturbations we
evaluate the usual gauge-invariant quantity
\begin{equation}
\label{defzeta} \zeta \equiv {\cal R}-{H\over\dot\rho}\delta\rho
\,,
\end{equation}
which reduces to the curvature perturbation, ${\cal R}$, on
uniform density hypersurfaces ($\delta\rho=0$). This is conserved
on large scales for purely adiabatic perturbations, as a
consequence of energy conservation (independently of the field
equations)~\cite{wmll}. The curvature perturbation on uniform
density hypersurfaces is given in terms of the scalar field
fluctuations on spatially flat hypersurfaces, $\delta\phi$, by
\begin{equation}
\zeta = H\,{\delta\phi\over\dot\phi} \,. \label{9}
\end{equation}
The field fluctuations at Hubble crossing ($k=aH$) in the
slow-roll limit are given by
$\langle\delta\phi^2\rangle\approx\left({H/2\pi} \right)^2$, a
result for a massless field in de Sitter space that is also
independent of the gravity theory~\cite{wmll}. For a single scalar
field the perturbations are adiabatic and hence the curvature
perturbation $\zeta$ can be related to the density perturbations
when modes re-enter the Hubble scale during the matter dominated
era which is given by $A_{\rm s}^2 = 4\langle \zeta^2 \rangle/25$.
Using the slow-roll equations and Eq.~(\ref{9}), this gives
\begin{equation}
\label{AS} A_{\rm s}^2 \approx \left . \left({512\pi\over75 M_{\rm
p} ^6}\, {V^3 \over V^{\prime2}}\right)\left[ {2\lambda + V \over
2\lambda} \right]^3 \right|_{k=aH}\,.
\end{equation}
Thus the amplitude of scalar perturbations is {\em increased}
relative to the standard result at a fixed value of $\phi$ for a
given potential.

The scale-dependence of the perturbations is described by the
spectral tilt
\begin{equation}
n_{\rm s}-1\equiv {d\ln A_{\rm s}^2 \over d\ln k} \approx
-6\epsilon + 2\eta \,,\label{15}
\end{equation}
where the slow-roll parameters are given in Eqs.~(\ref{epsilon})
and~(\ref{eta}). Because these slow-roll parameters are both
suppressed by an extra factor $\lambda/V$ at high energies, we see
that the spectral index is driven towards the Harrison-Zel'dovich
spectrum, $n_{\rm s}\to1$, as $V/\lambda\to\infty$; however, as
explained below, this does not necessarily mean that the
brane-world case is closer to scale-invariance than the general
relativity case.

As an example, consider the simplest chaotic inflation model
$V={1\over2}m^2\phi^2$. Equation~(\ref{efold}) gives the
integrated expansion from $\phi_{\rm i}$ to $\phi_{\rm f}$ as
\begin{equation}
N\approx {2\pi\over M_{\rm p}^2}\left(\phi_{\rm i}^2-\phi_{\rm
f}^2\right)+{\pi^2m^2\over3 M_\vd^6}\left(\phi_{\rm i}^4-\phi_{\rm
f}^4\right)\,. \label{20}
\end{equation}
The new high-energy term on the right leads to more inflation for
a given initial inflaton value $\phi_{\rm i}$.

The standard chaotic inflation scenario requires an inflaton mass
$m\sim 10^{13}$~GeV to match the observed level of anisotropies in
the cosmic microwave background (see below). This corresponds to
an energy scale $\sim 10^{16}$~GeV when the relevant scales left
the Hubble scale during inflation, and also to an inflaton field
value of order $3M_{\rm p}$. Chaotic inflation has been criticised
for requiring super-Planckian field values, since these can lead
to nonlinear quantum corrections in the potential.

If the brane tension $\lambda$ is much below $10^{16}$~GeV,
corresponding to $M_\vd<10^{17}$~GeV, then the terms quadratic in
the energy density dominate the modified Friedmann equation. In
particular the condition for the end of inflation given in
Eq.~(\ref{endinf}) becomes $\dot\phi^2<{2\over5}V$. In the
slow-roll approximation [using Eqs.~(\ref{7}) and~(\ref{8})]
$\dot\phi\approx-M_\vd^3/2\pi\phi$ and this yields
\begin{equation}
\phi_{\rm end}^4 \approx {5 \over 4\pi^2}\left({M_\vd\over
m}\right)^2M_\vd^4 \,.
\end{equation}
In order to estimate the value of $\phi$ when scales corresponding
to large-angle anisotropies on the microwave background sky left
the Hubble scale during inflation, we take $N_{\rm cobe}\approx55$
in Eq.~(\ref{20}) and $\phi_{\rm f}=\phi_{\rm end}$. The second
term on the right of Eq.~(\ref{20}) dominates, and we obtain
\begin{equation}
\label{phi55} \phi_{\rm cobe}^4 \approx {165\over\pi^2}
\left({M_\vd \over m}\right)^2M_\vd^4 \,. 
\end{equation}
Imposing the COBE normalization on the curvature perturbations
given by Eq.~(\ref{AS}) requires
\begin{equation}
A_{\rm s}\approx \left({8\pi^2\over45}\right){m^4\phi_{\rm
cobe}^5\over M_\vd^6} \approx 2\times10^{-5}\,.\label{22}
\end{equation}
Substituting in the value of $\phi_{\rm cobe}$ given by
Eq.~(\ref{phi55}) shows that in the limit of strong brane
corrections, observations require
\begin{equation}
m \approx 5\times 10^{-5}\, M_\vd\,,~~\phi_{\rm cobe}\approx
3\times 10^2\,M_\vd\,. \label{23}
\end{equation}
Thus for $M_\vd<10^{17}$~GeV, chaotic inflation can occur for
field values below the 4D Planck scale, $\phi_{\rm cobe}<M_{\rm
p}$, although still above the 5D scale $M_5$. The relation
determined by COBE constraints for arbitrary brane tension is
shown in Fig.~5, together with the high-energy approximation used
above, which provides an excellent fit at low brane tension
relative to $M_4$.

It must be emphasized that in comparing the high-energy
brane-world case to the standard 4D case, we implicitly require
the same potential energy. However, precisely because of the
high-energy effects, large-scale perturbations will be generated
at different values of $V$ than in the standard case, specifically
at lower values of $V$, closer to the reheating minimum. Thus
there are two competing effects, and it turns out that the shape
of the potential determines which is the dominant
effect~\cite{lidsmi}. For the quadratic potential, the lower
location on $V$ dominates, and the spectral tilt is slightly
further from scale invariance than in the standard case. The same
holds for the quartic potential. Data from WMAP and 2dF can be
used to constrain inflationary models via their deviation from
scale invariance, and the high-energy brane-world versions of the
quadratic and quartic potentials are thus under more pressure from
data than their standard counterparts~\cite{lidsmi}, as shown in
Fig.~6.

\begin{figure}[!bth]\label{plotnsr2}
\begin{center}
\includegraphics[height=3.5in,width=4.5in]{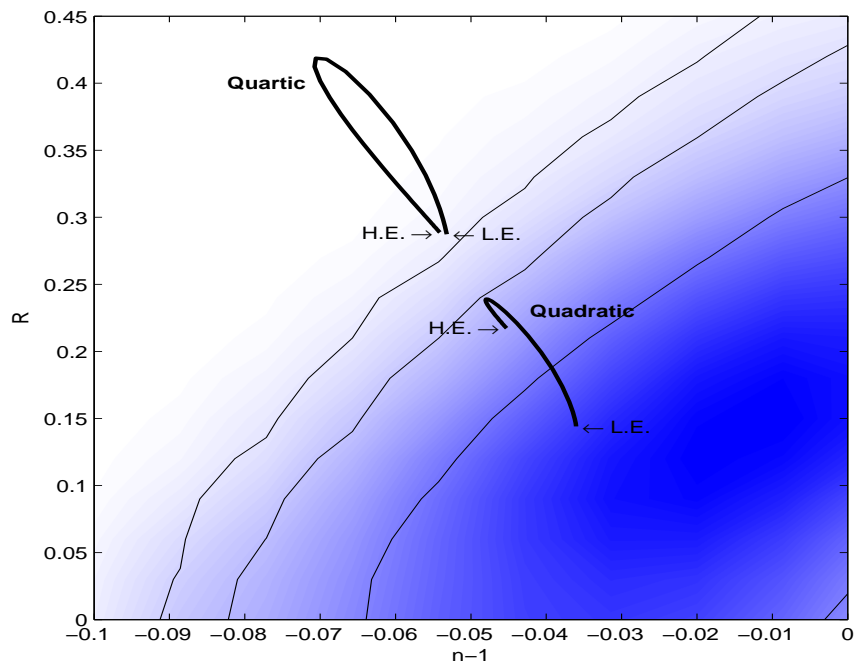}
\caption{Constraints from WMAP data on inflation models with
quadratic and quartic potentials, where $R$ is the ratio of tensor
to scalar amplitudes and $n$ is the scalar spectral index. The
high energy (H.E.) and low energy (L.E.) limits are shown, with
intermediate energies in between, and the 1-$\sigma$ and
2-$\sigma$ contours are also shown. (From~\cite{lidsmi}.)}
\end{center}
\end{figure}

Other perturbation modes have also been investigated:
\begin{itemize}
\item High-energy inflation on the brane also generates a
zero-mode (4D graviton mode) of tensor perturbations, and
stretches it to super-Hubble scales, as will be discussed below.
This zero-mode has the same qualitative features as in general
relativity, remaining frozen at constant amplitude while beyond
the Hubble horizon. Its amplitude is enhanced at high energies,
although the enhancement is much less than for scalar
perturbations~\cite{lmw}:
 \bea
A_{\rm t}^2 &\approx& \left({32V\over 75 M_{\rm p}^2}\right)
\left[ {3V^2 \over 4\lambda^2}\right],\label{higw}\\  {A_{\rm
t}^2\over A_{\rm s}^2} &\approx& \left({M_{\rm p}^2\over
16\pi}\,{V'^2\over V^2}\right) \left[ {6\lambda\over
V}\right].\label{ten}
 \eea
Equation~(\ref{ten}) means that brane-world effects suppress the
large-scale tensor contribution to CMB anisotropies. The tensor
spectral index at high energy has a smaller magnitude than in
general relativity,
 \be
n_{\rm t}=-3\epsilon\,,
 \ee
but remarkably the same consistency relation as in general
relativity holds~\cite{hulid1}:
 \be
n_{\rm t} = -2{A_{\rm t}^2\over A_{\rm s}^2}\,.
 \ee
This consistency relation persists when $Z_2$ symmetry is
dropped~\cite{hulid2} (and in a two-brane model with stabilized
radion~\cite{gklr}). It holds only to lowest order in slow-roll,
as in general relativity, but the reason for this~\cite{seetay}
and the nature of the corrections~\cite{cal} are not settled.

The massive KK modes of tensor perturbations remain in the vacuum
state during slow-roll inflation~\cite{lmw,grs}. The evolution of
the super-Hubble zero mode is the same as in general relativity,
so that high-energy brane-world effects in the early universe
serve only to rescale the amplitude. However, when the zero mode
re-enters the Hubble horizon, massive KK modes can be excited.

\item Vector perturbations in the bulk metric can support vector
metric perturbations on the brane, even in the absence of matter
perturbations (see the next section). However, there is no
normalizable zero mode, and the massive KK modes stay in the
vacuum state during brane-world inflation~\cite{bmwv}. Therefore,
as in general relativity, we can neglect vector perturbations in
inflationary cosmology.

\end{itemize}

Brane-world effects on large-scale isocurvature perturbations in
2-field inflation have also been considered~\cite{abd}.
Brane-world (p)reheating after inflation is discussed
in~\cite{preheat}.

\subsection{Brane-world instanton}

\begin{figure}[!bth]\label{fig1}
\begin{center}
\includegraphics[height=3.5in,width=4.5in]{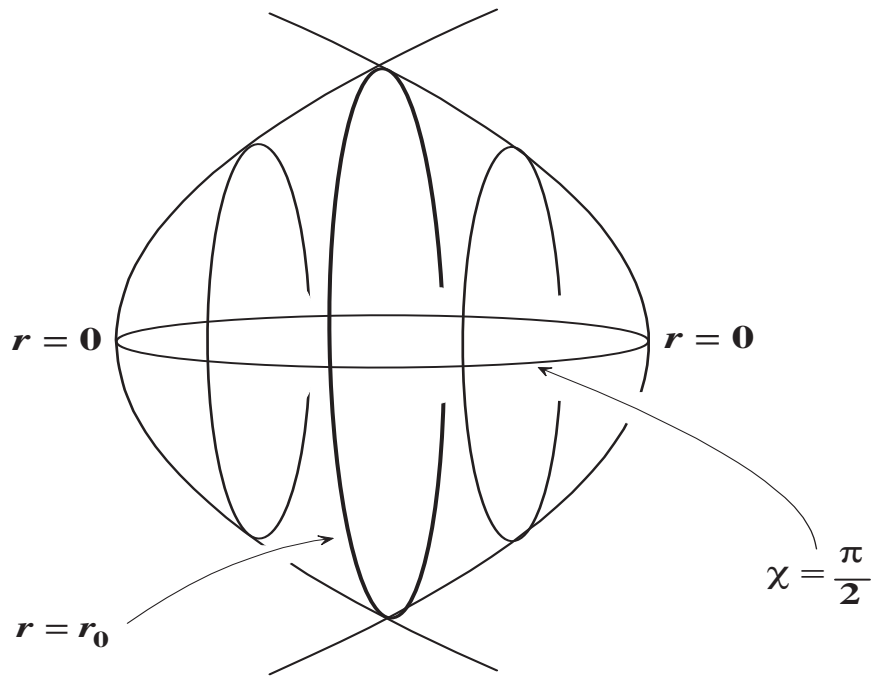}
\caption{Brane-world instanton. (From~\cite{gs}.)}
\end{center}
\end{figure}

The creation of an inflating brane-world can be modelled as a de
Sitter instanton in a way that closely follows the 4D instanton,
as shown in~\cite{gs}. The instanton consists of two identical
patches of AdS$_5$ joined together along a dS$_4$ brane with
compact spatial sections. The instanton describes the ``birth" of
both the inflating brane and the bulk spacetime, which are
together ``created from nothing", i.e. the point at the south pole
of the de Sitter 4-sphere. The Euclidean AdS$_5$ metric is
 \be
{}^\vu ds^2_{\rm euclid} = dr^2+\ell^2\sinh^2(r/\ell)\left[
d\chi^2+\sin^2 \chi\, d\Omega^2_{(3)} \right]\,,
 \ee
where $d\Omega^2_{(3)}$ is a 3-sphere, and $r\leq r_0$. The
Euclidean instanton interpolates between $r=0$ (``nothing") and
$r=r_0$ (the created universe), which is a spherical brane of
radius
 \be
H_0^{-1} \equiv \ell \sinh (r_0/\ell)\,.
 \ee
After creation, the brane-world evolves according to the
Lorentzian continuation, $\chi \to iH_0t+\pi/2$,
 \be
{}^\vu ds^2 = dr^2+(\ell H_0)^2\sinh^2(r/\ell)\left[-dt^2+H_0^{-2}
\cosh^2(H_0t)\, d\Omega^2_{(3)} \right]\,.
 \ee
(See Fig.~7.)

\subsection{Models with non-empty bulk}

The single-brane cosmological model can be generalized to include
stresses other than $\Lambda_\vd$ in the bulk.

\begin{itemize}
\item The simplest example arises from considering a charged bulk
black hole, leading to the Reissner-Nordstr\"om AdS$_5$ bulk
metric~\cite{bv}. This has the form of Eq.~(\ref{sads}), with
 \be\label{rnads}
F(R) = K+ {R^2\over \ell^2} -{m \over R^2}+{q^2\over R^4}\,,
 \ee
where $q$ is the ``electric" charge parameter of the bulk black
hole. The metric is a solution of the 5D Einstein-Maxwell
equations, so that ${}^\vu T_{AB}$ in Eq.~(\ref{5efe}) is the
energy-momentum tensor of a radial static 5D ``electric" field. In
order for the field lines to terminate on the boundary brane, the
brane should carry a charge $-q$. Since the RNAdS$_5$ metric is
4-isotropic, it is still possible to embed a FRW brane in it,
which is moving in the coordinates of Eq.~(\ref{sads}).

The effect of the black hole charge on the brane arises via the
junction conditions and leads to the modified Friedmann
equation~\cite{bv},
 \be\label{mf2}
H^2 = \frac{\kappa^2}{3} \rho\left(1+{\rho\over 2\lambda}\right)
+{m\over a^4}-{q^2\over a^6}+ \frac{1}{3} \Lambda - \frac{K}{a^2}
\,.
 \ee
The field lines that terminate on the brane imprint on the brane
an effective negative energy density $-3q^2/(\kappa^2a^6)$, which
redshifts like stiff matter ($w=1$). The negativity of this term
introduces the possibility that at high energies it can bring the
expansion rate to zero and cause a turn-around or bounce (but
see~\cite{hovmy} for problems with such bounces).

Apart from negativity, the key difference between this ``dark
stiff matter" and the dark radiation term $m/a^4$, is that the
latter arises from the bulk Weyl curvature via the ${\cal
E}_{\mu\nu}$ tensor, while the former arises from non-vacuum
stresses in the bulk via the ${\cal F}_{\mu\nu}$ tensor in
Eq.~(\ref{e:einstein1}). The dark stiff matter does not arise from
massive KK modes of the graviton.

\item Another example is provided by the Vaidya-AdS$_5$ metric,
which can be written after transforming to a new coordinate
$v=T+\int dR/F$ in Eq.~(\ref{sads}), so that $v=\,$const are null
surfaces, and
 \bea
{}^\vu ds^2 &=& -F(R,v)dv^2+2dvdR+R^2\left({dr^2 \over 1-
Kr^2}+r^2d\Omega^2\right)\,,\\ F(R,v) &=& K+ {R^2\over \ell^2}
-{m(v) \over R^2}\,.
 \eea
This model has a moving FRW brane in a 4-isotropic bulk (which is
not static), with either a radiating bulk black hole ($dm/dv<0$),
or a radiating brane ($dm/dv>0$)~\cite{ckn}. The metric satisfies
the 5D field equations~(\ref{5efe}) with a null-radiation
energy-momentum tensor,
 \be\label{vem}
{}^\vu T_{AB} = \psi k_Ak_B\,,~k_Ak^A=0\,,~k_Au^A=1\,,
 \ee
where $\psi\propto dm/dv$. It follows that
 \be\label{vf}
{\cal F}_{\mu\nu}= \kappa_\vd^{-2} \psi h_{\mu\nu}\,.
 \ee
In this case, the same effect, i.e. a varying mass parameter $m$,
contributes to both ${\cal E}_{\mu\nu}$ and ${\cal F}_{\mu\nu}$ in
the brane field equations. The modified Friedmann equation has the
standard 1-brane RS-type form, but with a dark radiation term that
no longer behaves strictly like radiation:
 \be\label{mf3}
H^2 = \frac{\kappa^2}{3} \rho\left(1+{\rho\over 2\lambda}\right)
+{m(t)\over a^4}+ \frac{1}{3} \Lambda - \frac{K}{a^2} \,.
 \ee

By Eqs.~(\ref{cong}) and (\ref{vem}), we arrive at the matter
conservation equations,
 \be
\nabla^\nu T_{\mu\nu}=-2\psi u_\mu \,.
 \ee
This shows how the brane loses ($\psi>0$) or gains ($\psi<0$)
energy in exchange with the bulk black hole. For an FRW brane,
this equation reduces to
 \be
\dot\rho+3H(\rho+p)=-2\psi\,.
 \ee
The evolution of $m$ is governed by the 4D contracted Bianchi
identity, using Eq.~(\ref{vf}):
 \be \label{nlcv}
\nabla^\mu{\cal
E}_{\mu\nu}={6\kappa^2\over\lambda}\,\nabla^\mu{\cal
S}_{\mu\nu}+{2\over3}\left[ \kappa_\vd^2\left(\dot\psi+\Theta\psi
\right) -3\kappa^2 \psi\right]u_\mu +{2\over3} \kappa_\vd^2\D_\mu
\psi\,.
 \ee
For an FRW brane, this yields
 \be
\dot{\rho}_{\cal E}+4H\cu=2\psi-{2\over3}{\kappa_\vd^2
\over\kappa^2}\left(\dot\psi+3H\psi\right)\,,
 \ee
where $\cu=3m(t)/(\kappa^2a^4)$.

\item A more complicated bulk metric arises when there is a
self-interacting scalar field $\Phi$ in the bulk~\cite{mw,sca}. In
the simplest case, when there is no coupling between the bulk
field and brane matter, this gives
 \be\label{bsf}
{}^\vu T_{AB}=\Phi_{,A}\Phi_{,B}-{}^\vu g_{AB}\left[ V(\Phi)+
{1\over 2}\,{}^\vu g^{CD}\Phi_{,C}\Phi_{,D} \right]\,,
 \ee
where $\Phi(x,y)$ satisfies the 5D Klein-Gordon equation,
 \be
{}^\vu \Box\Phi-V'(\Phi)=0\,.
 \ee
The junction conditions on the field imply that
 \be
\partial_y\Phi(x,0)=0\,.
 \ee
Then Eqs.~(\ref{cong}) and (\ref{bsf}) show that matter
conservation continues to hold on the brane in this simple case:
 \be
\nabla^\nu T_{\mu\nu}=0\,.
 \ee
From Eq.~(\ref{bsf}) one finds that
 \be
{\cal F}_{\mu\nu}={1\over
4\kappa_\vd^2}\left[4\phi_{,\mu}\phi_{,\nu}- g_{\mu\nu}\left\{
3V(\phi)+ {5\over 2} g^{\alpha\beta}\phi_{,\alpha}\phi_{,\beta}
\right\} \right]\,,
 \ee
where
 \be
\phi(x)=\Phi(x,0)\,,
 \ee
so that the modified Friedmann equation becomes
 \be\label{mf4}
H^2 = \frac{\kappa^2}{3} \rho\left(1+{\rho\over 2\lambda}\right)
+{m\over a^4}+ {\kappa_\vd^2\over6}\left[{1\over2}\dot\phi^2
+V(\phi) \right]+ \frac{1}{3} \Lambda - \frac{K}{a^2} \,.
 \ee

When there is coupling between brane matter and the bulk scalar
field, then the Friedmann and conservation equations are more
complicated~\cite{mw,sca}.

\end{itemize}

\section{Brane-world cosmology: perturbations}

The background dynamics of brane-world cosmology are simple
because the FRW symmetries simplify the bulk and rule out nonlocal
effects. But perturbations on the brane immediately release the
nonlocal KK modes. Then the 5D bulk perturbation equations must be
solved in order to solve for perturbations on the brane. These 5D
equations are partial differential equations for the 3D Fourier
modes, with both initial and boundary conditions needed.

The theory of gauge-invariant perturbations in brane-world
cosmology has been extensively investigated and
developed~\cite{m1,lmsw,bm,mwbh,steep,gm,lmw,grs,mu,pert,bmw,l1,l2}
and is qualitatively well understood. The key remaining task is
integration of the coupled brane-bulk perturbation equations.
Special cases have been solved, where these equations effectively
decouple~\cite{lmsw,bm,l1,l2}, and approximation schemes have
recently been developed~\cite{sod,koy,rbbd,kkt,elmw,mbb} for the
more general cases where the coupled system must be solved. From
the brane viewpoint, the bulk effects, i.e., the high-energy
corrections and the KK modes, act as source terms for the brane
perturbation equations. At the same time, perturbations of matter
on the brane can generate KK modes (i.e., emit 5D gravitons into
the bulk) which propagate in the bulk and can subsequently
interact with the brane. This nonlocal interaction amongst the
perturbations is at the core of the complexity of the problem. It
can be elegantly expressed via integro-differential
equations~\cite{mu}, which take the form (assuming no incoming 5D
gravitational waves)
 \be \label{ide}
A_k(t)=\int dt'\,{\cal G}(t,t') B_k(t')\,,
 \ee
where ${\cal G}$ is the bulk retarded Green's function evaluated
on the brane, and $A_k, B_k$ are made up of brane metric and
matter perturbations and their (brane) derivatives, and include
high-energy corrections to the background dynamics. Solving for
the bulk Green's function, which then determines ${\cal G}$, is
the core of the 5D problem.

We can isolate the KK anisotropic stress $\cp_{\mu\nu}$ as the
term that must be determined from 5D equations. Once
$\cp_{\mu\nu}$ is determined in this way, the perturbation
equations on the brane form a closed system. The solution will be
of the form, expressed in Fourier modes:
\begin{equation}\label{e:soln}
\pi^{\cal E}_k(t) \propto \int d t'\,\,{\cal G}(t,t') F_k(t') \,,
\end{equation}
where the functional $F_k$ will be determined by the covariant
brane perturbation quantities and their derivatives. It is known
in the case of a Minkowski background~\cite{ssm}, but not in the
cosmological case.

The KK terms act as source terms modifying the standard general
relativity perturbation equations, together with the high-energy
corrections. For example, the linearization of the shear
propagation equation~(\ref{pe5}) yields
 \be
\dot{\sigma}_{\mu\nu}+2H\sigma_{\mu\nu}+
E_{\mu\nu}-{\kappa^2\over2} \pi_{\mu\nu}- \D_{\langle
\mu}A_{\nu\rangle} = {\kappa^2\over 2}\cp_{\mu\nu}- {\kappa^2\over
4}(1+3w){\rho\over\lambda}\,\pi_{\mu\nu}\,.
 \ee
In 4D general relativity, the right hand side is zero. In the
brane-world, the first source term on the right is the KK term,
the second term is the high-energy modification. The other
modification is a straightforward high-energy correction of the
background quantities $H$ and $\rho$ via the modified Friedmann
equations.

As in 4D general relativity, there are various different, but
essentially equivalent, ways to formulate linear cosmological
perturbation theory. First I describe the covariant brane-based
approach.

\subsection{1+3-covariant perturbation equations on the brane}

In the 1+3-covariant approach~\cite{m1,l1,mjap}, perturbative
quantities are projected vectors, $V_\mu =V_{\langle \mu\rangle}$,
and projected symmetric tracefree tensors, $W_{\mu\nu}=W_{\langle
\mu \nu\rangle}$, which are gauge-invariant since they vanish in
the background. These are decomposed into (3D) scalar, vector and
tensor modes as
\begin{eqnarray}
 V_\mu  &=& \D_\mu  V+\bar{V}_\mu \,, \\
 W_{\mu\nu} &=& \D_{\langle\mu}\D_{\nu\rangle}{W}
+\D_{\langle\mu}\bar{W}_{\nu\rangle}+\bar{W}_{\mu\nu}\,,
\end{eqnarray}
where $\bar{W}_{\mu\nu}=\bar{W}_{\langle \mu\nu\rangle}$ and an
overbar denotes a (3D) transverse quantity,
 \be
\D^\mu\bar{V}_\mu =0= \D^\nu \bar{W}_{\mu\nu}\,.
 \ee
In a local inertial frame comoving with $u^\mu$, i.e.,
$u^\mu=(1,\vec 0)$, all time components may be set to zero:
$V_\mu=(0,V_i)$, $W_{0\mu}=0$,
$\vec\nabla_\mu=(0,\vec\nabla_i)$.

Purely scalar perturbations are characterized by the fact that
vectors and tensors are derived from scalar potentials, i.e.,
 \be
\bar{V}_\mu =\bar{W}_\mu =\bar{W}_{\mu\nu}=0\,.
 \ee
Scalar perturbative quantities are formed from the potentials via
the (3D) Laplacian, e.g., ${\cal V}=\D^\mu\D_\mu  V\equiv \D^2 V$.
Purely vector perturbations are characterized by
 \be\label{vec}
V_\mu =\bar{V}_\mu \,,~W_{\mu\nu}=\D_{\langle\mu}
\bar{W}_{\nu\rangle}\,,~\curl\D_\mu  f=-2\dot{f}\omega_\mu \,,
 \ee
where $\omega_\mu $ is the vorticity, and purely tensor by
 \be
\D_\mu  f=0=V_\mu \,,~ W_{\mu\nu}=\bar{W}_{\mu\nu}\,.
 \ee

The KK energy density produces a scalar mode $\D_\mu {\cu}$ (which
is present even if $\cu=0$ in the background). The KK momentum
density carries scalar and vector modes, and the KK anisotropic
stress carries scalar, vector and tensor modes:
\begin{eqnarray}
{\cq_\mu }&=&\D_\mu \cq+{\bcq_\mu }\,,\label{q*}\\
{\cp_{\mu\nu}}&=&\D_{\langle\mu}\D_{\nu\rangle}\cp +\D_{\langle
\mu}{\bcp_{\nu\rangle}}+{\bcp_{\mu\nu}}\,. \label{p*}
\end{eqnarray}

Linearizing the conservation equations for a single adiabatic
fluid, and the nonlocal conservation equations, we obtain
\begin{eqnarray}
&&\dot\rho+\Theta(\rho+p)=0\,,\label{ecl}
\\ &&c_{\rm s}^2\D_\mu +(\rho+p)A_\mu =0\,,\label{momcl}\\
&& \dot{\rho}_{\cal E}+{4\over3}\Theta{\cu}+\D^\mu{\cq_\mu }=0\,,
\label{nlc1}
\\&& \dot{q}^{\cal E}_\mu +4H{\cq_\mu }
+{{1\over3}}\D_\mu {\cu}+{{4\over3}}{\cu}A_\mu +\D^\nu
{\cp_{\mu\nu}} =-{ (\rho+p)\over\lambda}\D_\mu  \rho
\,.\label{nlc2}
\end{eqnarray}

Linearizing the remaining propagation and constraint equations
leads to
\begin{eqnarray}
&&\dot{\Theta}+{{1\over3}}\Theta^2 -\D^\mu  A_\mu
+{{1\over2}}\kappa^2(\rho + 3p) -\Lambda =
-{{\kappa^2\over2}}(2\rho+3p){\rho\over\lambda}- \kappa^2\cu\,,
\label{prl}\\ && \dot{\omega}_{\mu} +2H\omega_\mu
+{{1\over2}}\curl A_\mu  =0 \,,\label{pe4l}\\ &&
\dot{\sigma}_{\mu\nu} +2H\sigma_{\mu\nu} +E_{\mu\nu }-\D_{\langle
\mu}A_{ \nu\rangle } ={\kappa^2\over 2}\cp_{\mu\nu}\,,
\label{pe5l}\\ && \dot{E}_{\mu\nu} +3H E_{\mu\nu} -\curl
H_{\mu\nu} +{{\kappa^2\over2}}(\rho+p)\sigma_{\mu\nu} =
-{{\kappa^2\over2}}(\rho+p){\rho\over\lambda}
\sigma_{\mu\nu}\nonumber\\&&~~{}
-{\kappa^2\over6}\left[4\cu\sigma_{\mu\nu}+3\dot{\pi}^{\cal
E}_{\mu\nu} +3H\cp_{\mu\nu}
+3\D_{\langle\mu}\cq_{ \nu\rangle} \right] \,, \label{pe6l}\\
&&\dot{H}_{\mu\nu} +3H H_{\mu\nu} +\curl E_{\mu\nu}
={\kappa^2\over 2} \curl \cp_{\mu\nu} \,, \label{pe7l}
\\&&\D^\mu \omega_\mu  =0\,,\label{pcc1l}\\
&&\D^\nu \sigma_{\mu\nu}-\curl\omega_\mu  -{{2\over3}}\D_\mu
\Theta = -\cq_\mu
 \,,\label{pcc2l}\\ &&
\curl\sigma_{\mu\nu}+\D_{\langle \mu}\omega_{ \nu\rangle  }
 -H_{\mu\nu}=0 \,,\label{pcc3l}\\ && \D^\nu  E_{\mu\nu}
 -{{\kappa^2\over3}}\D_\mu \rho
={\kappa^2\over3}{ \rho\over \lambda} \D_\mu \rho +{\kappa^2\over
6 }\left[2\D_\mu \cu-4H \cq_\mu -3\D^\nu \cp_{\mu\nu}\right]
 \,,\label{pcc4l}\\
 &&\D^\nu  H_{\mu\nu}
-\kappa^2(\rho+p)\omega_\mu  = \kappa^2(\rho+ p){\rho\over\lambda}
\omega_\mu  + {\kappa^2\over 6}\left[8 \cu \omega_\mu
-3\curl\cq_\mu \right] \,.\label{pcc5l}
\end{eqnarray}
Equations~(\ref{ecl}), (\ref{nlc1}) and (\ref{prl}) do not provide
gauge-invariant equations for perturbed quantities, but their
spatial gradients do.

These equations are the basis for a 1+3-covariant analysis of
cosmological perturbations from the brane observer's viewpoint,
following the approach developed in 4D general
relativity~\cite{covp}. The equations contain scalar, vector and
tensor modes, which can be separated out if desired. They are not
a closed system of equations until $\cp_{\mu\nu}$ is determined by
a 5D analysis of the bulk perturbations. An extension of the
1+3-covariant perturbation formalism to 1+4 dimensions would
require a decomposition of the 5D geometric quantities along a
timelike extension $u^A$ into the bulk of the brane 4-velocity
field $u^\mu$, and this remains to be done. The 1+3-covariant
perturbation formalism is incomplete until such a 5D extension is
performed. The metric-based approach does not have this drawback.

\subsection{Metric-based perturbations}

An alternative approach to brane-world cosmological perturbations
is an extension of the 4D metric-based gauge-invariant
theory~\cite{metp}. A review of this approach is given
in~\cite{bmw}. In an arbitrary gauge, and for a flat FRW
background, the perturbed metric has the form
\begin{equation}
\label{pertmetric} \delta\, ^\vu g_{AB} = \left[
\begin{array}{ccc|c}
-2N^2\psi & & A^2(\partial_i{\cal B}-S_i)& N\alpha \\ &&& \\
A^2(\partial_j{\cal B}-S_j) & & A^2\left\{2{\cal R} \delta_{ij} +
2\partial_i \partial_j{\cal C}
+ 2\partial_{(i}F_{j)}+f_{ij} \right\} &
A^2(\partial_i\beta-\chi_i)\\&&&\\
\hline &&& \\ N\alpha && A^2(\partial_j\beta-\chi_j) & 2\nu
\end{array}
\right] \,,
\end{equation}
where the background metric functions $A,N$ are given by
Eqs.~(\ref{gnm1}) and (\ref{gnm2}). The scalars $\psi,{\cal
R},{\cal C}, \alpha, \beta, \nu$ represent scalar perturbations.
The vectors $S_i,F_i$ and $\chi_i$ are transverse, so that they
represent 3D vector perturbations, and the tensor $f_{ij}$ is
transverse traceless, representing 3D tensor perturbations.

In the Gaussian normal gauge, the brane coordinate-position
remains fixed under perturbation,
 \be
{}^\vu ds^2=\left[ {g}^{(0)}_{\mu\nu}(x,y)+\delta
g_{\mu\nu}(x,y)\right]dx^\mu dx^\nu +dy^2 \,,
 \ee
where ${g}^{(0)}_{\mu\nu}$ is the background metric,
Eq.~(\ref{gnm}). In this gauge,
 \be
\alpha= \beta=\nu=\chi_i=0\,.
 \ee

In the 5D longitudinal gauge,
 \be
-{\cal B}+\dot{\cal C}=0= -\beta+{\cal C}'\,.
 \ee
In this gauge, and for an AdS$_5$ background, the metric
perturbation quantities can all be expressed in terms of a
``master variable" $\Omega$ which obeys a wave equation~\cite{mu}.
In the case of scalar perturbations, we have for example,
 \be
{\cal R}={1\over 6A}\left(\Omega''-{1\over N^2}\,\ddot{\Omega}
-{\Lambda_\vd \over 3}\,\Omega \right) \,,
 \ee
with similar expressions for the other quantities. All of the
metric perturbation quantities are determined once a solution is
found for the wave equation
 \be
\left({1\over NA^3}\,\dot\Omega\right)^{\displaystyle{\cdot}} +
\left({\Lambda_\vd \over 6}+{k^2\over A^2} \right) {N \over
A^3}\,\Omega = \left({N \over A^3}\, \Omega'\right)'\,.
 \ee

The junction conditions Eq.~(\ref{ext}) relate the off-brane
derivatives of metric perturbations to the matter perturbations:
 \be \label{metjun}
\partial_y\,\delta g_{\mu\nu}=-\kappa^2_\vd\left[\delta
T_{\mu\nu} +{1\over 3}\left\{\lambda- T^{(0)}\right\} \delta
g_{\mu\nu}- {1\over3} {g}^{(0)}_{\mu\nu}\delta T\right],
 \ee
 where
 \bea
&& \delta T^0{}_0 = -\delta\rho\,,~ \delta T^0{}_i=a^2 q_i\,,\\ &&
\delta T^i{}_j = \delta p\, \delta^i{}_j+\delta\pi^i{}_j\,.
 \eea
For scalar perturbations in the Gaussian normal gauge, this gives
 \bea
\partial_y \psi(x,0) &=& {\kappa^2_\vd\over6}(2\delta\rho+3\delta
p)\,,\\ \partial_y {\cal B}(x,0) &=& \kappa^2_\vd\delta p\,,\\
\partial_y {\cal C}(x,0) &=& -{\kappa^2_\vd\over2}\delta\pi\,,
\\\partial_y {\cal R} (x,0) &=&- {\kappa^2_\vd\over6}\delta\rho-
\partial_i\partial^i\,{\cal C}(x,0)\,,
 \eea
where $\delta\pi$ is the scalar potential for the matter
anisotropic stress,
 \be
\delta\pi_{ij}=\partial_i\partial_j
\delta\pi-{1\over3}\delta_{ij}\,
\partial_k\partial^k \delta\pi\,.
 \ee
The perturbed KK energy-momentum tensor on the brane is given by
 \bea
&& \delta {\cal E}^0{}_0 = \kappa^2\delta\cu\,,~ \delta {\cal
E}^0{}_i
=-\kappa^2 a^2 \cq_i\,,\\
&& \delta {\cal E}^i{}_j = -{\kappa^2 \over3}\delta\cu\,
\delta^i{}_j - \delta\pi^{{\cal E}i}{}_j\,.
 \eea

The evolution of the bulk metric perturbations is determined by
the perturbed 5D field equations in the vacuum bulk,
 \be
\delta\, {}^\vu G^A{}_B=0\,.
 \ee
Then the matter perturbations on the brane enter via the perturbed
junction conditions, Eq.~(\ref{metjun}).

For example, for scalar perturbations in Gaussian normal gauge,
 \bea
\delta\, {}^\vu G^y{}_i &=&\partial_i\left\{-\psi'+\left({A'\over
A}-{N'\over N}\right) \psi-2{\cal R}'\right. \nonumber \\
&&~~~~{} \left. -{A^2\over 2N^2}\left[\dot{\cal B}'+\left( 5{\dot
A\over A}-{\dot N \over N}\right){\cal B}'\right]\right\} \,.
 \eea
For tensor perturbations (in any gauge), the only nonzero
components of the perturbed Einstein tensor are
 \bea
\delta\, {}^\vu G^i{}_j &=&-{1\over2} \left\{- {1\over N^2}
\ddot{f}^i{}_j+f''^i{}_j-{k^2 \over A^2} f^i{}_j \right. \nonumber
\\
&&~~{} \left. +{1\over N^2}\left({\dot N \over N}-3{\dot A \over
A}\right)\dot{f}^i{}_j+ \left({N' \over N}+3{A' \over
A}\right)f'^i{}_j \right\}
 \,.\label{tensor}
 \eea

~\\

In the following, I will discuss various perturbation problems,
using either a 1+3-covariant or a metric-based approach.

\subsection{Density perturbations on large scales}

In the covariant approach, we define matter density and expansion
(velocity) perturbation scalars, as in 4D general relativity,
 \be
\Delta={a^2\over\rho}\D^2\rho\,,~Z=a^2\D^2\Theta\,.
 \ee
Then we can define dimensionless KK perturbation
scalars~\cite{m1},
 \be\label{kkpert}
{   U}={a^2\over\rho}\D^2{\cu}\,,~{   Q}={a\over\rho} \D^2
\cq\,,~{  \Pi }={1\over \rho}\D^2\cp\,,
 \ee
where the scalar potentials $\cq$ and $\cp$ are defined by
Eqs.~(\ref{q*}) and (\ref{p*}). The KK energy density (dark
radiation) produces a scalar fluctuation $U$ which is present even
if $\cu=0$ in the background, and which leads to a non-adiabatic
(or isocurvature) mode, even when the matter perturbations are
assumed adiabatic~\cite{gm}. We define the total effective
dimensionless entropy $S_{\rm tot}$ via
 \be
p_{\rm tot}\,S_{\rm tot}=a^2\D^2 p_{\rm tot}-c_{\rm tot}^2a^2
\D^2\rho_{\rm tot}\,,
 \ee
where $c_{\rm tot}^2=\dot{p}_{\rm tot}/\dot{\rho}_{\rm tot}$ is
given in Eq.~(\ref{vh2}). Then
\begin{eqnarray}
S_{\rm tot}={9\left[c_{\rm s}^2 - {1\over 3} +\left({2\over
3}+w+c_{\rm s}^2\right){ {\rho/\lambda}}\right]\over
[3(1+w)(1+\rho/\lambda)+4\cu/\rho] [3w+3(1+2w)\rho/
2\lambda+\cu/\rho]}\, \left[ {4\over 3}{\cu\over \rho}\,\Delta
-(1+w)U \right]\,.\label{ent}
\end{eqnarray}

If $\cu=0$ in the background, then $U$ is an isocurvature mode:
$S_{\rm tot}\propto (1+w)U$. This isocurvature mode is suppressed
during slow-roll inflation, when $1+w\approx 0$.

If $\cu\neq0$ in the background, then the weighted difference
between $U$ and $\Delta$ determines the isocurvature mode: $S_{\rm
tot}\propto (4\cu/ 3\rho)\Delta -(1+w)U$. At very high energies,
$\rho\gg\lambda$, the entropy is suppressed by the factor
$\lambda/\rho$.

The density perturbation equations on the brane are derived by
taking the spatial gradients of Eqs.~(\ref{ecl}), (\ref{nlc1}) and
(\ref{prl}), and using Eqs.~(\ref{momcl}) and (\ref{nlc2}). This
leads to~\cite{gm}
\begin{eqnarray}
\dot{\Delta} &=&3wH\Delta-(1+w)Z\,, \\  \dot{Z}
&=&-2HZ-\left({c_{\rm s}^2\over 1+w}\right)
\D^2\Delta-\kappa^2\rho {   U}-{{1\over2}}\kappa^2 \rho\left[1+
(4+3w){ {\rho\over\lambda}}- \left({4c_{\rm s}^2\over
1+w}\right){\cu\over\rho}\right] \Delta \,,\\  {   \dot{U}} &=&
(3w-1)H{   U} + \left({4c_{\rm s}^2\over 1+w}\right)\left({{\cu
}\over\rho}\right) H\Delta -\left({4{\cu }\over3\rho}\right)
Z-a\D^2{ Q}\,,\\  {   \dot{Q}} &=&(3w-1)H{ Q}-{1\over3a}{
U}-{{2\over3}} a{ \D^2\Pi}+{1\over3\mu}\left[ \left({4c_{\rm
s}^2\over 1+w}\right){{\cu}\over\rho}-3(1+w) {
{\rho\over\lambda}}\right]\Delta\,.
\end{eqnarray}
The KK anisotropic stress term $\Pi$ occurs only via its
Laplacian, ${ \D^2\Pi}$. If we can neglect this term on large
scales, then the system of density perturbation equations closes
on super-Hubble scales~\cite{m1}. An equivalent statement applies
to the large-scale curvature perturbations~\cite{lmsw}. KK effects
then introduce two new isocurvature modes on large scales
(associated with $U$ and $Q$), as well as modifying the evolution
of the adiabatic modes~\cite{gm,l1}.

Thus on large scales the system of brane equations is closed, and
we can determine the density perturbations without solving for the
bulk metric perturbations.

\begin{figure}[!bth]\label{sp}
\begin{center}
\includegraphics{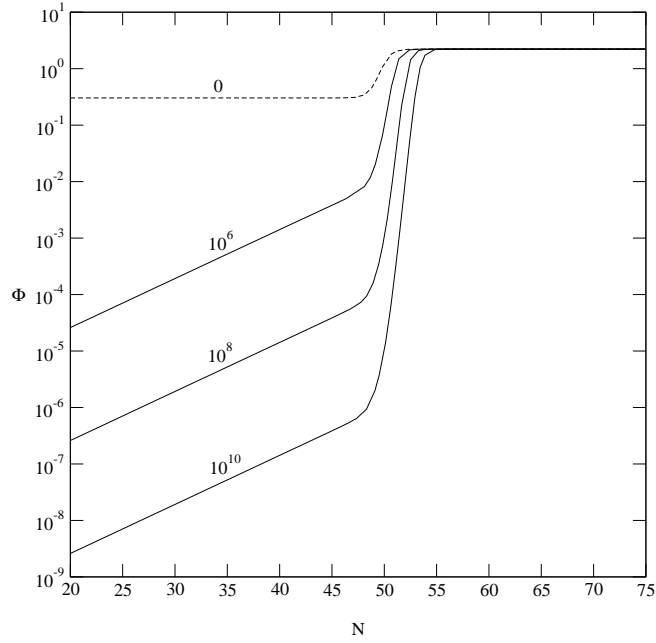}
\caption{The evolution of the covariant variable $\Phi$, defined
in Eq.~(\ref{cphi}) (and not to be confused with the Bardeen
potential), along a fundamental world-line. This is a mode that is
well beyond the Hubble horizon at $N =0$, about 50 e-folds before
inflation ends, and remains super-Hubble through the radiation
era. A smooth transition from inflation to radiation is modelled
by $w={1\over3}[(2- {3\over2}\epsilon)\tanh(N-50)-(1-
{3\over2}\epsilon)]$, where $ \epsilon$ is a small positive
parameter (chosen as $\epsilon=0.1$ in the plot). Labels on the
curves indicate the value of $\rho_0/\lambda$, so that the general
relativistic solution is the dashed curve ($\rho_0/\lambda=0$).
(From~\cite{gm}.) }
\end{center}
\end{figure}

We can simplify the system as follows. The 3-Ricci tensor defined
in Eq.~(\ref{gc2}) leads to a scalar covariant curvature
perturbation variable,
 \be
C\equiv a^4\D^2R^\perp = -4a^2HZ+2\kappa^2a^2\rho\left( 1+ {\rho
\over 2\lambda} \right)\Delta+ 2\kappa^2a^2 \rho U\,.
 \ee
It follows that $C$ is locally conserved (along $u^\mu$ flow
lines):
 \be
C=C_0\,,~ \dot{C}_0=0\,.
 \ee
We can further simplify the system of equations via the variable
 \be\label{cphi}
\Phi=\kappa^2a^2\rho \Delta\,.
 \ee
This should not be confused with the Bardeen metric perturbation
variable $\Phi_H$, although it is the covariant analogue of
$\Phi_H$ in the general relativity limit. In the brane-world,
high-energy and KK effects mean that $\Phi_H$ is a complicated
generalization of this expression~\cite{l1}, involving $\Pi$, but
the simple $\Phi$ above is still useful to simplify the system of
equations. Using these new variables, we find the closed system
for large-scale perturbations:
\begin{eqnarray}
\dot{\Phi}&=& -H\left[1+(1+w){\kappa^2\rho\over 2H^2}\left(1+
{\rho\over \lambda}\right)\right]\Phi  -
\left[(1+w){a^2\kappa^4\rho^2\over 2 H}\right]U +\left[(1+w)
{\kappa^2 \rho\over 4H}\right]C_0\,, \label{p1'}\\ \dot{U} &=&
-H\left[1-3w+{2\kappa^2{\cu}\over 3H^2}\right]U -{2 {\cu}\over 3
a^2 H\rho}\left[1+{\rho\over\lambda} - {6 c_{\rm s}^2H^2\over
(1+w)\kappa^2\rho}\right]\Phi+ \left[{{\cu}\over 3a^2H\rho}\right]
C_0\,. \label{p3}
\end{eqnarray}

If there is no dark radiation in the background, $\cu=0$, then
 \be
U=U_0\exp-\int(1-3w)dN\,,
 \ee
and the above system reduces to a single equation for $\Phi$. At
low energies, and for constant $w$, the non-decaying attractor is
the general relativity solution,
 \be\label{low}
\Phi_{\rm low} \approx {3(1+w) \over 2(5+3w)}\, C_0\,.
 \ee
At very high energies, for $w\geq -{1\over3}$,
 \be
\Phi_{\rm high} \to  {3\over 2}{\lambda \over \rho_0}(1+w) \left[
{C_0 \over 7+6w}-{2\tilde{U}_0 \over 5+6w} \right]\,,
 \ee
where $\tilde{U}_0=\kappa^2a_0^2\rho_0U_0$, so that the
isocurvature mode has an influence on $\Phi$. Initially, $\Phi$ is
suppressed by the factor $\lambda/\rho_0$, but then it grows,
eventually reaching the attractor value in Eq.~(\ref{low}). For
slow-roll inflation, when $1+w\sim\epsilon$, with $0<\epsilon \ll
1$ and $H^{-1}|\dot\epsilon|=|\epsilon'|\ll 1$,
 \be
\Phi_{\rm high} \sim {3\over 2}\epsilon {\lambda \over \rho_0}C_0
e^{3\epsilon N}\,,
 \ee
where $N=\ln(a/a_0)$, so that $\Phi$ has a growing-mode in the
early universe. This is different from general relativity, where
$\Phi$ is constant during slow-roll inflation. Thus more
amplification of $\Phi$ can be achieved than in general
relativity, as discussed above. This is illustrated for a toy
model of inflation-to-radiation in Fig.~8. The early (growing) and
late time (constant) attractor solutions are seen explicitly in
the plots.

\begin{figure}[!bth]\label{phirad}
\begin{center}
\includegraphics[height=4in, width=5in]{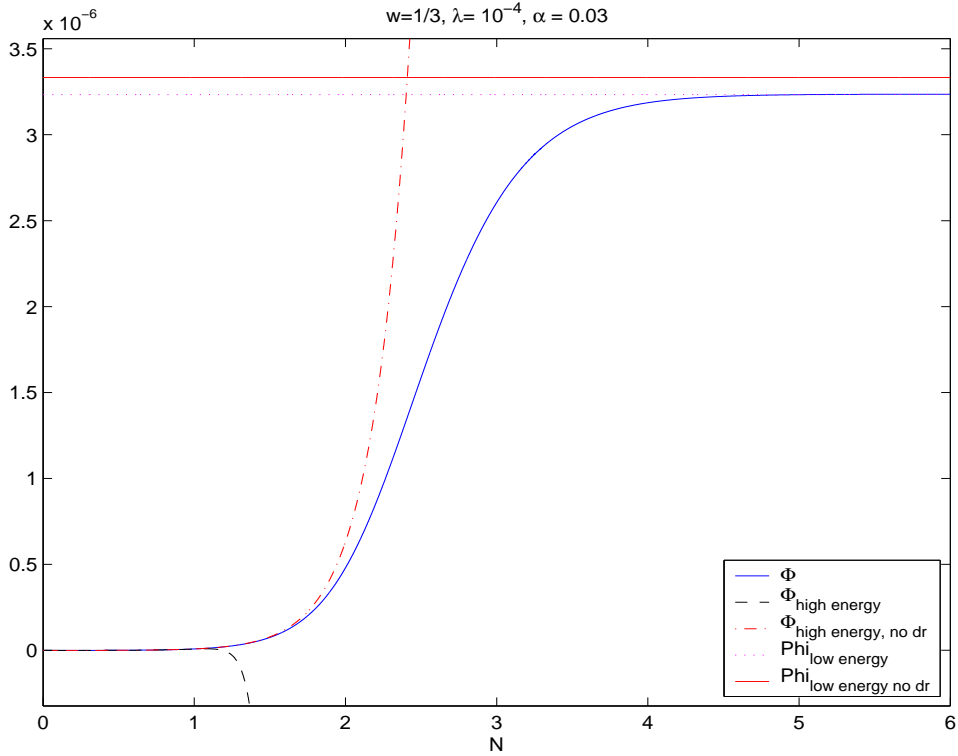}
\caption{The evolution of $\Phi$ in the radiation era, with dark
radiation present in the background. (From~\cite{ggm}.) }
\end{center}
\end{figure}

The presence of dark radiation in the background introduces new
features. In the radiation era ($w={1\over3}$), the non-decaying
low-energy attractor becomes~\cite{ggm}
 \bea
\Phi_{\rm low} & \approx & {C_0\over 3}(1-\alpha)\,,\\ \alpha &=&
{\cu\over \rho} \lesssim 0.05\,.
 \eea
The dark radiation serves to reduce the final value of $\Phi$,
leaving an imprint on $\Phi$, unlike the $\cu=0$ case,
Eq.~(\ref{low}). In the very high energy limit,
 \bea
\Phi_{\rm high} & \to & {\lambda\over \rho_0}\left[{2\over9} C_0
-{4\over7}\tilde{U}_0\right]+16\alpha \left({\lambda\over
\rho_0}\right)^2\left[{C_0\over273}-{4\tilde{U}_0 \over
539}\right] \,.
 \eea
Thus $\Phi$ is initially suppressed, then begins to grow, as in
the no-dark-radiation case, eventually reaching an attractor which
is less than the no-dark-radiation attractor. This is confirmed by
the numerical integration shown in Fig.~9.

\subsection{Curvature perturbations and the Sachs-Wolfe effect}

The curvature perturbation ${\cal R}$ on uniform density surfaces
is defined in Eq.~(\ref{pertmetric}). The associated
gauge-invariant quantity
 \be
\zeta={\cal R}+ {\delta\rho \over 3(\rho+p)}\,,
 \ee
may be defined for matter on the brane. Similarly, for the Weyl
``fluid" if $\cu\neq0$ in the background, the curvature
perturbation on hypersurfaces of uniform dark energy density is
 \be
\zeta_{\cal E}={\cal R}+ {\delta\rho_{\cal E} \over 4\cu}\,.
 \ee
On large scales, the perturbed dark energy conservation equation
is~\cite{lmsw}
 \be
(\delta\cu)^{\displaystyle{\cdot}}+4H\delta\cu+ 4\cu \dot{\cal
R}=0\,,
 \ee
which leads to
 \be
\dot{\zeta}_{\cal E}=0\,.
 \ee
For adiabatic matter perturbations, by the perturbed matter energy
conservation equation,
 \be
(\delta\rho)^{\displaystyle{\cdot}}+3H(\delta\rho+\delta p)+
3(\rho +p) \dot{\cal R}=0\,,
 \ee
we find
 \be
\dot\zeta=0\,.
 \ee
This is independent of brane-world modifications to the field
equations, since it depends on energy conservation only. For the
total, effective fluid, the curvature perturbation is defined as
follows~\cite{lmsw}: if $\cu\neq0$ in the background,
 \bea
{\zeta}_{\rm tot} &=& {\zeta}+\left[{4\cu \over 3(\rho+ p)(1+
\rho/\lambda)+4\cu}\right] \left(\zeta_{\cal E}-\zeta\right)\,,
 \eea
and if $\cu=0$ in the background,
 \bea
{\zeta}_{\rm tot} &=& {\zeta}+ {\delta\cu \over
3(\rho+p)(1+\rho/\lambda)} \\ \delta\cu &=& {\delta C_{\cal E}
\over a^4} \,,
 \eea
where $\delta C_{\cal E}$ is a constant. It follows that the
curvature perturbations on large scales, like the density
perturbations, can be found on the brane without solving for the
bulk metric perturbations.

Note that $\dot{\zeta}_{\rm \,tot}\neq 0$ even for adiabatic
matter perturbations; for example, if $\cu=0$ in the background,
 \be
\dot{\zeta}_{\rm \,tot}=H\left(c_{\rm tot}^2 -{1\over
3}\right){\delta\cu \over (\rho+p)(1+\rho/\lambda)}\,.
 \ee
The KK effects on the brane contribute a non-adiabatic mode,
although $\dot{\zeta}_{\rm \,tot}\to 0$ at low energies.

Although the density and curvature perturbations can be found on
super-Hubble scales, the Sachs-Wolfe effect requires
$\cp_{\mu\nu}$ in order to translate from density/ curvature to
metric perturbations. In the 4D longitudinal gauge of the metric
perturbation formalism, the gauge-invariant curvature and metric
perturbations on large scales are related by
 \bea
\zeta_{\rm tot} &=& {\cal R}-{H \over \dot H}\left( {\dot{\cal R}
\over H}-\psi\right) \,, \label{curv}
\\ {\cal R}+\psi &=& -\kappa^2a^2\delta \pi_{\cal E}\,,\label{metcurv}
 \eea
where the radiation anisotropic stress on large scales is
neglected, as in general relativity, and $\delta \pi_{\cal E}$ is
the scalar potential for $\cp_{\mu\nu}$, equivalent to the
covariant quantity $\Pi$ defined in Eq.~(\ref{kkpert}). In 4D
general relativity, the right hand side of Eq.~(\ref{metcurv}) is
zero. The (non-integrated) Sachs-Wolfe formula has the same form
as in general relativity:
 \be \label{swd}
{\delta T\over T}\Big|_{\rm now}=(\zeta_{\rm rad}+\psi-{\cal
R})|_{\rm dec}\,.
 \ee
The brane-world corrections to the general relativistic
Sachs-Wolfe effect are then given by~\cite{lmsw}
 \be\label{sachsw}
{\delta T\over T} = \left({\delta T\over T}\right)_{\rm gr}
-{8\over 3}\left({\rho_{\rm rad}\over \rho_{\rm
cdm}}\right)S_{\cal E}-\kappa^2a^2\delta \pi_{\cal E}
+{2\kappa^2\over a^{5/2}}\int da\,\, a^{7/2}\,\delta\pi_{\cal E}
\,,
 \ee
where $S_{\cal E}$ is the KK entropy perturbation (determined by
$\delta\cu$). The KK term $\delta\pi_{\cal E}$ cannot be
determined by the 4D brane equations, so that $\delta T/T$ cannot
be evaluated on large scales without solving the 5D equations.
[Equation~(\ref{sachsw}) has been generalized to a 2-brane model,
in which the radion makes a contribution to the Sachs-Wolfe
effect~\cite{ksw}.]

The presence of the KK (Weyl, dark) component has essentially two
possible effects.
\begin{itemize}
\item A contribution from the KK entropy perturbation $S_{\cal E}$
that is similar to an extra isocurvature contribution. \item The
KK anisotropic stress $\delta\pi_{\cal E}$ also contributes to the
CMB anisotropies.  In the absence of anisotropic stresses, the
curvature perturbation $\zeta_{\rm tot}$ would be sufficient to
determine the metric perturbation ${\cal R}$ and hence the
large-angle CMB anisotropies, via Eqs.~(\ref{curv}),
(\ref{metcurv}) and (\ref{swd}). However bulk gravitons generate
anisotropic stresses which, although they do not affect the
large-scale curvature perturbation $\zeta_{\rm tot}$, can affect
the relation between $\zeta_{\rm tot}$, ${\cal R}$ and $\psi$, and
hence can affect the CMB anisotropies at large angles.
\end{itemize}

A simple phenomenological approximation to $\delta\pi_{\cal E}$ on
large scales is discussed in~\cite{bm} and the Sachs-Wolfe effect
is estimated as
 \be
{\delta T\over T}\sim \left({\delta\pi_{\cal
E}\over\rho}\right)_{\!\rm in} \left({t_{\rm eq}\over t_{\rm
dec}}\right)^{2/3} \left[{\ln(t_{\rm in}/t_4)\over \ln (t_{\rm
eq}/t_4)} \right],
 \ee
where $t_4$ is the 4D Planck time and $t_{\rm in}$ is the time
when the KK anisotropic stress is induced on the brane, which is
expected to be of the order of the 5D Planck time.

A self-consistent approximation is developed in~\cite{koy}, using
the low-energy 2-brane approximation~\cite{sod} to find an
effective 4D form for ${\cal E}_{\mu\nu}$ and hence for
$\delta\pi_{\cal E}$. This is discussed below.

\subsection{ Vector perturbations}

The vorticity propagation equation on the brane is the same as in
general relativity,
 \be\label{vor}
\dot{\omega}_\mu +2H\omega_\mu =-{1\over2}\curl A_\mu \,.
 \ee
Taking the curl of the conservation equation~(\ref{c2}) (for the
case of a perfect fluid, $q_\mu =0=\pi_{\mu\nu}$), and using the
identity in Eq.~(\ref{vec}), one obtains
 \be
\curl A_\mu =-6Hc_{\rm s}^2\omega_\mu \,,
 \ee
as in general relativity, so that Eq.~(\ref{vor}) becomes
 \be
\dot{\omega}_\mu +\left(2-3c_{\rm s}^2\right) H\omega_\mu =0\,,
 \ee
which expresses the conservation of angular momentum. In general
relativity, vector perturbations vanish when the vorticity is
zero. By contrast, in brane-world cosmology, bulk KK effects can
source vector perturbations even in the absence of
vorticity~\cite{m2}. This can be seen via the divergence equation
for the magnetic part $H_{\mu\nu}$ of the 4D Weyl tensor on the
brane:
\begin{eqnarray}
\D^2\bar{H}_{\mu} = 2\kappa^2(\rho+p)\left[1+{
{\rho\over\lambda}}\right] \omega_\mu  + {4\over3} \kappa^2{\cu}
\omega_\mu  - {1\over2}\kappa^2 \curl{\bcq_\mu }\,,
\end{eqnarray}
where $H_{\mu\nu}=\D_{\langle\mu}\bar{H}_{\nu\rangle}$. Even when
$\omega_\mu =0$, there is a source for gravimagnetic terms on the
brane from the KK quantity $\curl{\bcq_\mu }$.

We define covariant dimensionless vector perturbation quantities
for the vorticity and the KK gravi-vector term:
 \be
\bar{\alpha}_\mu =a\,\omega_\mu \,,~ \bar{ \beta}_\mu
={a\over\rho}\curl{\bcq_\mu }\,.
 \ee
On large scales, we can find a closed system for these vector
perturbations on the brane~\cite{m2}:
\begin{eqnarray}
\dot{\bar{\alpha}}_\mu +\left(1-3c_{\rm s}^2\right)
H\bar{\alpha}_\mu &=&0 \,,\\ \dot{\bar{ \beta}}_\mu
+(1-3w)H\bar{\beta}_\mu  &=& {2\over 3}H\left[4\left( 3c_{\rm
s}^2-1\right) {{\cu}\over\rho}- 9(1+w)^2{
{\rho\over\lambda}}\right]\bar{\alpha}_\mu \,.\label{v2}
\end{eqnarray}
Thus we can solve for $\bar{\alpha}_\mu $ and $\bar{\beta}_\mu $
on super-Hubble scales, as for density perturbations. Vorticity in
the brane matter is a source for the KK vector perturbation
$\bar{\beta}_\mu $ on large scales. Vorticity decays unless the
matter is ultra-relativistic or stiffer ($w\geq {1\over3}$), and
this source term typically provides a decaying mode. There is
another pure KK mode, independent of vorticity, but this mode
decays like vorticity. For $w\equiv p/\rho=\,$const, the solutions
are
 \bea
\bar{\alpha}_\mu  &=& b_\mu  \left({a\over a_0}\right)^{3w-1}\,,\\
\bar{\beta}_\mu  &=& c_\mu \left({a\over a_0}\right)^{3w-1}+ b_\mu
\left[ {8\rho_{{\cal E}\,0} \over 3 \rho_0}\left({a\over
a_0}\right)^{2(3w-1)}+2(1+w){\rho_0\over \lambda} \left({a\over
a_0}\right)^{-4}\,\right]\,,
 \eea
where $\dot{b}_\mu =0=\dot{c}_\mu $.

Inflation will redshift away the vorticity and the KK mode.
Indeed, the massive KK vector modes are not excited during
slow-roll inflation~\cite{bmw}.

\subsection{ Tensor perturbations}

The covariant description of tensor modes on the brane is via the
shear, which satisfies the wave equation~\cite{m2}
\begin{eqnarray}
&& \D^2{\bar{\sigma}}_{\mu\nu}-\ddot{\bar{\sigma}}_{\mu\nu}
-5H\dot{\bar{\sigma}}_{\mu\nu}-\left[2\Lambda+{{1\over2}}
\kappa^2\left\{\rho-3p- (\rho+3p){ {\rho\over\lambda}}\right\}
\right]{\bar{\sigma}}_{\mu\nu}\nonumber\\ &&~~{}=-
{\kappa^2}\left( { \dot{\bar{\pi}}^{\cal E}_{\mu\nu}}+2H
{\bcp_{\mu\nu}} \right).
\end{eqnarray}
Unlike the density and vector perturbations, there is no closed
system on the brane for large scales. The KK anisotropic stress
$\bcp_{\mu\nu}$ is an unavoidable source for tensor modes on the
brane. Thus it is necessary to use the 5D metric-based formalism.
This is the subject of the next section.

\section{Gravitational wave perturbations in brane-world
cosmology}

The tensor perturbations are given by Eq.~(\ref{pertmetric}), i.e.
(for a flat background brane),
 \be
^\vu
ds^2=-N^2(t,y)dt^2+A^2(t,y)\left[\delta_{ij}+f_{ij}\right]dx^idx^j
+dy^2\,.
 \ee
The transverse traceless $f_{ij}$ satisfies Eq.~(\ref{tensor}),
which implies, on splitting $f_{ij}$ into Fourier modes with
amplitude $f(t,y)$,
 \be \label{tenwav}
{1\over N^2}\left[\ddot f+\left(3{\dot A \over A}- {\dot N \over
N} \right)\dot f \right]+{k^2 \over A^2} f = f''+ \left(3{ A'
\over A}+ { N' \over N} \right)f'\,.
 \ee
By the transverse traceless part of Eq.~(\ref{metjun}), the
boundary condition is
 \be
f'_{ij}\big|_{\rm brane} = \bar{\pi}_{ij}\,,
 \ee
where $\bar{\pi}_{ij}$ is the tensor part of the anisotropic
stress of matter-radiation on the brane.

The wave equation~(\ref{tenwav}) cannot be solved analytically
except if the background metric functions are separable, and this
only happens for maximally symmetric branes, i.e. branes with
constant Hubble rate $H_0$. This includes the RS case $H_0=0$,
already treated in Sec.~II. The cosmologically relevant case is
the de Sitter brane, $H_0>0$. We can calculate the spectrum of
gravitational waves generated during brane
inflation~\cite{lmw,grs,fk,gwtan}, if we approximate slow-roll
inflation by a succession of de Sitter phases. The metric for a de
Sitter brane, dS$_4$, in AdS$_5$ is given by
Eqs.~(\ref{gnm})--(\ref{gnm2}) with
 \bea
&& N(t,y)=n(y)\,,~~ A(t,y)=a(t) n(y)\,,\\&& n(y)= \cosh\mu
y-\left(1+{\rho_0\over\lambda}\right) \sinh\mu|y|\,,\\ &&
a(t)=a_0\exp{H_0(t-t_0)}\,,~~H_0^2={\kappa^2\over3}\rho_0\left(
1+{ \rho_0\over2\lambda}\right)\,,
 \eea
where $\mu=\ell^{-1}$.

\begin{figure}[!bth]\label{pot}
\begin{center}
\includegraphics{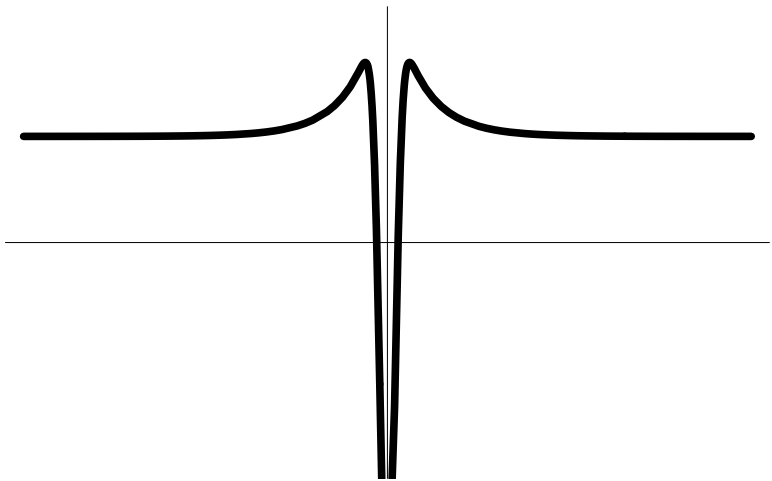}
\caption{Graviton ``volcano" potential around the dS$_4$ brane,
showing the mass gap (from~\cite{lan}).
 }
\end{center}
\end{figure}

The linearized wave equation~(\ref{tenwav}) is separable. As
before, we separate the amplitude as $f=\sum \varphi_m(t) f_m(y)$
where $m$ is the 4D mass, and this leads to:
 \bea
&& \ddot{\varphi}_m +3H_0\dot{\varphi}_m+\left[ m^2+{k^2\over
a^2}\right] \varphi_m =0\,, \label{varphieom}\\ && f_m''+4{n'\over
n}f_m'+ {m^2\over n^2}f_m = 0\,. \label{Heom}
 \eea
The general solutions for $m>0$ are
 \bea
\varphi_m(t) &=& \exp\left(-{3\over2}H_0t\right)B_\nu\left({
k\over H_0}e^{-H_0t}\right)\,,\label{vp2}\\ f_m(y) &=&
n(y)^{-3/2}L^\nu_{3/2} \left(\sqrt{1+{\mu^2\over
H_0^2}n(y)^2}\right)\,,
 \eea
where $B_\nu$ is a linear combination of Bessel functions,
$L^\nu_{3/2}$ is a linear combination of associated Legendre
functions, and
 \be
\nu=i\sqrt{{m^2\over H_0^2}-{9\over4}}\,.
 \ee

It is more useful to reformulate Eq.~(\ref{Heom}) as a Schr\"
odinger-type equation,
 \be \label{SE}
 {d^2\Psi_m\over dz^2} - V(z)\Psi_m =-m^2 \Psi_m \,,
\end{equation}
using the conformal coordinate
 \be
z=z_{\rm b} +\int_0^y {d{\tilde y}\over n({\tilde y})}\,,~~z_{\rm
b} = {1\over H_0}\sinh^{-1}\left({H_0\over \mu}\right)\,,
 \ee
and defining $\Psi_m\equiv n^{3/2}f_m$. The potential is given by
(see Fig.~10)
 \be
V(z)= {15H_0^2 \over 4\sinh^2(H_0 z)} + {{9\over4}}H_0^2 -
3\mu\left[1+{\rho_0\over\lambda}\right] \delta(z-z_{\rm b}) \,,
 \ee
where the last term incorporates the boundary condition at the
brane. The ``volcano" shape of the potential shows how the 5D
graviton is localized at the brane at low energies. (Note that
localization fails for an AdS$_4$ brane~\cite{nonloc}.)

The non-zero value of the Hubble parameter implies the existence
of a mass gap~\cite{gs},
 \be
\Delta m={3\over2}H_0\,,
 \ee
between the zero mode and the continuum of massive KK modes. This
result has been generalized: for dS$_4$ brane(s) with bulk scalar
field, a universal lower bound on the mass gap of the KK tower
is~\cite{fk}
 \be
\Delta m \geq \sqrt{3\over2}H_0\,.
 \ee
The massive modes decay during inflation, according to
Eq.~(\ref{vp2}), leaving only the zero mode, which is effectively
a 4D gravitational wave. The zero mode, satisfying the boundary
condition,
 \be
f'_0(x,0)=0\,,
 \ee
is given by
 \be
f_0= \sqrt{\mu}\,\,F\!\left({H_0/\mu}\right)\,,
 \ee
where the normalization condition
 \be
2 \int_{z_{\rm b}}^\infty |\Psi_0^2| dz=1\,,
 \ee
implies that the function $F$ is given by~\cite{lmw}
\begin{equation}
F\!\left(x\right) =\left\{ \sqrt{1+x^2} - x^2 \ln \left[ {1\over
x}+\sqrt{1+{1\over x^2}} \right] \right\}^{\!\!-1/2}
\!.\label{dtot}
\end{equation}
At low energies ($H_0\ll \mu$), we recover the general relativity
amplitude: $F\to 1$. At high energies, the amplitude is
considerably enhanced:
 \be
H_0\gg \mu ~\Rightarrow~ F\approx \sqrt{3H_0\over 2\mu}\,.
 \ee
The factor $F$ determines the modification of the gravitational
wave amplitude relative to the standard 4D result:
 \be
A_{\rm t}^2=\left[{8 \over M_\f^2}\left({H_0 \over 2\pi}\right)^2
\right]F^2(H_0/\mu).
 \ee
The modifying factor $F$ can also be interpreted as change in the
effective Planck mass~\cite{fk}.

This enhanced zero mode produced by brane inflation remains frozen
outside the Hubble radius, as in general relativity, but when it
re-enters the Hubble radius during radiation or matter domination,
it will no longer be separated from the massive modes, since $H$
will not be constant. Instead, massive modes will be excited
during re-entry. In other words, energy will be lost from the zero
mode as 5D gravitons are emitted into the bulk, i.e., as massive
modes are produced on the brane. A phenomenological model of the
damping of the zero mode due to 5D graviton emission is given
in~\cite{l2}. Self-consistent low-energy approximations to compute
this effect are developed in~\cite{kkt,elmw}.

\begin{figure}[!bth]\label{gw}
\begin{center}
\includegraphics[height=4in, width=4in]{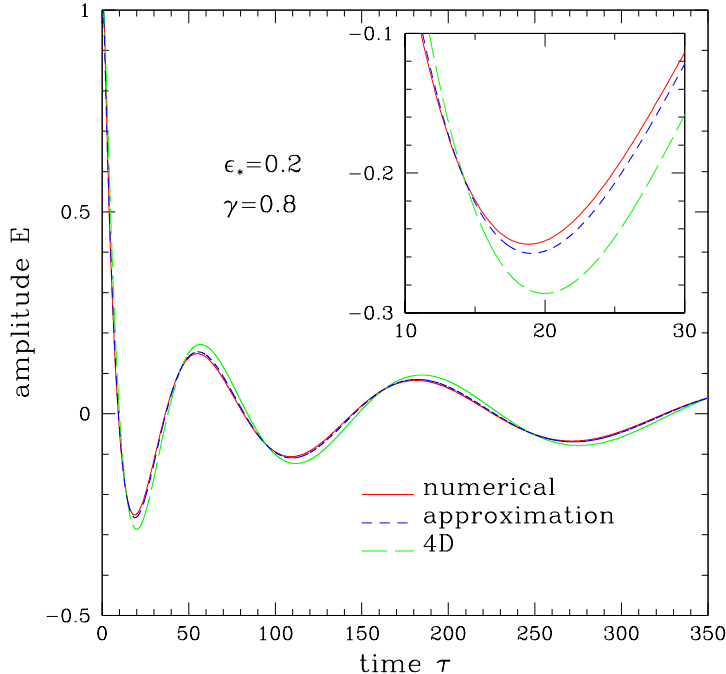}
\caption{Damping of brane-world gravity waves on horizon re-entry
due to massive mode generation. The solid curve is the numerical
solution, the short-dashed curve the low-energy approximation, and
the long-dashed curve the standard general relativity solution.
$\epsilon_{\cal E}=\rho_0/\lambda$ and $\gamma$ is a parameter
giving the location of the regulator brane. (From~\cite{kkt}.) }
\end{center}
\end{figure}

At zero order, the low-energy approximation is based on the
following~\cite{mbb}. In the radiation era, at low energy, the
background metric functions obey
 \be
A(t,y)\to a(t)e^{-\mu y}\,,~ N(t,y) \to e^{-\mu y}\,.
 \ee
To lowest order, the wave equation therefore separates, and the
mode functions can be found analytically~\cite{mbb}. The massive
modes in the bulk, $f_m(y)$, are the same as for a Minkowski
brane. On large scales, or at late times, the mode functions on
the brane are given in conformal time by
 \bea
\varphi^{(0)}_m(\eta)=\eta^{-1/2}B_{1/4}\left({ma_{\rm
h}^2\,\mu\over \sqrt{2}}\eta^2 \right)\,,
 \eea
where $a_{\rm h}$ marks the start of the low-energy regime
($\rho_{\rm h}=\lambda$) and $B_\nu$ denotes a linear combination
of Bessel functions. The massive modes decay on super-Hubble
scales, unlike the zero-mode. Expanding the wave equation in
$\rho_0/\lambda$, one arrives at the first order, where
mode-mixing arises. The massive modes $\varphi^{(1)}_m(\eta)$ on
sub-Hubble scales are sourced by the initial zero mode that is
re-entering the Hubble radius~\cite{elmw}:
 \be
\left({\partial^2 \over \partial \eta^2}-{\partial^2_\eta a \over
a}\right)a \varphi^{(1)}_m+ {k^2}a \varphi^{(1)}_m+ m^2 a^3
\varphi^{(1)}_m= -4{\rho_0 \over \lambda}I_{m0} {k^2
}a\varphi^{(0)}_0\,,
 \ee
where $I_{m0}$ is a transfer matrix coefficient. The numerical
integration of the equations~\cite{kkt} confirms the effect of
massive mode generation and consequent damping of the zero-mode,
as shown in Fig.~11.

\section{CMB anisotropies in brane-world cosmology}

For the CMB anisotropies, one needs to consider a multi-component
source. Linearizing the general nonlinear expressions for the
total effective energy-momentum tensor, we obtain
\begin{eqnarray}
\rho_{\text{tot}} &=& \rho\left(1 +\frac{\rho}{2\lambda} +
\frac{\rho_{\cal E}}{\rho} \right)\;, \\ \label{e:pressure1}
p_{\text{tot }}
&=& p  + \frac{\rho}{2\lambda} (2p+\rho)+\frac{\rho_{\cal E}}{3}\;, \\
q^{\text{tot }}_\mu  &=& q_\mu \left(1+  \frac{\rho}{\lambda}
\right) +q^{\cal E}_\mu \;, \\ \label{ep2} \pi^{\text{tot
}}_{\mu\nu} &=& \pi_{\mu\nu}
\left(1-\frac{\rho+3p}{2\lambda}\right)+\pi^{\cal E}_{\mu\nu}\;,
\end{eqnarray}
where
 \be
\rho=\sum_i\rho_{(i)}\,,~ p=\sum_ip_{(i)}\,,~q_\mu
=\sum_iq^{(i)}_\mu  \,,
 \ee
are the total matter-radiation density, pressure and momentum
density, and $\pi_{\mu\nu}$ is the photon anisotropic stress
(neglecting that of neutrinos, baryons and CDM).

The perturbation equations in the previous section form the basis
for an analysis of scalar and tensor CMB anisotropies in the
brane-world. The full system of equations on the brane, including
the Boltzmann equation for photons, has been given for
scalar~\cite{l1} and tensor~\cite{l2} perturbations. But the
systems are not closed, as discussed above, because of the
presence of the KK anisotropic stress  $\cp_{\mu\nu}$, which acts
a source term.

In the tight-coupling radiation era, the scalar perturbation
equations may be decoupled to give an equation for the
gravitational potential $\Phi$, defined by the electric part of
the brane Weyl tensor (not to be confused with ${\cal
E}_{\mu\nu}$):
 \be
E_{\mu\nu}=\D_{\langle\mu}\D_{\nu\rangle}\Phi\,.
 \ee
In general relativity, the equation in $\Phi$ has no source term,
but in the brane-world there is a source term made up of
$\cp_{\mu\nu}$ and its time-derivatives. At low energies
($\rho\ll\lambda$), and for a flat background ($K=0$), the
equation is~\cite{l1}
 \bea
&& 3x\Phi_k''+12\Phi_k'+x\Phi_k \nonumber\\ &&~~~{}={\mbox{const}
\over \lambda}\,\left[ \pi_k^{{\cal E}\prime\prime}-{1\over x}\,
{\pi_k^{{\cal E}\prime}} +\left({2\over x^3}- {3 \over x^2}+
{1\over x} \right) \cp_k\right],
 \eea
where $x=k/(aH)$, a prime denotes $d/dx$, and $\Phi_k$, $\cp_k$
are the Fourier modes of $\Phi$ and $\cp_{\mu\nu}$. In general
relativity the right hand side is zero, so that the equation may
be solved for $\Phi_k$, and then for the remaining perturbative
variables, which gives the basis for initializing CMB numerical
integrations. At high energies, earlier in the radiation era, the
decoupled equation is fourth order~\cite{l1}:
 \bea
&& 729 x^2\Phi_k''''+3888x\Phi_k'''+(1782+54x^2) \Phi_k'' +144x
 \Phi_k'+(90+x^2)\Phi_k  \nonumber\\ &&~~~{} ={\mbox{const}
}\,\left[243\left(\! {\cp_k\over\rho}\!
\right)^{\!\prime\prime\prime\prime} -{ 810 \over x}
\left(\!{\cp_k\over\rho}\!\right)^{\!\prime\prime\prime} +
{18(135+2x^2) \over x^2}
\left(\!{\cp_k\over\rho}\!\right)^{\!\prime\prime}\right.
\nonumber\\ &&~~~~~{} \left.- {30(162+x^2)\over x^3}
\left(\!{\cp_k\over\rho}\!\right)^{\!\prime} +
{x^4+30(162+x^2)\over x^4}
\left(\!{\cp_k\over\rho}\!\right)\right].
 \eea

The formalism and machinery are ready to compute the temperature
and polarization anisotropies in brane-world cosmology, once a
solution, or at least an approximation, is given for
$\cp_{\mu\nu}$. The resulting power spectra will reveal the nature
of the brane-world imprint on CMB anisotropies, and would in
principle provide a means of constraining or possibly falsifying
the brane-world models. Once this is achieved, the implications
for the fundamental underlying theory, i.e. M~theory, would need
to be explored.

However, the first step required is the solution for
$\cp_{\mu\nu}$. This solution will be of the form given in
Eq.~(\ref{e:soln}). Once ${\cal G}$ and $F_k$ are determined or
estimated, the numerical integration in Eq.~(\ref{e:soln}) can in
principle be incorporated into a modified version of a CMB
numerical code. The full solution in this form represents a
formidable problem, and one is led to look for approximations.

\subsection{The low-energy approximation}

The basic idea of the low-energy approximation~\cite{sod} is to
use a gradient expansion to exploit the fact that, during most of
the history of the universe, the curvature scale on the observable
brane is much greater than the curvature scale of the bulk
($\ell<1~$mm):
 \bea
&& L\sim |R_{\mu\nu\alpha\beta}|^{-1/2} \gg \ell \sim |\,^\vu
R_{ABCD}|^{-1/2} \nonumber \\ && \Rightarrow |\nabla_\mu|\sim
L^{-1} \ll |\partial_y|\sim \ell^{-1}\,. \label{grad}
 \eea
These conditions are equivalent to the low energy regime, since
$\ell^2\propto \lambda^{-1}$ and $|R_{\mu\nu\alpha\beta}|\sim
|T_{\mu\nu}|$:
 \be
{\ell^2 \over L^2} \sim {\rho \over \lambda} \ll 1\,.
 \ee
Using Eq.~(\ref{grad}) to neglect appropriate gradient terms in an
expansion in $\ell^2/L^2$, the low-energy equations can be solved.
However, two boundary conditions are needed to determine all
functions of integration. This is achieved by introducing a second
brane, as in the RS 2-brane scenario. This brane is to be thought
of either as a regulator brane, whose backreaction on the
observable brane is neglected (which will only be true for a
limited time), or as a shadow brane with physical fields, which
have a gravitational effect on the observable brane.

The background is given by low-energy FRW branes with tensions
$\pm\lambda$, proper times $t_\pm$, scale factors $a_\pm$, energy
densities $\rho_\pm$ and pressures $p_\pm$, and dark radiation
densities $\rho_{{\cal E}\,\pm}$. The physical distance between
the branes is $\ell \bar{d}(t)$, and
 \be
{d\over dt_-}=e^{\bar{d}}\,{d \over dt_+}\,,~
a_-=a_+e^{-{\bar{d}}}\,,~ H_-=e^{\bar{d}}\left(H_+- \dot
{\bar{d}}\right)\,,~ \rho_{{\cal E}\,-}=e^{4{\bar{d}}}\rho_{{\cal
E}\,+}\,.
 \ee
Then the background dynamics is given by (see~\cite{lang2} for the
general background, including the high-energy regime):
 \bea
&& H_\pm^2 = \pm{\kappa^2 \over 3} \left(\rho_\pm \pm\rho_{{\cal
E}\,\pm} \right)\,,
\\
&& \ddot {\bar{d}}+3H_+\dot {\bar{d}}-\dot{{\bar{d}}}^2 =
{\kappa^2 \over 6}\left[ \rho_+-3p_+ +e^{2{\bar{d}}}(\rho_--3p_-)
\right]\,.\label{dbar}
 \eea
The dark energy obeys $\rho_{{\cal E}\,+} =C/a_+^4$, where $C$ is
a constant. From now on, we drop the +-subscripts which refer to
the physical, observed quantities.

The perturbed metric on the observable (positive tension) brane is
described, in longitudinal gauge, by the metric perturbations
$\psi$ and ${\cal R}$, and the perturbed radion is $d= {\bar{d}}+
N$. The approximation for the KK (Weyl) energy-momentum tensor on
the observable brane is
 \bea
{\cal E}^{\mu}{}_{\nu}& =& \frac{2}{e^{2d}-1} \left[
-\frac{\kappa^2}{2}\left( T^{\mu}_{\nu}+e^{-2 d}T^{\mu}_{-~
\nu}\right) \right.\nonumber \\ &&\left.~~{} -
\nabla^{\mu}\nabla_{\nu} d + \delta^{\mu}_{\nu} \nabla^2 d
-\left\{ \nabla^{\mu}d \nabla_{\nu} d + \frac{1}{2}
\delta^{\mu}_{\nu} (\nabla d)^2 \right\} \right], \label{solE}
 \eea
and the field equations on the observable brane can be written in
scalar-tensor form as
\begin{eqnarray}
G^\mu{}_\nu & = & \frac{\kappa^2}{\chi}
T^\mu_{\nu}+\frac{\kappa^2(1-\chi)^2}{\chi} T^\mu_{-~\nu}
\nonumber \\
& &~{} +\frac{1}{\chi}\left(\nabla^\mu \nabla_\nu
\chi-\delta^\mu_\nu \nabla^2\chi \right)
+\frac{\omega(\chi)}{\chi^2}\left[\nabla^\mu \chi \nabla_\nu
\chi-\frac{1}{2}\delta^\mu_\nu (\nabla \chi)^2 \right],
\end{eqnarray}
where
 \be
\chi=1-e^{-2d}\,,~\omega(\chi)=\frac{3}{2}\frac{\chi}{1-\chi}\,.
 \ee

The perturbation equations can then be derived as generalizations
of the standard equations. For example, the $\delta G^0{}_0$
equation is~\cite{koyn}
\begin{eqnarray}
H^2 \psi &-& H \dot{\cal R}- \frac{1}{3}{ k^2 \over a^{2}} {\cal
R}
\nonumber\\
&=&  - \frac{1}{6} \kappa^2 \frac{e^{2 \bar d}}{e^{2 \bar d}-1}
\left(\delta \rho + e^{-4 \bar d} \delta \rho_-\right) +
\frac{2}{3} \kappa^2 \frac{e^{2 \bar d}}{e^{2 \bar d}-1} \cu\, N
\nonumber\\
&&{} - \frac{1}{e^{2\bar d}-1} \left[ \left(\dot{\bar d}-H \right)
\dot{N} + \left(\dot{\bar d}-H \right)^2 N -\dot{\bar d}^2 \psi +
2 H \dot{\bar d} \psi - \dot{\bar d} \dot{\cal R}- \frac{1}{3}{
k^2\over a^{2}} N \right].
\end{eqnarray}
The trace part of the perturbed field equation shows that the
radion perturbation determines the crucial quantity,
$\delta\pi_{\cal E}$:
 \be\label{rplusp}
{\cal R}+\psi = -{2 \over e^{2\bar d}-1}N=-\kappa^2a^2\delta
\pi_{\cal E}\,,
 \ee
where the last equality follows from Eq.~(\ref{metcurv}). The
radion perturbation itself satisfies the wave equation
\begin{eqnarray}
\ddot{N} &+& \left(3H -2 \dot{\bar d}\right) \dot{N} -\left(2
\dot{H}+4 H^2+2 \dot{\bar d}^2-6 H \dot{\bar d}-2 \ddot{\bar
d}\right)N
+ {k^2 \over a^{2}} N \nonumber\\
&&{} -\dot{\bar d} \dot{\psi}+3 \dot{\bar d} \dot{\cal R}+
\left(-2 \ddot{\bar d}-6 H \dot{\bar d}+2 \dot{\bar d}^2\right)
\psi \nonumber\\
&&{}= \frac{\kappa^2}{6} \left[ \delta \rho -3 \delta p + e^{-2
\bar d}(\delta \rho_--3 \delta p_-) \right].
\end{eqnarray}

A new set of variables $\varphi_\pm, E$ turns out be very
useful~\cite{ksw,koy}:
\begin{eqnarray}
{\cal R} &=&  -\varphi_+ - {a^{2}\over k^2} H \dot{E} +
\frac{1}{3} E\,,
\nonumber\\
\psi &=& - \varphi_+ - {a^{2}\over k^2} (\ddot{E}+ 2H \dot{E})\,,
\nonumber\\
N &=& \varphi_- - \varphi_+ -  {a^{2}\over k^2}\dot{\bar d}
\dot{E}\,.
\end{eqnarray}
Equation~(\ref{rplusp}) gives
\begin{equation}
\ddot{E}+ \left( 3 H + \frac{2 \dot{\bar d}}{e^{2 \bar d}-1}
\right) \dot{E} - \frac{1}{3} {k^2 \over a^2} E = -\frac{2 e^{2
\bar d}}{e^{2 \bar d}-1} {k^2 \over a^2} \left(\varphi_+ -
e^{-2\bar d} \varphi_-\right)\,.
\end{equation}
The variable $E$ determines the metric shear in the bulk, whereas
$\varphi_\pm$ give the brane displacements, in transverse
traceless gauge. The latter variables have a simple relation to
the curvature perturbations on large scales~\cite{ksw,koy}
(restoring the +-subscripts):
 \be
\zeta_{{\rm tot}\,\pm} =-\varphi_\pm +{H_\pm^2 \over \dot H_\pm}
\left({\dot{\varphi}_\pm \over H_\pm}+ \varphi_\pm \right)\,,
 \ee
where $\dot{f}_\pm\equiv df_\pm/dt_\pm$.

\subsection{The simplest model}

The simplest model is the one in which
 \be
\cu=0=\dot{\bar d}
 \ee
in the background, with $p_-/\rho_-=p/\rho$. The regulator brane
is assumed to be far enough away that its effects on the physical
brane can be neglected over the timescales of interest. By
Eq.~(\ref{dbar}), it follows that
 \be
\rho_-=-\rho e^{2\bar d}\,,
 \ee
i.e., the matter on the regulator brane must have fine-tuned and
negative energy density to prevent the regulator brane from moving
in the background. With these assumptions, and further assuming
adiabatic perturbations for the matter, there is only one
independent brane-world parameter, i.e., the parameter measuring
dark radiation fluctuations:
 \be
\delta C_{*}= {\delta\cu \over \rho_{\rm rad}}\,.
 \ee

\begin{figure}[bth]\label{koyfig}
\begin{center}
\includegraphics[height=4in, width=5in]{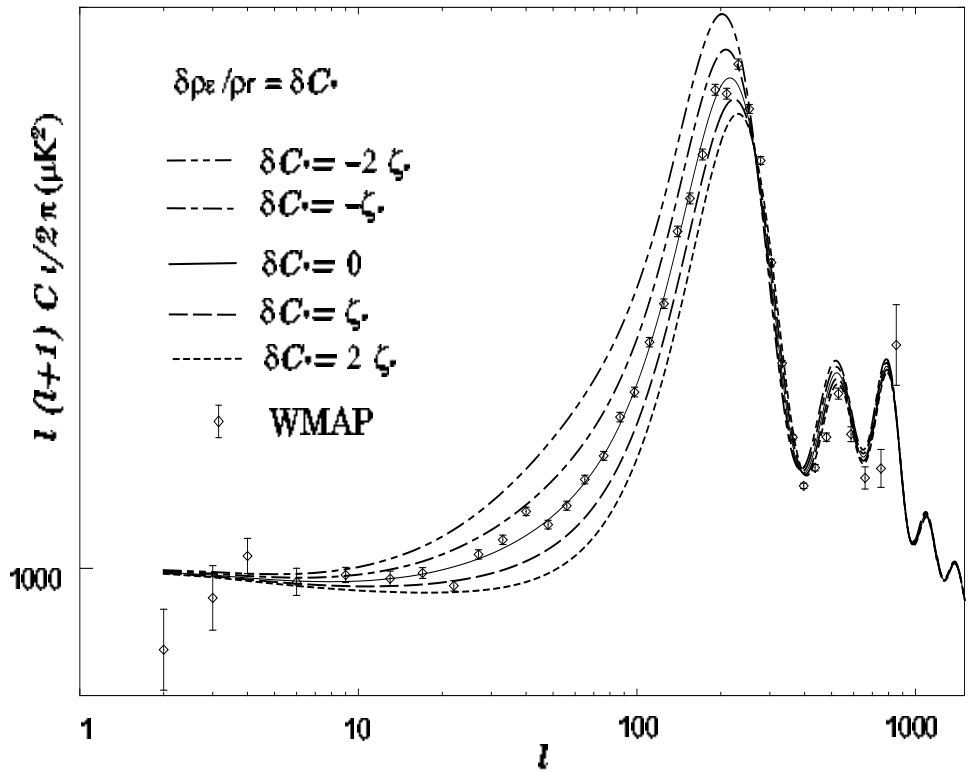}
\caption{The CMB power spectrum with brane-world effects, encoded
in the dark radiation fluctuation parameter $\delta C_*$ as a
proportion of the large-scale curvature perturbation for matter
(denoted $\zeta_*$ in the plot). (From~\cite{koy}.) }
\end{center}
\end{figure}

This assumption has a remarkable consequence on large scales: the
Weyl anisotropic stress $\delta\pi_{\cal E}$ terms in the
Sachs-Wolfe formula Eq.~(\ref{sachsw}) cancel the entropy
perturbation from dark radiation fluctuations, so that there is no
difference on the largest scales from the standard general
relativity power spectrum. On small scales, beyond the first
acoustic peak, the brane-world corrections are negligible. On
scales up to the first acoustic peak, brane-world effects can be
significant, changing the height and the location of the first
peak. These features are apparent in Fig.~12. However, it is not
clear to what extent these features are general brane-world
features (within the low-energy approximation), and to what extent
they are consequences of the simple assumptions imposed on the
background. Further work remains to be done.

A related low-energy approximation, using the moduli space
approximation, has been developed for certain 2-brane models with
bulk scalar field~\cite{rbbd}. The effective gravitational action
on the physical brane, in the Einstein frame, is
 \be\label{mod}
S_{\rm eff}={1\over 2\kappa^2}\int\, d^4x\sqrt{-g}\left[
R-{12\alpha^2 \over 1+2\alpha^2}(\partial \phi)^2-{6 \over
1+2\alpha^2} (\partial \chi)^2-V(\phi,\chi)\right],
 \ee
where $\alpha$ is a coupling constant, and $\phi$ and $\chi$ are
moduli fields (determined by the zero-mode of the bulk scalar
field and the radion). Figure~13 shows how the CMB anisotropies
are affected by the $\chi$-field.

\begin{figure}[!bth]\label{cmbb}
\begin{center}
\includegraphics[height=5.5in, width=4.in, angle=-90]{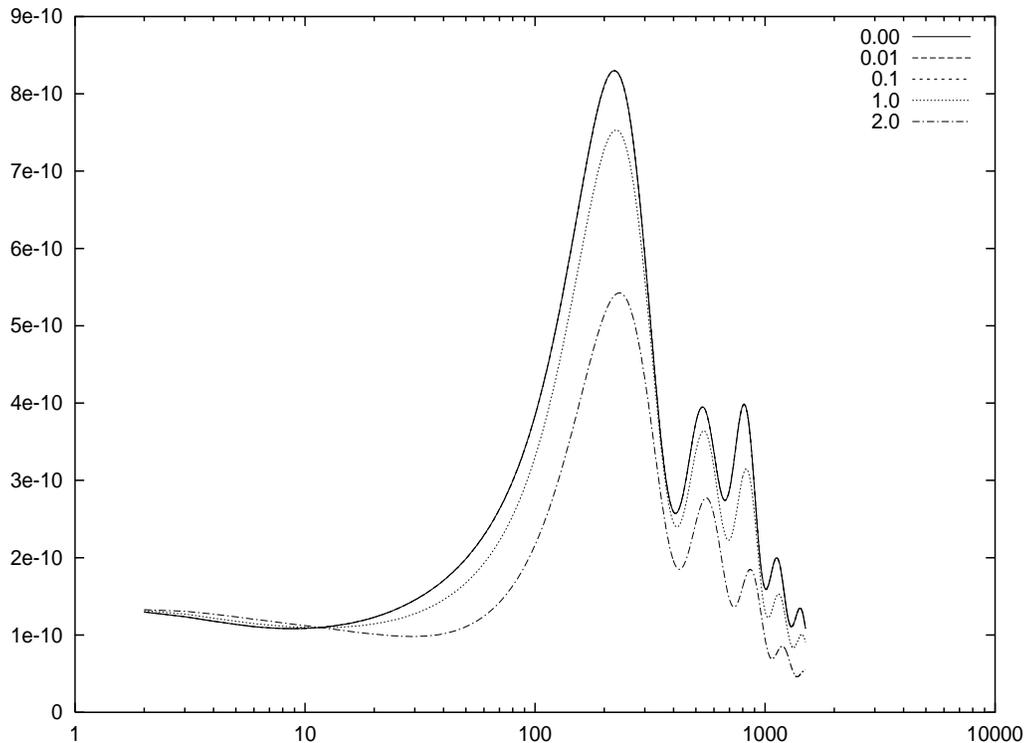}
\caption{The CMB power spectrum with brane-world moduli effects
from the field $\chi$ in Eq.~(\ref{mod}). The curves are labelled
with the initial value of $\chi$. (From~\cite{rbbd}.) }
\end{center}
\end{figure}

\section{Conclusion}

Simple brane-world models of RS type provide a rich phenomenology
for exploring some of the ideas that are emerging from M~theory.
The higher-dimensional degrees of freedom for the gravitational
field, and the confinement of standard model fields to the visible
brane, lead to a complex but fascinating interplay between
gravity, particle physics and geometry, that enlarges and enriches
general relativity in the direction of a quantum gravity theory.

This review has attempted to show some of the key features of
brane-world gravity from the perspective of astrophysics and
cosmology, emphasizing a geometric approach to dynamics and
perturbations. It has focused on 1-brane RS-type brane-worlds
which have some attractive features:
\begin{itemize}

\item they provide a simple 5D phenomenological realization of the
Horava-Witten supergravity solutions in the limit where the hidden
brane is removed to infinity and the moduli effects from the 6
further compact extra dimensions may be neglected;

\item they develop a new geometrical form of dimensional reduction
based on a strongly curved (rather than flat) extra dimension;

\item they provide a realization to lowest order of the AdS/CFT
correspondence;

\item they incorporate the self-gravity of the brane (via the
brane tension);

\item they lead to cosmological models whose background dynamics
are completely understood and reproduce general relativity results
with suitable restrictions on parameters.

\end{itemize}

The review has highlighted both the successes and the remaining
open problems of the RS models and their generalizations. The open
problems stem from a common basic difficulty, i.e., understanding
and solving for the gravitational interaction between the bulk and
the brane (which is nonlocal from the brane viewpoint). The key
open problems of relevance to astrophysics and cosmology are:

\begin{itemize}
\item to find the simplest realistic solution (or approximation to
it) for an astrophysical black hole on the brane, and settle the
questions about its staticity, Hawking radiation and horizon;

\item to develop realistic approximation schemes (building on
recent work~\cite{sod,koy,rbbd,kkt,elmw}) and manageable numerical
codes (building on~\cite{koy,rbbd,kkt,elmw}) to solve for the
cosmological perturbations on all scales, compute the CMB
anisotropies and large-scale structure, and impose observational
constraints from high-precision data.

\end{itemize}

The RS-type models are the simplest brane-worlds with curved extra
dimension that allow for a meaningful approach to astrophysics and
cosmology. One also needs to consider generalizations that attempt
to make these models more realistic, or that explore other aspects
of higher-dimensional gravity which are not probed by these simple
models. Two important types of generalization are the following.

\begin{itemize}

\item

{\bf The inclusion of dynamical interaction between the brane(s)
and a bulk scalar field,} so that the action is
 \bea
S &=& {1 \over 2\kappa_\vd^2} \int d^5x\sqrt{-{}^\vu
g}\left[{}^\vu R
-\kappa_5^2\partial_A\Phi\partial^A\Phi-2\Lambda_\vd(\Phi) \right]
\nonumber\\&&~~{} +\int_{\rm brane(s)}
d^4x\sqrt{-g}\left[-\lambda(\Phi)+{K\over \kappa_\vd^2}+L_{\rm
matter}\right]\,.
 \eea
(See~\cite{mw,sca,hs}.)

The scalar field could represent a bulk dilaton of the
gravitational sector, or a modulus field encoding the dynamical
influence on the effective 5D theory of an extra dimension other
than the large fifth dimension~\cite{ek5,rbbd,mod}. For two-brane
models, the brane separation introduces a new scalar degree of
freedom, the radion. For general potentials of the scalar field
which provide radion stabilization, 4D Einstein gravity is
recovered at low energies on either brane~\cite{2b}. (By contrast,
in the absence of a bulk scalar, low energy gravity is of
Brans-Dicke type~\cite{gt}.)

In particular, such models will allow some fundamental problems to
be addressed:
    \begin{itemize}
    \item
the hierarchy problem of particle physics;
    \item
an extra-dimensional mechanism for initiating inflation (or the
hot radiation era with super-Hubble correlations) via brane
interaction (building on the initial work
in~\cite{kss,ek,ek5,bub});
    \item
an extra-dimensional explanation for the dark energy (and possibly
also dark matter) puzzles: could dark energy or late-time
acceleration of the universe be a result of gravitational effects
on the visible brane of the shadow brane, mediated by the bulk
scalar field?
    \end{itemize}

\item {\bf The addition of stringy and quantum corrections to the
Einstein-Hilbert action,} including:
    \begin{itemize}
    \item
Higher-order curvature invariants, which arise in the AdS/CFT
correspondence as next-to-leading order corrections in the CFT.
The {\bf Gauss-Bonnet} combination in particular has unique
properties in 5D, giving field equations which are second-order in
the bulk metric (and linear in the second derivatives), and being
ghost-free. The action is
 \bea
S&=&{1  \over 2\kappa_\vd^2} \int d^5x\sqrt{-{}^\vu g}\left[{}^\vu
R-2\Lambda_\vd+\alpha\left\{ {}^\vu R^2-4{}^\vu R_{AB}{}^\vu
R^{AB}+{}^\vu R_{ABCD}{}^\vu R^{ABCD} \right\} \right]
\nonumber\\&&~~{} +\int_{\rm brane}
d^4x\sqrt{-g}\left[-\lambda+{K\over \kappa_\vd^2}+L_{\rm
matter}\right]\,,
 \eea
where $\alpha$ is the Gauss-Bonnet coupling constant, related to
the string scale. The cosmological dynamics of these brane-worlds
is investigated in~\cite{gbon}. In~\cite{bmsv} it is shown that
the black string solution of the form of Eq.~(\ref{bs1}) is ruled
out by the Gauss-Bonnet term. In this sense, the Gauss-Bonnet
correction removes an unstable and singular solution.

In the early universe, the Gauss-Bonnet corrections to the
Friedmann equation have the dominant form
 \be
H^2 \propto \rho^{2/3}\,,
 \ee
at the highest energies. If the Gauss-Bonnet term is a small
correction to the Einstein-Hilbert term, as may be expected if it
is the first of a series of higher-order corrections, then there
will be a regime of RS-dominance as the energy drops, when $H^2
\propto \rho^2$. Finally at energies well below the brane tension,
the general relativity behaviour is recovered.

    \item
Quantum field theory corrections arising from the coupling between
brane matter and bulk gravitons, leading to an induced 4D Ricci
term in the brane action. The original {\bf induced gravity}
brane-world~\cite{dgp} was put forward as an alternative to the RS
mechanism: the bulk is flat Minkowski 5D spacetime (and as a
consequence there is no normalizable zero-mode of the bulk
graviton), and there is no brane tension. Another viewpoint is to
see the induced-gravity term in the action as a correction to the
RS action:
 \bea
S&=& {1  \over 2\kappa_\vd^2} \int d^5x\sqrt{-{}^\vu
g}\left[{}^\vu R-2\Lambda_\vd \right] \nonumber\\&&~~{} +\int_{\rm
brane} d^4x\sqrt{-g}\left[\beta{R}-\lambda+{K\over
\kappa_\vd^2}+L_{\rm matter}\right]\,,
 \eea
where $\beta$ is a positive coupling constant. Unlike the other
brane-worlds discussed, these models lead to 5D behaviour on {\em
large scales} rather than small scales. The cosmological models
have been analyzed in~\cite{ind}. (Brane-world black holes with
induced gravity are investigated in~\cite{kpz}.)

The late-universe 5D behaviour of gravity can naturally produce a
late-time acceleration, even {\em without dark energy,} although
the fine-tuning problem is not evaded.

The effect of the induced-gravity correction at early times is to
restore the standard behaviour $H^2\propto \rho$ to lowest order
at the highest energies. As the energy drops, but is still above
the brane tension, there is an RS regime, $H^2\propto\rho^2$. In
the late universe at low energies, instead of recovering general
relativity, there are strong deviations from general relativity,
and late-time acceleration from 5D gravity effects (rather than
negative pressure energy) is typical.

Thus we have a striking result that both forms of correction to
the gravitational action, i.e., Gauss-Bonnet and induced gravity,
suppress the Randall-Sundrum type high-energy modifications to the
Friedmann equation when the energy reaches a critical level.
(Cosmologies with both induced-gravity and Gauss-Bonnet
corrections to the RS action are considered in~\cite{kmp}.)

    \end{itemize}
\end{itemize}
~\\

In summary, brane-world gravity opens up exciting prospects for
subjecting M~theory ideas to the increasingly stringent tests
provided by high-precision astronomical observations. At the same
time, brane-world models provide a rich arena for probing the
geometry and dynamics of the gravitational field and its
interaction with matter.

\[ \]
{\bf Acknowledgments}

I thank my many collaborators and friends for discussions and
sharing of ideas. My work is supported by PPARC.

\end{document}